\documentclass[aps, prd, twocolumn, superscriptaddress,showpacs, 10pt]{revtex4-1}
\usepackage{amsmath,amsfonts,amssymb,graphicx,color, bm, xcolor, enumerate}
\usepackage[colorlinks=true,citecolor=blue]{hyperref}
\usepackage[normalem]{ulem}
\usepackage{soul}
\setstcolor{red}

\usepackage{comment}

\newcommand{\el}{{\cal L}}
\newcommand{\de}{\partial}

\newcommand{\be}{\begin{equation}}
\newcommand{\ee}{\end{equation}}
\newcommand{\ba}{\begin{eqnarray}}
\newcommand{\ea}{\end{eqnarray}}

\newcommand{\tr}{\tilde{\rho}}

\newcommand{\tilteta}{\tilde{\theta}}

\usepackage{enumitem}

\begin{document}

\title{Binary superfluids: Low-energy properties and dissipative processes from   spontaneous emission of massive phonons}

\author{Silvia Trabucco}
\affiliation{Gran Sasso Science Institute, Viale Francesco Crispi, 7, 67100 L'Aquila, Italy.} 
\affiliation{INFN, Laboratori Nazionali del Gran Sasso, Via Giovanni Acitelli,
22, 67100 Assergi (AQ), Italy.}

\author{Luca Lepori}
\affiliation{Dipartimento di Scienze Matematiche, Fisiche e Informatiche, Universit\`a  di Parma, Parco Area delle Scienze, 53/A, I-43124 Parma, Italy.}
\affiliation{Gruppo Collegato di Parma, INFN-Sezione Milano-Bicocca, I-43124 Parma, Italy.}

\author{Maria Luisa Chiofalo}
\affiliation{Dipartimento di Fisica, Universit\`a di Pisa,
Polo Fibonacci, Largo Bruno Pontecorvo 3.}
\affiliation{INFN Sezione di Pisa,
Polo Fibonacci, Largo Bruno Pontecorvo 3, 56127 Pisa, Italy.}

\author{Massimo Mannarelli}
\affiliation{INFN, Laboratori Nazionali del Gran Sasso, Via Giovanni Acitelli,
22, 67100 Assergi (AQ), Italy.}
\affiliation{Gran Sasso Science Institute, Viale Francesco Crispi, 7, 67100 L'Aquila.}

\begin{abstract}  We discuss the low-energy properties of binary superfluids with density-dependent interactions.  Adding an intraspecies coupling  that induces an explicit  soft symmetry breaking, we determine the background pressure and we show that the low-energy spectrum consists of a massless Nambu-Goldstone boson and  a massive (pseudo) Nambu-Goldstone boson. 
 When the background velocities of the two superfluids are transonic, the system is characterized by two distinct acoustic horizons: the hydrodynamic analog of the black hole event horizon.  The Hawking-like  emission  occurring at these horizons produces an effective friction on the fluids. We compute the viscosity-to-entropy ratios close to the two acoustic horizons, finding that  the emission  of pseudo Nambu-Goldstone bosons violates the bound conjectured by Kovtun, Son, and Starinet.
\end{abstract}

\maketitle

\section{Introduction}
\label{sec:introduction}

The current generation of quantum technologies  opens up the possibility of investigating the physics of ultracompact objects, like  neutron stars or black holes~\cite{Barcelo:2005fc, Altman:2019vbv, Warszawski:2011vy, Warszawski:2012ns, Faccio:2013kpa, Poli:2023vyp, Bland:2024klj, YagoMalo24, Cipriani:2024bcc, Coviello:2024vht, Magierski:2024gvu, Tu:2024sam, Schutzhold:2025qna}, in analog experiments involving quantum fluids with a tunable degree of correlations. Concerning black holes, key aspects are the dissipative effects close to horizons, as characterized by the shear viscosity-to-entropy ratio $\eta/s$~\cite{Liberati:2005pr,Adams_2012}. This quantity is known to depend on the degree of correlations in the fluid and reaches a minimum in correspondence of a critical point~\cite{Adams_2012}. Interestingly enough, this minimum is of the order of $\hbar/k_B$ in strongly correlated quantum fluids, $k_B$ being the Boltzmann constant, with a value that is similar in quite diverse systems, as for instance the quark-gluon plasma and  ultracold Fermi gases, irrespective of their very different microscopic details. The lower bound $\eta/s =1/4\pi$ in units of $\hbar/k_B$, known from Kovtun, Son, and Starinet (KSS)~\cite{Kovtun:2004de}, has been derived in different frameworks, i.e. gravity-dual theories~\cite{Cremonini}, analogue gravity~\cite{Barcelo:2005fc}, kinetic theories~\cite{manuel2010,LuisaChiofalo:2022ykx}, and more generally from Kubo relations~\cite{Adams_2012,Barcelo:2005fc,Cremonini}. While the KSS bound is seen to hold for a remarkably large class of theories, specific situations show that it is nonuniversal~\cite{Adams_2012, Cherman:2007fj}. These include translationally invariant non-Fermi liquids in two dimensions~\cite{Ge}, the presence of higher curvature corrections in the dual gravitational theory and massive gravity in a
higher derivative theory, such as the Gauss–Bonnet theory~\cite{Sadeghi}. The latter introduces a lower bound $\eta/s_{GB} =(1/4\pi)(16/25)$, which in fact avoids violations of causality and of energy positivity~\cite{Adams_2012}. Overall, lower bounds $\eta/s$ establish that the low-shear viscosity systems, i.e. nearly perfect fluids, are also strongly correlated, bearing important information to fully exploit the analogy between black hole horizons and quantum fluids. Whether all theories predict that one universal $\eta/s$ bound exists, remains a relevant open question. It is therefore especially interesting to design a theoretical framework, amenable to realistic experimental implementations, to explore the degree of KSS-bound violation under the tuning of specific external parameters. 

Multicomponent quantum gases offer a versatile platform for investigating a wide range of phenomena, for instance due to the possibility to manage multiple levels simultaneously, and to the high tunability of both intraspecies and interspecies interactions.  These features have been shown to allow access to novel phases of matter\,\cite{CW, recati2022, Baroni_2024},  possibly supporting nontrivial topological defects\,\cite{Son:2001td, Kevrekidis_2004, Bakkali_Hassani_2021, Romero-Ros:2023mgp, Hamner_2011, Farolfi_2020,Richaud_2023}, as well as self-bound quantum droplets\,\cite{petrov2015, Semeghini_2018}. 
The first experimental realization of two overlapping quantum boson mixtures was achieved by sympathetic cooling of two integer spin states of $^{87}$Rb atoms\,\cite{Myatt:1997zz}. Subsequently, binary superfluids of fermions, of bosons as well as of a mixture of  fermion and  boson gases  have been explored to investigate various  relevant phenomena for elementary physics\,\cite{Tuoriniemi2002,PhysRevB.85.134529,  annabook,fallanibook,2014Sci...345.1035F,lvm2015, Lepori:2019vec}.
Among them, we mention the Andreev-Bashkin effect\,\cite{Nespolo_2017},  analog black holes and cosmology\,\cite{Fischer:2004bf, Visser:2005ss, Liberati_2006, berti2024}, and bubble formation in first-order phase transitions\,\cite{Zenesini:2023afv}. The possible phases of multicomponent quantum gases are also of relevance to characterize the interior of compact stars~\cite{YagoMalo24}. Inside them, superfluid neutrons should coexist with superconducting protons, moreover, if deconfined quark matter is present,  baryonic superfluidity may be realized along with  color superconductivity\,\cite{Shapiro:1983du, Alford:1998mk, Bedaque:2001je, Kaplan:2001qk, Alford:2007xm, Anglani:2013gfu}.

In this work, we  focus on a binary mixture of weakly interacting bosons  at vanishing temperature.
In each component,  superfluidity arises when the global symmetry associated with the conservation of number of particles is  spontaneously broken\,\cite{annett,pethick},  resulting in the presence of massless Goldstone bosons in the  low-energy spectrum, i.e.  the phonons. When an intraspecies Rabi coupling is added, such symmetry is explicitly broken\,\cite{Son:2001td, lepori2018} and the low-energy spectrum of the binary mixture is modified. If the Rabi coupling has a subleading effect, the low-energy spectrum   consists of  both a massless and a massive phonon\,\cite{ PhysRevLett.80.1130, PhysRevLett.80.1134}. These degrees of freedom may be interpreted as density waves\,\cite{Hall:1998zz}, and they can be described as massless and massive scalar particles propagating on top of an emergent acoustic metric\,\cite{Unruh:1980cg, Bilic:1999sq, Visser:2010xv, Brout:1995rd, Barcelo:2005fc, Volovik:2003ga}.  
This system is particularly interesting when the flow of one or both species is transonic, providing the hydrodynamic analog of black holes\,\cite{Fischer:2004bf, Liberati:2005pr,   berti2024}. Similarly to astrophysical black holes, Hawking-like emission of phonons occurs at the acoustic horizon, imprinting a signature in the density-density correlation functions\,\cite{Carusotto, Steinhauer, berti2024}. This spontaneous phonon emission is an irreversible process that occurs at the expense of the fluid kinetic energy, and can be described as effective shear and bulk  viscosities of the superfluid\,\cite{LuisaChiofalo:2022ykx}.

Within this framework, we compute both the shear-to-entropy density ratio and the bulk-to-entropy density ratio associated to the spontaneous emission at the acoustic horizon of binary transonic superfluid, extending the previous results for a single species\,\cite{LuisaChiofalo:2022ykx}.  We show how the saturation of the bound occurs for the spontaneous emission of massless phonons, while the bound is violated when a massive phonon is emitted. In fact, in the latter condition, the corresponding pressure and energy density decrease, reducing  the effective shear and bulk viscosities with respect to the massless case.
These findings are shown in the nonrelativistic limit and are directly applicable to homonuclear boson mixtures of alkali atoms with coupled hyperfine states\,\cite{Myatt:1997zz, Semeghini_2018}. 

This paper is organized as follows. In Sec.\,\ref{sec:onebec}, we  briefly review the analogue gravity set-up with a single-species  boson gas. In Sec.\,\ref{sec:two_fluids}, we describe the general features of binary boson systems, while in   Sec.\,\ref{sec:pressure} we  determine the ground-state pressure.   The  low-energy field theory of binary superfluids is developed in Sec.\,\ref{sec:fluctuations}. The Hawking-like  emission of massive phonons is studied in Sec.\,\ref{sec:massive:emission} and in Sec.\,\ref{sec:viscosity} we evaluate the resulting effective viscosity coefficients. We draw our conclusions in Sec.\,\ref{sec:conclusions}. Two distinct and general derivations of the effective low-energy Lagrangian are  given in the Appendix\,\ref{sec:appendix}. \\
Unless otherwise stated, we use natural units $\hbar = c = k_B = 1$.

\section{Effective theory of a single ultracold boson fluid}
\label{sec:onebec}
As a first step, we briefly recap some useful facts regarding  dilute single-species boson gas with a  $U(1)$  global symmetry.
In particular, we show how the acoustic metric emerges in the low-energy effective theory.

\subsection{General considerations}
\label{sec:onebecgeneral}
 We consider a complex scalar field, $\Phi$,  with nonvanishing chemical potential and  local interactions that do not explicitly break the $U(1)$ global symmetry associated to the conserved number of particles. We assume that the bulk temperature of the system is negligible.  For definiteness, we examine a simple model  with contact self-interaction; for a more general system see the Appendix\,\ref{sec:appendix}. 
The corresponding Lagrangian density, hereafter called the Lagrangian, is given by
\be
\label{eq:lag0}
{\cal L} = (D_\nu\Phi)^\ast D ^\nu \Phi - m^2 |\Phi|^2 -  U |\Phi|^4 \,,
\ee
where  $m$ is the mass of the bosons,  $0<U \ll1$ a perturbative dimensionless interaction constant,  and 
\be 
\label{eq:covder}
D_\nu  = \de_\nu + i \, \frac{\mu}{\gamma} \, v_\nu\,,
\ee
is the  covariant derivative,  depending on the medium properties, see for instance~\cite{son2002} for more details. We identify $\mu$ with the boson chemical potential, while $v_\nu = \gamma(1, \bm v)$ is the fluid four-velocity:  $\gamma$ is the Lorentz factor, such that $v^\mu v^\nu \eta_{\mu\nu}=1$, with $\eta_{\mu\nu}= \text{diag}(1,-1,-1,-1)$ the Minkowski metric. The nonrelativistic limit of Eq.\,\eqref{eq:lag0} is the  Gross-Pitaevskii 
(Hartree-Fock)  Lagrangian describing  a diluted and weakly interacting boson gas at vanishing temperature\,\cite{pethick,Dalfovo:1999zz}. 
The excitation spectrum  over the superfluid ground state can be calculated linearizing the Gross-Pitaevskii  Lagrangian, this approach being equivalent to the Bogolyubov formalism\,\cite{pitstring,recati2022}. 

Expanding the covariant derivative in Eq.\,\eqref{eq:covder}, we  have
\begin{align}
\label{eq:lag1}
{\cal L} =&\partial_\nu \Phi^*\partial^\nu \Phi - i\frac{\mu}{\gamma} v_\nu(\Phi^* \partial^\nu \Phi - \Phi \partial^\nu \Phi^*)  \nonumber \\ &- \left(m^2-\frac{\mu^2}{\gamma^2}\right) |\Phi|^2  -  U |\Phi|^4\,,
\end{align}
showing that the chemical potential gives two distinct contributions to the Lagrangian:  an explicit Lorentz symmetry breaking, due to the  term linear in the chemical potential, and the effective mass shift  $\mu^2/\gamma^2$.  
When  $\mu = \gamma \, m$ the effective mass of the scalar field  vanishes, signaling a spontaneous symmetry breaking.  For  $\mu > \gamma m$, the  $U(1)$ global symmetry is indeed spontaneously broken, and the minimum of the  potential 
\be
\label{eq:potential}
V= \left(m^2-\frac{\mu^2}{\gamma^2}\right) |\Phi|^2 + U |\Phi|^4\,,
\ee
is 
\be 
\label{eq:minimum}
 |\Phi|^2  = \frac{\mu ^2 -m ^2 \gamma^2}{2 U \gamma^2}\,.
 \ee
The fact that the potential depends only on   $|\Phi|$ and not on its phase,  invites us to use the Madelung representation 
 \be 
 \label{eq:madelung}
\Phi= \frac{\rho}{\sqrt{2}} e^{i\theta},
\ee
where the two real scalar fields $\rho$ and  $\theta$  are called the radial and phase fields, respectively. In the Madelung representation the Lagrangian in Eq.\,\eqref{eq:lag1} reads as
 \begin{align}
{\cal L}=& \frac{1}{2}\de _\nu \rho \de ^\nu \rho +\frac{1}{2}\rho ^2 \de _\nu \theta \de ^\nu \theta + \frac{ \rho^2 \mu }{\gamma}  v_\nu\de^\nu \theta \nonumber \\ &-\frac{m^2 \gamma^2 -\mu^2}{2 \gamma^2}  \rho  ^2   - \frac{U}{4}\rho^4  \,,
 \label{eq:lagr}
\end{align}
where the radial field is massive and with quartic self-interaction. On the other hand, the $\theta$ field is a phase, thus it only  appears in derivative terms. Since $\theta$ is a cyclic variable, the current
\be
\label{eq:j_mu_onebec}
J_\nu=\frac{\delta {\cal L}}{\delta \partial^\nu \theta  } = \rho^2 \left(\frac{\mu}{\gamma} v_\nu + \partial_\nu\theta \right)\,,
\ee
is conserved. In the ground state, the phase is homogeneous; therefore 
\be\label{eq:j_mu}
J_\nu = \rho^2 \frac{\mu}{\gamma} v_\nu= n v_\nu\,,
\ee
where 
\be\label{eq:n}
n= \frac{\mu}{ \gamma} \rho^2 \,,
\ee
is the number density.  From Eq.\,\eqref{eq:j_mu_onebec}, it is clear that we can introduce the chemical potential and the fluid velocity in an alternative way. If we  replace in Eq.\,\eqref{eq:lag0}
the covariant derivative with the standard derivative,  then in the expression of the current, see Eq.\,\eqref{eq:j_mu_onebec},  the first term in the round brackets vanishes. Then, we can  define a mean field value of the phase field  in such a way that 
\be
\label{eq:vnu}
\de_\nu\theta = \frac{\mu}{\gamma} v_\nu \, ,
\ee 
and thus the mean field value of the current is  equal to  Eq.\,\eqref{eq:j_mu}. This procedure implies that  the mean field value of the phase field  is equal to ${\mu} v_\nu x^\nu/\gamma$.
One can see that this is equivalent to a gauge transformation:  we can replace the covariant derivative in Eq.\,\eqref{eq:lag0} with a standard derivative and then  by the gauge transformation
\be
\Phi \to \Phi \, e^{i \frac{\mu}{\gamma} v_\nu x^\nu}\,,
\ee  
we obtain the  same Lagrangian reported in Eq.\,\eqref{eq:lag1}.  This procedure has the merit of treating the radial and the phase field on the same footing: the background medium properties are taken into account  by appropriate mean field values of $\rho$ and $\theta$. The  former is related to the  number density, while the latter to the chemical potential and the fluid four-velocity.\\ 
The radial field can be viewed as a variational parameter:  its mean field value, see Eq.\,\eqref{eq:minimum},
\be\label{eq:rho_mean_field}
\rho^2=\frac{\mu^2 -m^2 \gamma^2}{U \gamma^2} \quad \text{for}\,\, \mu \geq \gamma m\,,
\ee
 corresponds to the solution of the stationary condition  that minimizes the potential in Eq.\,\eqref{eq:lag1}. This is equivalent to maximizing the ground state Lagrangian. Since the grand potential per unit volume is given by
 \be 
 \Omega = -\frac{1}{{\rm V}}\int d^4 x {\cal L}\,,
 \ee 
where ${\rm V}$ is the total volume, the configuration that maximizes the Lagrangian minimizes the thermodynamic potential.  Given that $P = - \Omega $,  substitution of  
Eq.\,\eqref{eq:rho_mean_field} in Eq.\,\eqref{eq:lagr} yields the quantum pressure of the system: 
\be
\label{eq:pressure}
P = \frac{(\mu^2 - m^2 \gamma^2)^2}{4 U \gamma^4}\, .
\ee
Notice that this includes only the contribution of condensation and neglects    temperature as well as quantum fluctuations. The velocity dependence,  determined by the Lorentz factor,  gives the relativistic analog of the Bernoulli equation.
Indeed,  we have that
\be
P + \frac{\mu n v^2}{2} = \frac{(\mu^2 - m^2)^2}{4 U} + {\cal O}(v^4)\, ,
\ee
where the number density is defined in Eq.\,\eqref{eq:n}. We remark that 
  $\theta$ is not a variational parameter: its derivative  determines the velocity of the fluid. If we were to take it as a variational parameter, we should have considered the value that maximizes the pressure, that is the value for which the velocity vanishes\,\cite{Cipriani:2024bcc}.
    This approach is however incorrect, because the fluid current $J_\nu$ in Eq.\,\eqref{eq:j_mu_onebec}  is conserved. On the other hand, it indicates  that the addition of dissipative interactions in the Lagrangian in Eq.\,\eqref{eq:lagr} would have the effect of changing $\theta$ so that the velocity of the fluid is reduced and thus the pressure maximized.

\subsection{Effective low-energy theory}
\label{sec:general}

We now determine the effective Lagrangian for the low-energy excitations of the superfluid.
 We separate both the radial and phase fluctuations, $\tilde{\rho}$ and $\tilde{\theta}$, from their mean field solutions as
\be 
\rho \to \rho + \tr \,, \quad
\theta \to \theta + \tilteta \,,
\label{eq:rhoexp}
\ee
 where $\rho$ and $\theta$ now indicate the solutions of the classical equations of motion, corresponding to a stationary point of the action.
 \\
 Expanding around the mean field   we have that
 \begin{align}
{\cal L} =&  {\cal L}_0(\rho, \partial_\mu \rho,  \partial_\mu \theta) + \tr \left.\left(\frac{\delta \el}{\delta \rho} - \partial_\mu \frac{\delta \el}{\delta \partial_\mu\rho} \right)\right\vert_{\rho,\theta}\, \nonumber \\ &+\delta \partial_\mu\tilteta\left.\frac{\delta \el}{\delta \partial_\mu\theta}\right\vert_{\rho,\theta} + \el_2+ \dots \, ,
 \end{align}
 where the dots indicate  neglected cubic terms and higher, as well as surface terms. The term $ {\cal L}_0(\rho, \partial_\mu \rho,  \partial_\mu \theta) $ is the  background pressure previously evaluated in Eq.\,\eqref{eq:pressure}. The linear term vanishes, because we are evaluating it at the stationary point of the action.  The quadratic part of the Lagrangian reads as
\begin{align}
\label{eq:L2_4}
\el_2 =& \frac{1}2 \partial_\nu\tr \partial^\nu\tr-\frac{\tilde m^2}2 \tr^2 + V^\nu \tr \partial_\nu \tilteta + \frac{\rho^2}2 \partial_\nu \tilteta \partial^\nu \tilteta\,,
\end{align}
where 
\be\label{eq:mt} \tilde m = \sqrt{2 U \rho^2}\,,
\ee
is the radial field  mass. The phase fluctuation is  instead massless, and is the Nambu Goldstone boson (NGB) associated to the spontaneous breaking of the $U(1)$ symmetry. Hereafter, we will call it phonon. The vector coupling  between $\tilde \rho$ and $\de_\nu \tilteta$
\be 
V^{\nu} = 2 \rho \frac{\mu}{\gamma} v^\nu\,,
\ee
is determined by the background properties.  Since the radial field is massive, we can integrate it out using its equation of motion
\be
(\square + \tilde m^2) \tr = V^\mu \partial_\mu \tilteta \,.
\ee
The formal inversion of the previous equation yields
\be
\label{eq:substituten}
\tr = \frac{1}{\square + \tilde m^2} V^\mu \partial_\mu \tilteta\,.
\ee
For phonon momenta $p$ smaller than the radial field mass, the expansion of $(\square +\tilde m ^2)^{-1}$ gives terms suppressed as powers of $(p/\tilde m)^2$.  
Therefore, the relevant scale for the momentum  expansion is not $m$ but $\tilde m$, that is the curvature of the potential  at the minimum. Close to any second order phase transition point the curvature of the potential  vanishes, and the energy and momentum range of validity of the low-energy effective Lagrangian shrinks.

Upon substituting Eq.\,\eqref{eq:substituten} in Eq.\,\eqref{eq:L2_4},  we obtain
\begin{align}
\el_\text{eff} =& \frac{1}2  V^\mu  \partial_\mu \tilteta \frac{1}{\square + \tilde m^2} V^\nu \partial_\nu \tilteta + \frac{\rho^2 \eta^{\mu\nu}}2 \partial_\nu \tilteta \partial_\mu \tilteta \, ,
\end{align}
and the leading order in $p/\tilde m $ consists of replacing
\be
 \frac{1}{\square + \tilde m^2} \to  \frac{1}{\tilde m^2}\, .
\ee
We can finally rewrite the effective low-energy Lagrangian as
\begin{align}\label{eq:lag_effective_2}
\el_\text{eff}=&  \frac{\rho^2}2 \left(  \eta^{\mu\nu} + \frac{V^\mu V^\nu}{\rho^2\tilde m^2}\right)  \partial_\nu \tilteta \partial_\mu \tilteta \equiv \frac{\sqrt{- g} g^{\mu\nu}}2\partial_\nu \tilteta \partial_\mu \tilteta\,,
\end{align}
 where  the emergent acoustic metric for the phonon field is given by
\be
\sqrt{-g}g^{\mu \nu} = \rho^2\left(\eta^{\mu\nu} + \left(\frac{1}{c_s^2} -1\right) v^\mu v^\nu\right)\,,
\label{defg}
\ee
and
\be
c_s = \sqrt{\frac{ U \rho^2}{2 m^2 + 3 U \rho^2}}\,,
\ee
is the adiabatic speed of sound. This quantity vanishes for $U \rho^2 \to 0$, corresponding to the transition from the broken to the normal phase, while its upper bound is  $ \sqrt{1/3}$. 
The obtained effective field theory is valid for momenta lower than $\tilde{m}$, and it is equivalent to the hydrodynamic description of a superfluid,  holding at 
distances larger than the healing length\,\cite{Dalfovo:1999zz}
\be
\xi= \frac{1}{\sqrt{2} m c_s}\,.
\ee
By combining the nonrelativistic limit of the sound speed $c_s \simeq \sqrt{U \rho^2/(2 m^2)}$ with the  Eq.\,\eqref{eq:mt}, we have
\be
\xi \simeq \frac{\sqrt{2}}{ \tilde{m}}\,,
\ee
showing that distances larger than $\xi$ correspond to phonon momenta smaller than $\tilde{m}$.
We observe that retaining the leading order in $p/\tilde m $ in Eq.\,\eqref{eq:lag_effective_2} is equivalent to neglecting the kinetic term  $\partial_\nu\tr \partial^\nu\tr/2$ in Eq.\,\eqref{eq:L2_4}, and treat $\tr$ as a homogeneous field. This will be useful in the development of the low-energy theory of binary superfluids.
More general approaches to the effective low-energy theory  are reported in the Appendix \ref{sec:appendix}. Now, we move to consider binary mixture of superfluids.

\section{Binary boson superfluids}
\label{sec:two_fluids}
In the previous section, we have seen how it is possible to define an emergent acoustic metric for the long-wavelength excitations of a single superfluid. Here, we  extend the same approach to a mixture composed of two  {\it flavors} $A$ and $B$. In  experiments with homonuclear gases, they may correspond to two atomic hyperfine states. In astrophysics, they may indicate the superfluid components in the neutron star interior.  We take a scheme-independent approach and, as in the previous section, we employ a relativistic framework. We  assume  vanishing temperature and that the two fluids move with velocities $\bm v_A$ and $\bm v_B$ with respect to an external observer.

If the number of particles of each flavor is separately conserved, we have a $U(1)$ symmetry group associated to each flavor yielding to $U_A(1) \otimes U_{B}(1)$ as a symmetry group.
The spontaneous symmetry breaking of such group can produce one or two NGBs, depending on the interaction mechanism. 
We discuss in detail the case in which 
a $U(1)$  subgroup of $U_A(1) \otimes U_{B}(1)$  is  explicitly broken by a soft coupling term. Then, we expect that the spontaneous symmetry breaking mechanism could produce one genuine NGB and one pseudo-NGB, with a mass proportional to the soft coupling.

\subsection{Lagrangian of binary boson systems}
\label{sec:2bosons_lag}

The general form of the Lagrangian for two complex scalar fields, $\Phi _{A/B}$, is
\be 
\label{eq:implicit_lagr}
{\cal L}= \el_K(\Phi _A , \Phi _B) - \frac{1}{2}m _A^2 \Phi^\ast_A \Phi_A - \frac{1}{2}m _B^2 \Phi^\ast_B \Phi_B - V(\Phi _A , \Phi _B)\, , 
\ee
where  $m_{A/B}$ are the masses of the two fields. The  kinetic term  is
\be
\label{eq:K_AB}
{\cal L}_K=(D_{\mu, A} \bm \Phi_A)^* D^{\mu}_A \bm \Phi_A + (D_{\mu, B} \bm \Phi_B)^* D^{\mu}_B \bm \Phi_B  \, ,
\ee
with the covariant derivatives    defined as in Eq.\,\eqref{eq:covder}, that is
\be 
\label{eq:covder_2}
D_A^\nu  = \de^\nu + i \, \frac{\mu_A}{\gamma_A} \, v_A^\nu\,, \quad D_B^\nu  = \de^\nu + i \, \frac{\mu_B}{\gamma_B} \, v_B^\nu\,,
\ee
as each component has distinct chemical potential and velocity. Here,  $\gamma_{A/B} = (1 - v_{A/B}^2)^{-1/2}$ are the Lorentz factors of each fluid component. We consider only  quartic  intraspecies and interspecies interactions, $V_U$, to which a  Rabi term, $V_{\lambda}$, is conveniently added:
\begin{align}
\label{eq:potential2}
 V =V_U + V_\lambda = & \sum _{ i =A,B}  U_{ii } (\Phi ^\ast _i \Phi _i)^2 + 
\sum _{ i \neq j} \frac{1}{2}   U_{ij } \Phi ^\ast _i  \Phi _i \Phi ^\ast _j \Phi _j  \nonumber\\ 
& +\lambda \, (\Phi_A ^\ast \Phi_B +\Phi_B ^\ast \Phi_A ) \, ,
\end{align}
with $U_{ij}$ dimensionless coupling constants. The strength of the Rabi coupling, $\lambda$, has dimension of a  mass squared and it is  assumed to be so small that  it does not  sensibly modify the energy spectrum obtained for $\lambda = 0$. In this setting, the quartic interactions are the leading terms,  while $V_\lambda$ is assumed to be a subleading perturbation.

This aspect can be better clarified  considering the nonrelativistic limit of Eq.\,\eqref{eq:implicit_lagr}, relevant to current and realistic experiments with ultracold atom platforms\,\cite{annabook,fallanibook}. The above description takes the form of a two-species interacting Hamiltonian\,\cite{pethick,Dalfovo:1999zz}  given by
\begin{align}
H =&
 \sum_{l, i}   \left[ b_{l,i}^{\dagger} \left(-\frac{\hbar^2}{2 m_i} \frac{\partial^2}{\partial\mathbf{r}_l^2} + V_\text{trap}(\mathbf{r}_l) \right)  b_{l,i} + \tilde{U}_{ii} \, \big(b_{l,i}^{\dagger} b_{l,i} \big)^2 \right] 
 \nonumber \\
&+ \,  \sum_{i\neq i} \frac{1}2 \tilde{U}_{ij} \,  \sum_{l,m} \,  b_{l,i}^{\dagger} b_{l, i}  b_{m,j}^{\dagger} b_{m,j}  
\nonumber\\ & + 
\tilde{\lambda} \sum_{l}  \Big(b_{l , A}^{\dagger} b_{l , B} +  b_{l , B}^{\dagger} b_{l , A} \Big) \,, 
\end{align}
where $b_{l,i}$ are the ladder operators,  the indices $(i,j)$  label the two boson flavors,  that in principle can have different numbers or fillings, while the indices  $(l,m)$   refers  to the position   in a lattice or   in  continuous space. The added one-particle 
potential, $V_\text{trap}(\mathbf{r}_l)$, is typically harmonic and is necessary to provide particle trapping. \\
The above Hamiltonian can be obtained from the relativistic Lagrangian by rescaling the interaction couplings as
\begin{align}
\label{eqs:nonrela}
 \tilde{U}_{ij} = U_{ij} \, \sqrt{m_i \, m_j}\,, \quad 
\tilde{\lambda}  =\frac{\lambda}{\sqrt{m_i \, m_j}}\,,
\end{align}
as a result all couplings have the same dimension and can be compared.
 For harmonic $V_\text{trap}(\mathbf{r}_l)$, with isotropic trapping frequency $\Omega$, the requirement that all the couplings are perturbative and that the Rabi term has a subleading effect with respect to the quartic ones implies that
\be
\label{eq:estimated_lambda}
\tilde\lambda \ll  \tilde{U}_{ij} \ll \hbar \Omega \simeq 10^{-15} - 10^{-13}\, \text{eV}\, ,
\ee	
with $\hbar \Omega$ being  the scale of the energy difference between the ground state and the first excited state. The estimated value of $\hbar \Omega$ is obtained considering that in current ultracold atom experiments, typical stable ranges to avoid gravity effects (at lower frequencies) and three-body losses (at higher frequencies) are $\Omega \sim 10 - 10^3$ Hz, then $ \hbar \Omega \simeq 10^{-15} - 10^{-13}\, \text{eV}$.
This bound for the intensity of the Rabi coupling can drastically change  on a real space lattice, where low  values of $\tilde{\lambda}$ are needed to avoid the population of excited bands\,\cite{lepori2018}. 

Referring back to Eq.\,\eqref{eq:potential2}, the $U_{ij}$  quartic couplings  determine the strengths of the density-density boson interactions.  
Perturbation theory holds when $|U_{ij}| \ll 1$; moreover for 
\be
\label{eq:stability}
U_{AA} >0 \,, \quad  U_{BB} >0 \, , \quad
|U_{AB}| < 2 \sqrt{U_{AA}U_{BB}}\,,
\ee
the two species are miscible
and are homogeneously distributed in space~\cite{PhysRevLett.81.5718}.
 Without loss of generality, we take
\be\label{eq:Uaa_Ubb}
0 < U_{BB}  \leq U_{AA} \ll 1 \, ,
\ee
while the sign of the interflavor coupling, $U_{AB}$, is not fixed. 

As in the single fluid model, the quartic couplings  can produce  the condensation  of one or of both flavors. In other words, they are responsible of  the corresponding spontaneous symmetry breaking.
On the other hand, the Rabi term in Eq.~\eqref{eq:potential2} mixes the phases of the $\Phi _{A}$ and $\Phi _{B}$ fields, explicitly breaking the $U_A(1) \otimes U_{B}(1)$ global symmetry. 
 Given the relative richness of the set of   possible symmetry breaking patterns, we characterize them in detail in the following section.

\subsection{Symmetries} 
\label{sec:symmetries}

In order to discuss the  symmetries  of the two-fluid system, 
we introduce the vector  notation
\be
\bm \Phi = \left(\begin{array}{c} \Phi_A \\ \Phi_B \end{array} \right)\,,
\ee
that allows a more compact representation. We 
refer to the space spanned by the $\bm \Phi$ field as the flavor space.  
We begin with the discussion of the symmetries of the kinetic term in Eq.\,\eqref{eq:K_AB}. In the vector basis, it can be generalized to
\be
\label{eq:LK_compact}
{\cal L}_K=(D_\nu \bm \Phi)^\dagger (D^\nu \bm \Phi) \, ,
\ee
where the covariant derivative is
\be 
D_\nu = {\cal I}_2 \, \de_\nu  - i A_\nu  \,,
\ee
 with ${\cal I}_2$ the $2 \times 2 $ identity, and $A_\nu$ the Hermitian $2 \times 2$ matrix that represents a general external vector field. As in Eqs.\,\eqref{eq:K_AB} and\,\eqref{eq:covder_2}, we assume that it has only timelike components that are  diagonal in the flavor basis.
 Indicating with $\sigma_i$ the Pauli matrices, the only nonvanishing terms in $A_\nu $ can be cast as
\be
 A_\nu =  A^I_\nu {\cal I}_2 + A^3_\nu \sigma_3 =
\left(\begin{matrix} 
\mu_A/\gamma_A & 0 \\ 0 & \mu_B/\gamma_B
\end{matrix} \right) \delta_{\nu 0} \,.
\ee
 Given the expression of these covariant derivatives, the kinetic term in Eq.\,\eqref{eq:LK_compact} turns out to be invariant under the $U_A(1) \otimes U_{B}(1)$  symmetry group, which we rewrite as
 \be
 \label{eq:G} 
G= U(1)_{\cal V} \otimes U(1)_{\cal A}\,,
\ee 
 where  $U(1)_{\cal V}$ is associated to a global phase change,  generated by ${\cal I}_2$,  while  $U(1)_{\cal A}$   is the {\it axial} group generated by $\sigma_3$. If the two species have the same chemical potentials and velocities, then $A^3_\nu$ vanishes and  ${\cal L}_K$ is invariant under  the  symmetry group 
 \be 
 G_T = U(1)_{\cal V} \otimes SU(2) \, \supset G \,,
 \ee 
where $SU(2)$ generates flavor rotations.   The above relation simply states that for two fluids having the same chemical potentials and velocities, the kinetic term does not distinguish  the two fluids; thus it is invariant under a global phase redefinition  and  flavor rotations.

In the vector notation, the mass--or quadratic--term  contains also the Rabi interaction and reads as
\be
\frac{1}{2}m_A^2 \Phi^\ast_A \Phi_A + \frac{1}{2}m_B^2 \Phi^\ast_B \Phi_B + V_\lambda(\Phi_A , \Phi_B)=
 \frac{1}{2}{\bm \Phi}^\dagger M {\bm \Phi} \,,
\label{eq:massPhi}
\ee
with mass matrix
\be
\label{eq:massmatrix}
M = \left(\begin{array}{cc} m_A^2  & 2 \lambda \\ 2 \lambda & m_B^2 \end{array} \right) = m^2 {\cal I}_2 + \Delta m^2\, \sigma_3 + 2 \lambda\, \sigma_1 \,,
\ee
where $m^2_{A/B} = m^2  \pm  \Delta m^2 $. \\
Let us scrutinize the symmetries of this term in various cases:
\begin{enumerate}
\item If $\lambda =0$ and $ \Delta m^2 =0$, the  mass term symmetry is $G_T$: the two fields are indistinguishable and invariant under global phase and flavor rotations.
\item If $\lambda \neq 0$ and $ \Delta m^2 = 0$, the mass term is invariant under $U(1)_{\cal V}$ times the $U(1)$ group generated by $\sigma_1$. The mass eigenstates  are distinguishable:   they have   mass squared, $m^2 \pm 2 \lambda^2$.
\item If $\lambda =0$ and $ \Delta m^2 \neq 0$, the mass term symmetry is $G$:  the two fields are distinguishable  because they have different masses.
\item If both $\lambda \neq 0$ and $ \Delta m^2 \neq 0$, the mass term is only invariant under  the $U(1)_{\cal V}$ symmetry.
\end{enumerate}
In this work, we will focus on the last two cases.
For sufficiently small values of $\lambda$,  the perturbative Rabi term produces a soft  breaking of the axial symmetry.  As a consequence, in the spontaneously broken phase,  we expect to have one true NGB, originating from the  breaking of the  $U(1)_{\cal V}$ symmetry,  plus a pseudo-NBG associated to the breaking of the nonexact $U(1)_{\cal A}$ symmetry: a massless phonon and a massive phonon.  In the remainder of this paper, we will consider the corrections induced by the Rabi term only at order ${\cal O} (\lambda)$, ignoring the possible formation of topological solitons; see however\,\cite{Li:1971vr}.

We discuss now the density-density interactions in Eq.\,\eqref{eq:potential2}: in the vector basis both the interspecies and intraspecies interactions can be rewritten as
\begin{align}
 V_U=  \, & U_{AA} ({\bm \Phi}^\dagger {\cal P}_A {\bm \Phi})^2 + U_{BB} ({\bm \Phi}^\dagger {\cal P}_B {\bm \Phi })^2  \nonumber \\ 
&+ U_{AB} ( {\bm \Phi}^\dagger {\cal P}_A {\bm \Phi} )(  {\bm \Phi}^\dagger {\cal P}_B {\bm \Phi })\,, 
\end{align}
with
\be
{\cal P}_A = \frac{{\cal I}_2+\sigma_3}{2} \quad \text{and} \quad  {\cal P}_B = \frac{{\cal I}_2-\sigma_3}{2} \,,
\ee
the two flavor projectors. Since the interaction terms are proportional to  ${\cal I}_2$ and $\sigma _3$, they are invariant under $G$ but not under the enlarged group  $G_\text{T}$: they  distinguish  the two boson species.  The terms proportional to $\sigma _3$ vanish only when $U_{AA}$ = $U_{BB}$, and $ U_{AA} = 2 |U_{AB}|$, however, this condition violates the stability requirement in Eq.~\eqref{eq:stability}.

Let us  briefly recap the most  important aspects of the above analysis.  
For vanishing Rabi coupling,  the  Lagrangian has the $G$ symmetry given in Eq.\,\eqref{eq:G}.  If this symmetry is spontaneously broken the system is expected to have  two genuine NGBs, that is two massless phonons. If $\lambda\neq 0$, the system has only the $U(1)_{\cal V}$ symmetry and in the superfluid ground state only one NGB can appear. Furthermore, if the Rabi term can be treated as a soft explicit symmetry breaking,  the low-energy theory is expected to include a pseudo-NGB, with a square mass proportional to $\lambda$.

\section{Pressure of  binary superfluids}
\label{sec:pressure}

The considered system consists of two interacting gases having both intraspecies  and interspecies interactions, as well as a perturbative Rabi coupling.
In absence of both the interspecies and Rabi interactions, the two fluids would be noninteracting and the total pressure  of the system would be equal to the sum of two partial ones associated to each flavor. 
By adding the interactions between the two components, we might naively expect that the total pressure 
cannot be separated into two independent contributions.
Nevertheless, if the Rabi coupling vanishes and the $G$ symmetry holds, the ground-state Lagrangian can be cast as a sum of two terms, and the total pressure is the sum of two partial pressures. Furthermore, the same separation can be extended for perturbative Rabi coupling. 
In the following, we investigate which are the conditions for such separation.

\subsection{Flavor rotation in the ground state }
\label{sec:gsrotation}

As for a single fluid in the broken phase, it is useful to adopt the Madelung representation for both  $\Phi _A$ and $\Phi_B$. Introducing two pairs of real scalar fields, $\rho_i, \, \theta _i$,  such that
\be 
\Phi _i = \frac{\rho_i}{\sqrt{2}} e^{i \theta _i}\, , \qquad i = A, B\,,
\label{madrep}
\ee
  the  Lagrangian in Eq.\,\eqref{eq:implicit_lagr} with the potential in Eq.\,\eqref{eq:potential2} reads as
\begin{align} 
\label{eq:lagr2norm}
    \el  = \! \sum _{i = A,B }&  \left[ \frac{1}{2}\de _\nu \rho_i \de ^\nu \rho_i  + \frac{1}{2} \rho_i ^2   \de _\nu \theta_i \de ^\nu \theta_i  - \frac{1}{2}m_i ^2 \rho _i ^2 \right. \nonumber \\
   & \left.  + \, \frac{1}{2} \frac{\mu _i^2}{\gamma_i^2} \rho_i ^2 -\frac{1}{4} U _{ii}\rho _i ^4 \right] - \frac{U_{AB}}{4}  \rho _A ^2 \rho _B^2  \nonumber \\ 
   & - \, \lambda \, \rho _A \rho _B \cos ( \theta _B - \theta_A )\,,
\end{align}
where we have replaced the covariant derivatives with standard derivatives. As illustrated in Sec.~\ref{sec:onebec}, chemical potentials and fluid velocities can be introduced as expectation values of the phase fields. In the last line, the Rabi term explicitly violates the conservation of the currents
\be
\label{eq:j_mu_twobecs}
J^\nu_{i}=\frac{\delta {\cal L}}{\delta \partial_\nu \theta_{i}}\, , \qquad i={A,B}\,,
\ee
while conserving the total one $J^\nu_{B} + J^\nu_{A}$. Their difference satisfies
\be
\label{eq:relcurr}
\partial_\nu (J^\nu_{B}-  J^\nu_{A}) =2 \lambda \rho_A \rho_B \sin(\theta_B - \theta_A)\,,
\ee
which, for $\rho_A \simeq \rho_B$ both homogeneous, reduces to the sine-Gordon equation
\be\label{eq:sine_gordon}
\square(\theta_B - \theta_A) - 2 \lambda \sin(\theta_B - \theta_A) =0\,,
\ee
suggesting that solitonic solutions could be present\,\cite{PhysRevA.65.063621, PhysRevA.95.033614}. 
Unlike the  phase of a single fluid, the relative phase $(\theta_B - \theta_A)$ is an observable that is accessible in ultracold atom experiments\,\cite{Hall:1998zz}.\\

In the present section, we  focus on the background properties, while the effective action for the long-wavelength excitations will be discussed in Sec.\,\ref{sec:fluctuations}. 
Hereafter, we will indicate with $\rho_A$, $\rho_B$, $\theta_A$ and $\theta_B$ the mean field values. In analogy with Eq.\,\eqref{eq:vnu}, we write the gradients of the  phase fields 
as
\be
\label{eq:vAvB}
\partial^\nu \theta_{i} = \frac{\mu_{i}}{\gamma_i} v^\nu_{i}\,,
 \qquad i = A, B\,,
\ee
where $v^\mu_A $ and $v^\mu_B $ are the two fluid four-velocities.  Hence, from Eq.\,\eqref{eq:lagr2norm}, we obtain the background Lagrangian 
\be
\begin{aligned}
 \label{eq:L0_madelung}
{\cal L}_0 =& \frac{\rho_{A}^2}{2 \gamma_A^2}(\mu_A^2-m_A^2 \gamma_A^2) + \frac{\rho_{B}^2}{2 \gamma_B^2}(\mu_B^2-m_B^2 \gamma_B^2)   \\ 
&-\frac{1}{4} \rho_{A}^4 U_{AA} -\frac{1}{4} \rho_{B}^4 U_{BB}  -\frac{1}{4} \rho_{B}^2 \rho_{A}^2U_{AB} \\ 
&- \lambda  \, \rho_{A}\rho_{B} \cos(\theta_{B} -\theta_{A})\,.
\end{aligned}
\ee
Before estimating the mean field values of the radial fields, 
 we remark on a subtle aspect about the phase fields. They are not variational parameters, as they are connected by Eq.\,\eqref{eq:vAvB} with the fluids' flow (the same holds in the single-fluid case).  
 Since the Rabi coupling explicitly depends on the relative phase, it tends to fix such a difference. For instance,  
 in the broken phase, the mean field values $\rho_{A}$ and $\rho_{B}$ are positive; thus if $\lambda <0$ the background pressure is maximized for
\be 
\label{eq:thetaA}
\theta_{A} = \theta_{B}\,. \ee 
 For this reason we shall refer to the last term in Eq.\,\eqref{eq:L0_madelung}  as the {\it Rabi pressure}. If  at each space-time point the two phases are equal, Eq.\,\eqref{eq:vAvB} implies that
\be
\frac{\mu_A}{\gamma_A} v^\nu_A = \frac{\mu_B}{\gamma_B} v^\nu_B \,.
\ee
  Since for each flavor  $v^\nu_{i} = \gamma_{i} (1, -{\bm v}_i), \, i=A,B$,   Eq.\,\eqref{eq:thetaA} in turn implies that $\mu_A = \mu_B $ and $\bm v_A= \bm v_B$. Thus, the Rabi pressure tends to make the two fluids flow with the same velocity and to have equal chemical potentials. 
In the background Lagrangian of Eq.\,\eqref{eq:L0_madelung}, we consider $\theta_A \neq \theta_B $, because both the chemical potentials and the fluids' velocities are assumed to be externally fixed.\\
Following the same procedure of Sec\,\ref{sec:onebecgeneral}, we now
 determine the pressure by maximizing  ${\cal L}_0$  with respect to  $\rho_{A}$ and $\rho_{B}$,  which are the two independent variational parameters. 
We introduce the flavor vector notation
\be
{\bm \rho}_2=\left( \begin{array}{c} \rho^2_{A} \\ \rho^2_{B}\end{array} \right)\,, \quad {\bm \mu}_2=\left( \begin{array}{c} \mu^2_{A} \\ \mu^2_{B}\end{array} \right) \,, \quad {\bm m}_2=\left( \begin{array}{c} m^2_{A} \\ m^2_{B}\end{array} \right) \,,
\label{reparam}
\ee
where the subscripts indicate that they contain squared quantities.
For the sake of brevity, we define
\be
\label{eq:lambdaAB} 
\lambda_{AB} = \lambda \cos(\theta_{B} -\theta_{A}) \, ,
\ee
and we redefine the chemical potentials
\be 
\frac{\mu_i}{\gamma_i} \to  \mu_i\,.
\ee
The Lagrangian in Eq.~\eqref{eq:L0_madelung} now reads as
\begin{align}
\label{eq:LO}
{\cal L}_0 =& \frac{1}{2}{\bm\rho}_2\cdot ({\bm \mu}_2 -{\bm m}_2) -\frac{1}{4} {\bm\rho}_2^t{\bm U}{\bm\rho}_2- \lambda_{AB}  \rho_{A}\rho_{B} \,,
\end{align}
where 
\be 
{\bm U} = \begin{pmatrix} U_{AA} & U_{AB}/2\\
    U_{AB}/2 & U_{BB}
\end{pmatrix} \, ,
\label{defU}
\ee
and we left the last term as a function of $\rho_{A}$ and $\rho_{B}$. \\
Since we treat the Rabi term as a small perturbation,  we proceed first to diagonalize $\el_0$ in the sum of two independent terms neglecting the contribution of the Rabi. Any transformation
\be 
{\bm \rho}_2' = R(\alpha) \bm \rho_2\,, \quad
{\bm \mu}_2' = R(\alpha) \bm \mu_2 \,,\quad
{\bm m}'_2 = R(\alpha) \bm m_2\,,
\ee
where 
\be
R(\alpha) = \left( \begin{array}{cc} \cos\alpha & \sin\alpha  \\ \sin\alpha & -  \cos\alpha \end{array}\right) = \sigma_3 \cos\alpha + \sigma_1 \sin\alpha\,,
\label{rotback}
\ee
 is a  matrix in flavor space depending on the  mixing angle $\alpha$, leaves invariant the first term of the Lagrangian in Eq.\,\eqref{eq:LO}. This transformation clearly mixes the different flavors: the primed vectors  are a mixing of the  vector components in the not-primed basis. We refer to this transformation as {\it rotation} in flavor space.
 
Although the first term in Eq.\,\eqref{eq:LO} locks the flavor rotation angles of $\bm\rho_2$, $\bm\mu_2$ and $\bm m_2$ to be the same,  the value of $\alpha$ is  not fixed.  On the other hand, the diagonalization of  
\be
\label{eq:VUsq} 
V_U = \frac{1}4{\bm \rho}_2^t \bm U \bm \rho_2\,,\ee 
sets 
\begin{align}
\alpha = \pm \arcsin\!\left(\sqrt{\frac{1}{2}-\frac{\Delta U}{\sqrt{4\Delta U^2+U_{AB}^2}}}\right)\,,
\label{eq:cossin}
\end{align}
where
\be 
\Delta U = \frac{U_{AA}-U_{BB}}{2}\,.
\ee
For later convenience, we also define the following quantities 
\be
\label{def:U_and_Upm}
 U = \frac{U_{AA}+U_{BB}}{2}\,, \qquad U_{\pm} = U \pm \sqrt{\Delta U^2+\frac{U_{AB}^2}{4}}\,.
\ee
Given the choice of ordering of the couplings in Eq.\,\eqref{eq:Uaa_Ubb}, we have $\Delta U \geq 0$; moreover, 
in order to have positive-defined  squared radial fields, chemical potentials and masses, we take the positive sign in the right-hand side of Eq.~\eqref{eq:cossin}. \\
Indicating the components of the rotated vectors as
\be
{\bm \rho}'_2=\left( \begin{array}{c} \rho^2_{I} \\ \rho^2_{II}\end{array} \right)\,, \quad {\bm \mu}'_2=\left( \begin{array}{c} \mu^2_{I} \\ \mu^2_{II}\end{array} \right) \,,\quad {\bm m}'_2=\left( \begin{array}{c} m^2_{I} \\ m^2_{II}\end{array} \right) \,,
\ee
 the rotated squared radial fields are given by
\begin{align}\label{eq:rotation_I_II}
\rho_I^2 & = \rho_A^2  \cos\alpha +  \rho_B^2 \sin\alpha\,, \nonumber\\
\rho_{II}^2 & =  -\rho_B^2 \cos\alpha  + \rho_A^2  \sin\alpha \,,
\end{align}
and, with our choice of the interaction couplings, it follows that
 \be\label{eq:rhoI_rhoII}
\rho_I^2 \geq \rho_{II}^2\,.
 \ee
Similar expressions hold for the squared masses and chemical potentials.

Now, we consider some specific cases. 
For $U_{AA} = U_{BB} $, i.e. $\Delta U = 0$, the mixing angle is maximal $\alpha = \pi/4 \pm n \pi$, with $n$ integer,
and the rotation matrix is
\be
R = \pm \frac{1}{\sqrt{2}}\left( \begin{array}{cc} 1 & 1  \\ 1 & -  1 \end{array}\right) = \pm\frac{1}{\sqrt{2}} (\sigma_3 + \sigma_1 )\,.
\label{rotrabi}
\ee
Close to maximal mixing,
\be
\label{eq:closetomaximal}
\rho_I^2  \simeq  \frac{1}{\sqrt{2}} (\rho_A^2 +  \rho_B^2)\,, \quad
\rho_{II}^2 \simeq  \frac{1}{\sqrt{2}} (\rho_A^2 -  \rho_B^2)\,,
\ee
and for $\rho_A \simeq  \rho_B$,  this implies that $\rho_I^2 \gg \rho_{II}^2 $.

For general values of $\rho_A$ and $ \rho_B$, the rotation angle   \be\label{eq:mix_tan}
\alpha = \arctan\left(\frac{\rho_B^2}{\rho_A^2}\right)\,,
\ee
makes $\rho_{I}$  maximal while $\rho_{II}$ vanishes. Regarding the masses in the rotated basis, in analogy with Eqs.\,\eqref{eq:rhoI_rhoII} and \eqref{eq:mix_tan}, the mass of the radial field  $II$ is smaller than the mass of the field $I$, and it vanishes when
\be 
\alpha= \arctan\left(\frac{m_B ^2}{m_A^2}\right)\,.
\ee
We remark that the vanishing of $m_{II}$ is not associated to any specific phenomenon: it does not signal a phase transition.  The phase transition to the broken phase occurs when the rotated chemical potentials exceed the corresponding masses. This can be observed by rewriting the Lagrangian in Eq. \eqref{eq:LO} in the  ${I,II}$ flavor basis  
\begin{align}\label{eq:L0_w_lambda}
{\cal L}_0  = &  \frac{\rho_{I}^2}{2}(\mu_I^2-m_I^2) - \frac{\rho_{I}^4}{4} U_+ \nonumber\\   &+ \frac{\rho_{II}^2}{2}(\mu_{II}^2-m_{II}^2) - \frac{\rho_{II}^4}{4} U_- - V_\lambda
\,,
\end{align}
which, apart from the Rabi term, is the sum of two independent  pieces. 
We now determine the quantum pressure of the system for vanishing Rabi coupling; we will include it in Sec.\,\ref{rabipressure}.

\subsection{Vanishing Rabi coupling}

For vanishing Rabi coupling, the Lagrangian in Eq.\,\eqref{eq:L0_w_lambda} can be cast as that of two non-interacting fluids
\begin{align}
\label{eq:lag_noRabi}
{\cal L}_0  =  {\cal L}_{I} + {\cal L}_{II} =&\frac{\rho_{I}^2}{2}(\mu_I^2-m_I^2) - \frac{\rho_{I}^4}{4} U_+  \nonumber\\   &+\frac{\rho_{II}^2}{2}(\mu_{II}^2-m_{II}^2) - \frac{\rho_{II}^4}{4} U_- \,,
\end{align}
where the two Lagrangians, ${\cal L}_{I}$ and ${\cal L}_{II}$, are independent. We determine the expectation values of the radial fields in the same manner as in the single-fluid case discussed in Sec.\,\ref{sec:onebec}.  The values maximizing the  Lagrangian are 
\be
\label{eq:rho_mean_field_I}
\rho_{I\phantom{I}}^2 = \begin{cases}\frac{\mu_{I\phantom{I}}^2-m_{I\phantom{I}}^2}{U_+} & \text{if}\,\,\,  \mu_{I\phantom{I}} \geq m_{I\phantom{I}}\\
0 & \text{if}\,\,\,  \mu_{I\phantom{I}} \leq m_{I\phantom{I}}\end{cases} \, ,
\ee
and
\be
\label{eq:rho_mean_field_I:II}
\rho_{II}^2 = \begin{cases}\frac{\mu_{II}^2-m_{II}^2}{U_-} & \text{if}\,\,\,  \mu_{II} \geq m_{II}\\
0 & \text{if}\,\,\,  \mu_{II} \leq m_{II}\end{cases} \, ,
\ee
and are analogous to the expression for the single scalar field in Eq.\,\eqref{eq:rho_mean_field}.  
In the $I,\, II$ basis the conditions to have nonvanishing expectation values of the scalar  fields $\rho_I$ and $\rho_{II}$, are  $\mu_I > m_I$ and $\mu_{II} > m_{II}$, respectively.
When expressing these conditions in the $A,B$ basis, we have that   
\begin{align}
(\mu_A^2 -m_A^2)\cos\alpha +(\mu_B^2 -m_B^2)\sin\alpha >0\, , \nonumber \\
(\mu_A^2 -m_A^2)\sin\alpha -(\mu_B^2 -m_B^2)\cos\alpha >0 \,,
\end{align}
which  can be cast as
\begin{align}
\label{eq:tans}
    \tan \alpha  > - \frac{\mu_A^2-m_A^2}{\mu_B^2-m_B^2} \quad \text{and} \quad
\tan \alpha  >  \frac{\mu_B^2-m_B^2}{\mu_A^2-m_A^2}\,,
\end{align}
respectively. Since 
\be
\frac{\mu_B^2-m_B^2}{\mu_A^2-m_A^2} > - \frac{\mu_A^2-m_A^2}{\mu_B^2-m_B^2}\,,
\ee
it is  possible that $\rho_I>0$, while   $\rho_{II}=0$.  In the following, we will assume for simplicity that the second condition in \eqref{eq:tans} is satisfied; therefore  both $\rho_I$ and $\rho_{II}$ are positive.

The quantum pressure of the system is given by the value taken by the background Lagrangian at the stationary points and  it is the sum of two partial pressures
\be
\label{eq:Ptot}
P=P_I + P_{II}\,,
\ee
with
\be
\label{eq:PI_PII}
P_I = \frac{(\mu_I^2-m_I^2)^2}{4 U_+} \, \quad \text{and} \quad \quad P_{II} = \frac{(\mu_{II}^2-m_{II}^2)^2}{4 U_-} \,.
\ee
Using standard thermodynamic relations, we obtain the number density for each rotated component
\be
n_I = \mu_I \rho_{I}^2 \, , \qquad n_{II} =  \mu_{II} \rho_{II}^2 \,,
\ee
showing that when $|U_{AB}|= 2 \sqrt{U_{AA}U_{BB}}$, the number density $n_{II}$ diverges. This signals the transition to a dilute liquidlike droplet state\,\cite{petrov2015}.

\subsection{Including the Rabi coupling}
\label{rabipressure}

We examine the contribution of the Rabi interaction to the background pressure. In the ${I,\, II}$ basis, the Rabi term takes the following form: 
\begin{align}
\label{eq:L0-rabi}
V_{\lambda}  = \frac{\lambda_{AB}}{\sqrt{2}} \sqrt{(\rho_{I}^4-\rho_{II}^4)\sin(2 \alpha) - 2\rho_{I}^2\rho_{II}^2 \cos(2 \alpha)}\,,
\end{align}
and it is a nontrivial coupling between the  $\rho_I$ and $\rho_{II}$ fields.
If the explicit breaking is small, we can treat $V_{\lambda}$ as a perturbation. In this case, we expand the mean field solutions at leading order in $\lambda$ as follows
\be
\begin{aligned}
\label{eq:shift_rho_I_II}
\rho_{j} \to \rho_{j} + \frac{ \lambda_{AB}}{\rho_{j}^2} \delta\rho_{j} \, , \quad  j = I,II\\
\end{aligned}
\ee
where 
\be 
\label{cond:lambdas}
\frac{| \lambda_{AB}|}{ \rho_{I}^2} \ll 1 \quad \text{and} \quad \frac{| \lambda_{AB}|}{ \rho_{II}^2} \ll 1 \,,
\ee 
while $ \delta\rho_{I}$ and  $ \delta\rho_{II}$ are two variations to be determined.
Upon substituting these expressions in the equations for the stationary point of the Lagrangian, we find that the leading order contributions are
\be
 \delta\rho_{I} = -\frac{\rho_{I}^3 \sin(2\alpha) -\rho_{I} \rho_{II}^2 \cos(2 \alpha)}{  U_+\sqrt{2(\rho_{I}^4-\rho_{II}^4)\sin(2 \alpha) -4 \rho_{I}^2\rho_{II}^2 \cos(2 \alpha)}} \, ,
\ee
and
\be
 \delta\rho_{II} = +\frac{\rho_{II}^3 \sin(2\alpha) +\rho_{II} \rho_{I}^2 \cos(2 \alpha) }{  U_-\sqrt{2(\rho_{I}^4-\rho_{II}^4)\sin(2 \alpha) -4 \rho_{I}^2\rho_{II}^2 \cos(2 \alpha)}}\,.
\ee
Even when the Rabi coupling satisfies the conditions in Eq.\,\eqref{cond:lambdas}, the subsystems $I$ and $II$ are coupled:  $\delta\rho_{I}$ and $\delta\rho_{II}$ are  functions of both $\mu_I$ and $\mu_{II}$. \\
If we want to still write the pressure as the sum of two independent terms as in Eq.\,\eqref{eq:Ptot}, we have to impose an additional constraint:
\be 
\rho_{II} \ll \rho_{I}\,.  
\ee 
This, in turn, implies that $\rho_A \simeq \rho_B$ [see Eq.\,\eqref{eq:mix_tan}] and thus
\begin{align}
 \delta\rho_{I} & \simeq - \frac{\rho_{I} }{U_+} \sqrt{\frac{\sin(2\alpha)}2} \, , \nonumber \\
 \delta\rho_{II} & \simeq + \frac{\rho_{II} \cos(2\alpha)  }{U_- \sqrt{2\sin(2\alpha)} }\,,
 \label{rhorot}
\end{align}
meaning that  each radial field depends on the corresponding chemical potential, see Eqs.\,\eqref{eq:rho_mean_field_I} and \eqref{eq:rho_mean_field_I:II}. Upon replacing these expressions in Eq.\,\eqref{eq:shift_rho_I_II} 
we obtain the shifts of the radial fields due to the Rabi term 
\begin{align}\label{eq:shift_rho_In}
\rho_{I} & \to \rho_{I}\left( 1 + \frac{ \lambda_{AB}}{U_+ \rho_{I}^2 } \sqrt{\frac{\sin(2\alpha)}2} \right) \, ,\\
\label{eq:shift_rho_IIn}\rho_{II} & \to \rho_{II}\left( 1 + \frac{ \lambda_{AB}}{U_-\rho_{II}^2} \frac{\cos(2\alpha)  }{ \sqrt{2 \sin(2\alpha)} } \right)\,.
\end{align}

Comparing these expressions with the estimate in Eq.\,\eqref{eq:estimated_lambda}, we better comprehend the condition for treating perturbatively the Rabi term.
In the broken phase, the actual dimensionless expansion parameter is the ratio between the Rabi coupling and the expectation values of the rotated radial fields. Since $\rho_{I,II}$ are of the order of $m_{A,B}$, the condition  $\tilde\lambda \ll  \tilde{U}_{ij}$ holds. \\
Upon substituting  Eqs.\,\eqref{eq:shift_rho_In} and \eqref{eq:shift_rho_IIn} in
Eq.\,\eqref{eq:L0-rabi}, one obtains that the two subsystems are decoupled and the pressure can be expressed as the sum of two independent contributions:
\begin{align}
P_I &\to P_I (\lambda_{AB} =0) + \lambda_{AB} \,\delta P_I \, , \nonumber \\ P_{II} &\to P_{II}  (\lambda_{AB} =0) + \lambda_{AB}\, \delta P_{II}  \,,
\end{align}
where $\lambda_{AB}$ is defined in Eq.\,\eqref{eq:lambdaAB} and $\delta P_I$ and 
 $\delta P_{II} $ are the pressure variations depending only on $\mu_I$ and $\mu_{II}$, respectively.  In this case, each
 subsystem is characterized by its thermodynamic quantities that do not depend on the properties of the other subsystem. \\
Alternatively, a more straightforward approach can be used to derive the perturbative effect of the Rabi coupling on the pressure. Combining Eq.\,\eqref{eq:L0-rabi} with the decoupling condition $\rho_{II} \ll \rho_{I}$,  we have that
\be
V_{\lambda}  \simeq  \lambda_{AB} \rho_{I}^2 \sqrt{\sin(2\alpha)/2}  -  \lambda_{AB} \rho_{II}^2 \sqrt{\sin(2\alpha)/2} \cot(2\alpha)\,.
\ee
Consistently with the observations in Sec.\,\ref{sec:symmetries}, these terms
can be interpreted as small shifts of the $m_I$ and $m_{II}$ masses, that is
\begin{align}
m_I^2  & \to m_I^2 +  \lambda_{AB} \sqrt{\sin(\alpha)\cos(\alpha)} \, , \nonumber \\m_{II} & \to m_{II}^2 - \lambda_{AB} \sqrt{\sin(\alpha)\cos(\alpha)} \cot(2\alpha)\,.
\end{align}
In fact, the corresponding partial pressures can be readily obtained from those in Eq.\,\eqref{eq:PI_PII}, with the above shifted masses. In detail, we have that
\be 
P_I = \frac{(\mu_I^2-m_I^2)^2}{4 U_+} - \frac{\lambda_{AB} \sqrt{\sin(2\alpha)}}{2 \sqrt{2}  U_+}(\mu_I^2-m_I^2) \, , \ee
and
\be
 P_{II} = \frac{(\mu_{II}^2-m_{II}^2)^2}{4 U_-} +\frac{\lambda_{AB} \sqrt{\sin(2\alpha)} \cot(2\alpha)}{2 \sqrt{2} U_-}(\mu_{II}^2-m_{II}^2)\,. 
 \ee
We remark that for both nonvanishing $U_{AB}$ and $\lambda$,  in general, it is not possible  to simultaneously diagonalize $V_U$ and $V_\lambda$. They differ in the dependence in $\rho_A$ and $\rho_B$: quadratic for the interspecies interaction, and linear for the Rabi term. Consequently, the diagonalization of $V_U$ requires a non linear transformation of the fields, as in Eq.\,\eqref{eq:rotation_I_II}, leaving the Rabi term non diagonal. The specular situation occurs when the Rabi term dominates over the interspecies interaction. If we diagonalize the Rabi term, the transformed fields remain interacting due to a nonvanishing $U_{AB}$\,\cite{lepori2018}.

\section{Quantum fluctuations in binary superfluids}
\label{sec:fluctuations}

In the previous section, we characterized the mean field Lagrangian of the boson mixture. Following the same steps of the single-fluid model, we analyze the effective action for the long-wavelength excitations. We replace
\be
\begin{aligned}
\label{def:fluctuationsAB}
\rho_i \to \rho_{i}+ \tr_{i}\, , \quad \theta_i \to \theta_{i}+ \tilteta_{i} \, , \quad i = A,B\, , 
\end{aligned}
\ee
and throughout the remainder of this work, $\rho_{i}$ and $\theta_{i}$ indicate the mean field solutions, 
while $\tr_{i}$ and $\tilteta_{i}$ are their corresponding fluctuations. We expand the Lagrangian as
\be
{\cal L} = {\cal L}_0 + {\cal L}_2\,,
\ee
with ${\cal L}_0$ the mean-field one in Eq.\,\eqref{eq:L0_madelung}, while $ {\cal L}_2$ is quadratic in the fields' fluctuations.
In the following sections, we first discuss the case for vanishing Rabi interaction ($\lambda=0$), and then the nonvanishing case ($\lambda\neq0$). 
We will show that $\mathcal{L}_2$ can be written as the one of two noninteracting scalar fields, propagating on an emergent acoustic metric. The Rabi term will induce an effective mass in one of the scalar fields.

\subsection{Fluctuations with vanishing Rabi term}
\label{sec:fluctuations_no_Rabi}

Given Eq.\,\eqref{def:fluctuationsAB}, we compute the  Lagrangian, $\mathcal{L}_2$, quadratic in the fluctuations.
We consider a 
homogeneous and stationary background, with both fluids moving at the same velocity $v^\nu$. For $\lambda =0$, we have
\begin{align}
\label{eq:Lquad}
{\cal L}_2 =& \frac{1}2 \tilde\rho_A^2 \left[(\mu_A^2-m_A^2) - \frac{3}2 U_{AA}\rho_{A}^2-\frac{U_{AB}}{4} \rho_{B}^2   \right] \nonumber\\
&+ \frac{1}2 \tilde\rho_B^2 \left[(\mu_B^2-m_B^2) - \frac{3}2 U_{BB}\rho_{B}^2-\frac{U_{AB}}{4} \rho_{A}^2 \right] \nonumber\\
&-U_{AB} \rho_{A}\rho_{B} \tilde\rho_A \tilde\rho_B \nonumber\\
&+ \frac{1}2 \rho_{A}^2 \partial_\nu \tilde\theta_A  \partial^\nu \tilde\theta_A +  \frac{1}2 \rho_{B}^2 \partial_\nu \tilde\theta_B  \partial^\nu \tilde\theta_B  \nonumber\\
&+ 2 \rho_{A} \tilde\rho_A \mu_A v^\nu \partial_\nu \tilde\theta_A + 2 \rho_{B} \tilde\rho_B \mu_B v^\nu \partial_\nu \tilde\theta_B\,,
 \end{align}
and we disregarded the kinetic terms of the radial fluctuations because they give subleading corrections to the low-energy effective theory (see Sec.\,\ref{sec:general}). 
For later convenience, we introduce the rescaled fluctuations
\be
\label{eq:theta_scaling}
\hat {\rho_i} =  \mu_{i}  \tilde\rho_i \, , \quad   \hat {\theta_i} =  \rho_{i}  \tilde\theta_i \, , 
\quad i = A,B\,,
\ee
and the corresponding flavor vectors
\begin{align}
\hat{\bm \rho}=
\left( \begin{array}{c} \hat \rho_A \\  \hat \rho_B\end{array} \right)\,, \qquad
\hat {\bm \theta}=\left( \begin{array}{c} \hat \theta_A \\ \hat \theta_B\end{array} \right)\,. 
\end{align}
Thus, the quadratic Lagrangian in Eq.\,\eqref{eq:Lquad} can be cast as
\be
\label{eq:Lquadn}
{\cal L}_2 = \frac{1}{2} \hat{\bm\rho}^t H  \hat{\bm\rho} +2 v^\nu \hat{\bm\rho} \cdot \partial_\nu \hat{\bm \theta} + \frac{1}{2} \partial_\nu \hat{\bm \theta} \cdot  \partial^\nu \hat{\bm \theta} \,,
\ee
where $H$ is the $2\times 2$ symmetric matrix with elements
\begin{align}
\label{eq:H_components}
H_{11}&=   \frac{1}{\mu_A^2}\left[\mu_A^2-m_A^2 - \frac{3}2 U_{AA}\rho_{A}^2-\frac{U_{AB}}{4} \rho_{B}^2 \right] \, , \nonumber\\
H_{22}&= H_{11}(A\leftrightarrow B) \, , \nonumber\\
H_{12}&=-U_{AB} \frac{\rho_{A}\rho_{B}}{2\mu_A\mu_B}\,.
\end{align}
We diagonalize $\mathcal L_2$ with a suitable transformation in flavor space of the fields $\hat {\bm\rho}$ and $\hat{\bm\theta}$. 
The first and last term in the rhs of Eq.\,\eqref{eq:Lquadn} are separately invariant under any flavor mixing transformation of  $\hat {\bm\rho}$ and $\hat{\bm\theta}$, as the one in Eq.\,\eqref{rotback}. However, the invariance of the second term locks the mixing angle to be the same for $\hat {\bm\rho}$ and $\hat{\bm\theta}$.
The locking implies that the vectors in the new basis are
\be
\label{eq:rotation_gamma} 
\tilde{\bm \rho}' = R(\beta)  \hat {\bm\rho}  \quad \text{and} \quad  \tilde{\bm \theta}' = R(\beta)  \hat {\bm\theta}\,, 
\ee  
where $\beta$ is the common rotation angle. 
The flavor rotation angle is still arbitrary, and we can use such  freedom to rotate the  $\hat {\bm\rho}$ field to diagonalize $H$, similarly to the the procedure in Sec.\,\ref{sec:gsrotation}.
This is achieved choosing
\begin{align}
\beta = \pm \arcsin\left(\frac{1}{\sqrt{2}}\sqrt{1-\frac{H_{11}-H_{22}}{\sqrt{(H_{11}-H_{22})^2+4 H_{12}^2}}}\right)\label{eq:cossin2}\,.
\end{align}
We remark that the flavor rotation angle required to disentangle the radial fluctuations, differs from the one  in Eq.\,\eqref{rotback} used to diagonalize the background. 
As a consequence, the components $\tilde \rho_{a/b}$ and $ \tilde\theta_{a/b}$ of the rotated fields
\be
\tilde{\bm \rho}'=\left( \begin{array}{c} \tilde \rho_{a} \\ \tilde\rho_{b}\end{array} \right)  \,,\quad \tilde{\bm \theta}'=\left( \begin{array}{c} \tilde\theta_{a} \\ \tilde\theta_{b}\end{array} \right)\,,
\ee
are not the fluctuations of  $ \rho_{I/II}$ and  $ \theta_{I/II}$. 
However, they can be written as a linear combination of  $\tilde \rho_{a/b}$ and $ \tilde\theta_{a/b}$, respectively.

In the ${a,b}$ basis  the matrix $H$ is diagonal; thus  no mixing exists. In other words
\be
\label{eq:Lab}
{\cal L}_2 = {\cal L}_{2 a} + {\cal L}_{2 b}\,,
\ee
where 
\be
{\cal L}_{2 a} = -\frac{1}2\tilde m_{a}^2 \tr_a^2 + 2  v^\nu \tr_a \partial_\nu \tilteta_a + \frac{1}2 \partial_\nu \tilteta_a \partial^\nu \tilteta_a\,,
\ee
and ${\cal L}_{2 b}$ is given by an analogous expression. They  have the same formal expression obtained in the single-fluid case [see  Eq.\,\eqref{eq:L2_4}] when neglecting the kinetic term of the radial mode. Note that the radial modes have dimension $2$; therefore their {\it masses} 
\begin{align}
\tilde m_a^2 &= -\frac{H_{11}+H_{22}}{2}+ \frac{1}{2} \sqrt{(H_{11}-H_{22})^2 + 4 H_{12}^2 } \, , \nonumber \\
\tilde m_{b}^2 &= -\frac{H_{11}+H_{22}}{2}- \frac{1}{2} \sqrt{(H_{11}-H_{22})^2 + 4 H_{12}^2 }\,,
\end{align}
are dimensionless.   Upon integrating out the radial modes, the effective Lagrangian can  be written as
\be
{\cal L}_2=\frac{1}{2} \left(\eta^{\mu\nu} + \left(\frac{1}{c_{sa}^2} -1\right) v^\mu v^\nu\right)\partial_{\mu} \tilde\theta_{a} \partial_{\nu} \tilde\theta_{a} +( a\to b) \, ,
\label{l2}
\ee
where the two sound speeds are given by
\be
\label{eqs:soundspeed}
c_{sa}^2 = \frac{2 \tilde m_a^2}{1+\tilde m_a^2}\,, \qquad c_{sb}^2 = \frac{2 \tilde m_{b}^2}{1+\tilde m_{b}^2}\, ,
\ee
and given the expression of the $H$ matrix components in \eqref{eq:H_components}, each sound mode depends on the intraspecies and  interspecies couplings, as well as on the chemical potentials $\mu_A$  and $\mu_B$  and the masses $m_A$  and $m_B$.

\subsection{Including the Rabi term in the fluctuations}
\label{sec:fluctuations_with_Rabi}

Including fluctuations in the $A,B$ flavor basis, the Rabi term is given by
\be
 V_{\lambda} =  \lambda \, (\rho_A + \tilde \rho_A) (\rho_B+ \tilde \rho_A) \cos(\theta_A -\theta_B +\tilde\theta_A -\tilde\theta_B)\,,
\ee
and it  produces a plethora of terms. The linear terms in the fluctuations vanish at the stationary point.  There are two quadratic terms in the fluctuations, which have distinct effects. One term is proportional to $\tilde\rho_A\tilde\rho_B$ and thus contributes to the $H_{12}$ matrix element, see Eq.\,\eqref{eq:H_components}. This produces a trivial small shift of the effective masses of the radial fluctuations. The second quadratic term in the fluctuations is 
\be
\label{eq:Rabi_AB} 
{\cal L}_{\lambda\theta^2} = \frac{\lambda_{AB}}{2} \rho_A\rho_B  (\tilde\theta_A -\tilde\theta_B)^2 \, ,
\ee
where $\lambda_{AB}$ is defined in Eq.\,\eqref{eq:lambdaAB}.
This is an effective mass term for the mode  corresponding to the relative phase fluctuations. However, this term is expressed in the ${A,B}$ basis, but  to diagonalize the 
Lagrangian in the fluctuations  and write it as in Eq.\,\eqref{eq:Lab}, we  performed a rotation in flavor space by the angle $\beta$ in Eq.\,\eqref{eq:cossin2}. Therefore, we need to express the Rabi quadratic term  \,\eqref{eq:Rabi_AB}  in the rotated basis ${a,b}$.

Before doing that, we observe that to have the total pressure as the sum of two partial pressures in the background analysis, we  had to assume that $\rho_{I} \gg \rho_{II}$  and, correspondingly, a background flavor rotation angle close to maximal mixing, that is $\alpha \simeq \pi/4$. We expect that the same conditions would lead to an effective theory with two decoupled modes, one massless and one massive. To this end, we follow the same procedure discussed for vanishing Rabi coupling. In particular, we rescale the phases as in  Eq.\,\eqref{eq:theta_scaling} and obtain that, close to maximal background mixing, the condition  $\rho_{I} \gg \rho_{II}$ implies  $\rho_{B} \simeq \rho_{A}$, see Eq.\,\eqref{eq:closetomaximal}.  Then, Eq.\,\eqref{eq:Rabi_AB} can be rewritten as
\be
{\cal L}_{\lambda\theta^2} \simeq \lambda_{AB} \frac{ \rho_I^2} {\rho_{A}^2} \sqrt{\frac{\sin 2\alpha}{2}}  (\hat \theta_B -\hat \theta_A)^2\,,
\ee
thus it corresponds to   the mass term for the relative phase mode in the rescaled basis, see Eq.\,\eqref{eq:theta_scaling}. Now we perform a flavor rotation to the ${a,b}$ basis as in Eq.\,\eqref{eq:rotation_gamma}. This is a necessary operation, because to 
 integrate out the radial mode  we have first  to obtain a Lagrangian that is diagonal in flavor basis. 
Note that we are forced to do this flavor rotation for the following  chain of reasons:  we need to rotate the radial fluctuations to diagonalize  the $H$ matrix, and the radial and phase fluctuations are locked by the interaction term  between radial and phase fluctuations. 
In the rotated basis
\be
\hat{ \theta}_B -\hat{\theta}_A = \tilde\theta_a (\sin\beta - \cos\beta) - \tilde\theta_{b} (\sin\beta + \cos\beta)\,,
\ee 
  the approximations $\rho_{A} \simeq \rho_{B}$  and of maximal background mixing lead to $H_{11} \simeq H_{22}$, thus the fluctuation flavor mixing angle $\beta$ is close to maximal as well. In this case,  we can neglect the mass of the  $\tilde\theta_a $ field and rewrite the Rabi coupling as  
\be
{\cal L} \simeq 2 \, \lambda_{AB} \, \tilde\theta_{b}^2 \, .
\label{lrabi}
\ee
Upon integrating out the radial fluctuations using the same procedure discussed for the single-fluid case in Sec.\,\ref{sec:general}, we obtain the quadratic effective Lagrangian  in the phase fluctuations 
\begin{align}
{\cal L}_\text{eff}= {\cal L}_\text{eff,a} +   {\cal L}_\text{eff,b}\,, 
\end{align}
where
\begin{align}
{\cal L}_\text{eff,a}= &\frac{1}{2} \left(\eta^{\mu\nu} + \left(\frac{1}{c_{sa}^2} -1\right) v^\mu v^\nu\right)\partial_{\mu} \tilde\theta_{a} \partial_{\nu} \tilde\theta_{a}\,,
\label{l2_Rabi_a}
\end{align}
and
\begin{align}
{\cal L}_\text{eff,b}= &
\frac{1}{2} \left(\eta^{\mu\nu} + \left(\frac{1}{c_{sb}^2} -1\right) v^\mu v^\nu\right)\partial_{\mu} \tilde\theta_{b} \partial_{\nu} \tilde\theta_{b} -\frac{1}{2} m_b^2 c_{sb}^2\,\tilde\theta_{b}^2 \, ,
\label{l2_Rabi}
\end{align}
with
\be
m_b^2 = - \frac{4 \, \lambda_{AB}}{c_{sb}^2} = - \frac{4 \, \lambda} {c_{sb}^2} \cos(\theta_A -\theta_B)\,,
\label{mb}
\ee
the mass squared of the $\tilde\theta_{b}$ phonon. A similar expression was obtained in\,\cite{Liberati:2005pr} with a different method. The above expression implies that  $\lambda \cos(\theta_A -\theta_B)$ must be negative; otherwise the $\tilde\theta_{b}$ field would be tachyonic. This is  the same  condition  needed to have a positive contribution of the Rabi term to the background pressure, see Eq.\,\eqref{eq:L0_madelung}. Taking $\lambda <0$, the Rabi pressure is maximal for $\theta_A = \theta_B$ [see Eq.\,\eqref{eq:thetaA}], which is also the condition that  maximizes the mass of the $\tilde\theta_{b}$ field. Alternatively, the favored state may be a crystal of sine-Gordon kinks [see Eq.\,\eqref{eq:sine_gordon}], and the negative value of the squared mass would actually indicate the presence of phonon excitations of the crystal~\cite{Takayama:1992eu}.

 An interesting point is that, from Eq.\,\eqref{eq:vAvB}, the difference of the background phases is related to a difference in the chemical potentials and fluid velocities. Since in the derivation we have  assumed that the  two velocities are equal  (it was not a necessary condition, but it helped to simplify the analysis),  the condition to have a positive defined  squared mass can be related to a condition on the chemical potentials. 
As we have previously discussed, it is  possible to prepare the system in such a way that 
\be
\theta_A = \theta_B + \Theta (x) \, ,
\ee
where $ \Theta(x)$ is determined by the chemical potential difference between the two fluids: from Eq. \eqref{eq:vAvB}, we have that for equal velocities of the two fluids
\be
\mu_A = \mu_B +  \gamma \, v_\mu \de^\mu  \Theta (x) \,,
\ee
where $\gamma$ is their common Lorentz factor.
  Considering for simplicity the nonrelativistic limit,  we have that 
\be
\Theta = \Theta_0 + (\mu_A-\mu_B) \, t \, ,
\ee
where $\Theta_0$ is some constant phase that can be reabsorbed in the definition of the background phases. Then, the Rabi coupling will tend to make the two chemical potentials equal; indeed any chemical potential difference would produce a  mode that oscillates in time between a stable and a tachyonic configuration, signaling that the system is not in the ground state.

\section{Spontaneous phonon emission}
\label{sec:massive:emission}

Having obtained the low-energy effective theory for  binary superfluids,  we  consider the effect of  a background velocity profile. In particular, we are interested in determining the effect of the phonon mass on the spontaneous emission process at the acoustic horizon~\cite{Liberati:2005pr, Weinfurtner:2006wt}. Given the two Lagrangians in Eqs.\,\eqref{l2_Rabi_a} and\,\eqref{l2_Rabi} for the massless and massive phonons, respectively, the corresponding acoustic horizons are in the positions, $x^H_a$ and $x^H_b$, such that
\be
\label{eqs:hor}
v(x^H_a) = c_{sa} \quad \text{and} \quad v(x^H_b) = c_{sb}\,,
\ee
where we assumed that the flow is along the $x$ direction and thus the acoustic horizons are two planes  centered at  $x^H_a$ and  $x^H_b$. More precisely, the horizon positions can be  determined with precision  of the order of the healing length, that is the threshold scale length required  for the validity of  the hydrodynamic description of  realistic superfluids\,\cite{Dalfovo:1999zz}.
Close to the acoustic horizons, we consider that the fluid velocity is given by $\bm v = - v  \bm \hat x $ with velocity profile 
\be\label{eq:v}
v =  c_s - C x \,,
\ee
with $C$ a constant and $c_s = (c_{sa}+c_{sb})/2 $. Then  we have that
\be\label{eq:relative}
x^H_a = \frac{c_{sb}-c_{sa}}{2 C} = - x^H_b\,,
\ee
are the positions of the horizons of the two acoustic holes.  Notice that in our three-dimensional setup, the horizon must be intended as a two-dimensional surface orthogonal to the fluid flow.

For each acoustic hole it is possible to define the corresponding  Hawking temperature following the same reasoning used in one component superfluid, see for instance\,\cite{Barcelo:2005fc}. Since the two species $a$ and $b$ are independent, in principle they have distinct Hawking temperatures
\be
\label{eq:hawking}
T_{a,b} = \frac{1}{2 \pi} \left.\left(\frac{ |c_{sa,b}-v|}{1-|c_{sa,b} v| }\right)'\ \right\vert_H
 \,,
\ee
 where the prime indicates the derivative with respect to the direction orthogonal to the horizon. However, in the nonrelativistic limit the two  temperatures are equal and  given by 
\be\label{eq:T_hawking}
T \simeq \frac{C}{2 \pi}\,,
\ee
thus they only depend on the space gradient of the velocity field. Given the expression in Eq.\,\eqref{eq:relative}, and considering Eqs.\,\eqref{eqs:soundspeed}, in the nonrelativistic limit  the relative distance between the two horizons is
\be
x^H_a -x^H_b\simeq \frac{\tilde m_b -\tilde m_a}{\sqrt{2} \pi T}\,.
\ee
Since the discussion of the Hawking emission of  massless phonons is equivalent to the one of a one component superfluid, hereafter  we focus on the massive mode. For the sake of simplicity in the notation, in the following we drop the $b$ index on the corresponding quantities.

\subsection{Dispersion close to the acoustic horizon}

The  emission   of massive particles at the acoustic horizon can be studied by one of the  approaches used to explain the Hawking  radiation, for instance one may determine how the tunneling probability depends on the mass of the emitted particles\,\cite{Parikh:1999mf, LuisaChiofalo:2022ykx}.

Here we present a qualitative description of the phenomenon, based on semiclassical arguments.
For this purpose, it is useful to analyze first the dispersion law of the phonons  around the acoustic horizon. The equation of  motion of  phonons with mass $m_\text{ph}$ is
\be
g_{\mu \nu} p^\mu p^\nu = m_\text{ph}^2 c_s^2\,,
\ee
then, taking $p_\mu = (E, -p,0,0)$ and the metric from Eqs. \eqref{defg} and \eqref{l2_Rabi}, we obtain  the dispersion law 
\be
E_\pm = \frac{-(1-c_s^2) \, v \, p_x \pm \frac{1}\gamma \sqrt{c_s^2p_x^2+(1-c_s^2 v^2)m_\text{ph}^2 c_s^4}}{1 - c_s^2 v^2}\,,
\ee
which is valid in the collinear regime, meaning that both the fluid velocity   and the phonon momentum are along  the $x$ axis.  In the following, we restrict ourselves, to positive energy states  in  the nonrelativistic limit ($v\ll 1 $ and $c_s \ll 1$). In this case the  dispersion law of phonons simplifies:
\be\label{eq:massive_phonon}
E_+ =-v p_x + \sqrt{c_s^2p_x^2+m_\text{ph}^2 c_s^4}\,,
\ee
which has still a relativisticlike form, because  we did not assume that $|p_x| < m_\text{ph} c_s$. In this way we can take into account the case of an arbitrarily small  phonon mass.
 The corresponding  group velocity is given by
\be\label{eq:vg}
v_g = \frac{\mathrm{d} E_+}{\mathrm{d} p_x} = -v + \frac{c_s p_x}{\sqrt{p_x^2+m_\text{ph}^2 c_s^2}}\,,
\ee
indicating that at the acoustic horizon, corresponding to $v=c_s$, the group velocity of massive phonons is negative:   they  are  trapped unless their momenta are larger than the threshold momentum
\be\label{eq:threshold_p}
p_\text{c} = \frac{m_\text{ph} c_s v}{\sqrt{c_s^2-v^2}} \simeq \frac{m_\text{ph} c_s^{3/2}}{\sqrt{2 C x}} \simeq  \frac{m_\text{ph} c_s^{3/2}}{\sqrt{4 \pi T x}}\,,
\ee
where in the last equation we exploited Eq.\, \eqref{eq:T_hawking} ($x$ being  the distance from the horizon).   Clearly, for $m_\text{ph}=0$ we have  $p_\text{c} =0$, and this is independent of $x$. Instead, for nonvanishing phonon mass,  the above expression implies that the largest threshold momentum is obtained at the shorter distance from the horizon, that is for   $x\simeq L_c$, where $L_c$ is a distance, already introduced after Eq.\,\eqref{eq:T_hawking}, of the order of the healing length\,\cite{Dalfovo:1999zz} (see also\,\cite{Son:2001td} for a discussion of healing lengths in binary superfluids with Rabi coupling).  The threshold momentum $p_c$ in Eq.\,\eqref{eq:threshold_p} must be lower than the inverse of the healing length,
\be
p_c \ll \frac{1}{\xi} \, ,
\label{disxi}
\ee
to ensure the validity of the hydrodynamic description.  In typical superfluid experiments $T \sim 0.1 \, m c_s^2$\,\cite{Steinhauer, MunozdeNova:2018fxv}, and using Eq.\,\eqref{eq:threshold_p} we have that the above inequality can be rewritten as
\be
m_\text{ph} c_s^2 \ll   10\, T\,,
\ee
which is satisfied because $m_\text{ph} c_s^2 \lesssim 2 \sqrt{m \tilde \lambda} c_s < T$, where we used $m \simeq 100$ GeV and   $\tilde \lambda \simeq 10^{-16}$ eV.

Upon substituting Eq.\,\eqref{eq:threshold_p} in Eq.\eqref{eq:massive_phonon}, 
it follows that the threshold energy  is
\begin{align}\label{eq:E_c}
E_c =& - p_c v + \sqrt{p_c^2 c_s^2+ m_\text{ph}^2 c_s^4} \, \simeq \nonumber\\ &\simeq m_\text{ph} c_s^{2}\left( \frac{\sqrt{k}}{2} + \sqrt{1+\frac{1}{k}} -\frac{1}{\sqrt{k}} \right) \,,
\end{align}
where \be\label{eq;k} k = \frac{4 \pi\, T\, L_c}{c_s}\,. \ee 
Therefore the threshold energy is ruled not only by the Rabi coupling, entering in the phonon mass, but also by  $L_c$, the Hawking temperature, as well as by the speed of sound $c_s$.

\subsection{Thermodynamics close to the horizon}
\label{sec:thermodynamics}

At a distance from the horizon smaller than the phonon mean free path, phonons can be treated as an ideal Bose gas at the Hawking temperature\,\cite{Mannarelli:2020ebs}. Therefore, in this limit, the energy density of massive phonons is 
\be
\label{eq:epsilon}
\epsilon_\text{ph} = \int E f(p) \frac{d^3 p}{(2 \pi)^3} \simeq \frac{1}{2 \pi L_c^2} \int_{p_c}^{\infty} \frac{E}{e^{E/T}-1} dp_x \,,
\ee
where the  distribution function 
\be
f(p) =  \frac{1}{e^{E/T}-1} \, \frac{2 \pi}{L_c} \delta(p_y) \, \frac{2 \pi}{L_c} \delta(p_x) \, \theta(p_x-p_c)\,,
\ee
takes into account that phonons are only emitted in the direction orthogonal to the horizon\,\cite{Mannarelli:2021olc, LuisaChiofalo:2022ykx}, that is along the $x$ axis,  and that  the phonon  momenta  must be larger than the threshold momentum, given in Eq.\,\eqref{eq:threshold_p}.
Upon changing the variable, $p_x \to E$,
and then defining $z=E/T$, this expression can be rewritten as
\be
\label{eq:energy_density}
\epsilon_\text{ph}(m_\text{ph}) = \frac{T^2}{2 \pi L_c^2(c_s^2-v^2)}  \int_{z_c}^{\infty} \frac{v z + \frac{c_s z^2}{\sqrt{z^2-z_c^2}}}{e^{z}-1} dz \,,
\ee
where $z_c = E_c/T$. Close to the acoustic horizon, that is at a distance $x \sim L_c$,  the approximation $c_s^2 - v^2 \simeq 4 \pi T c_s L_c $ can be used, where we employed Eqs.\,\eqref{eq:v} and\,\eqref{eq:T_hawking}. Then, at
the leading order in  $E_c/T$ we find that
\be
\epsilon_\text{ph}(m_\text{ph}) \simeq \epsilon_\text{ph}(0) \left(1 - \frac{3 E_c}{\pi^2 T} \right)\,,
\ee
where
\be
\epsilon_\text{ph}(0) = \frac{T }{24 L_c^3}\,,
\ee
is the energy density of a massless phonon gas\,\cite{Mannarelli:2020ebs}.  
Naively, one may have expected a   correction to the energy density   of the order of $m_\text{ph} c_s^2/T$, but it is instead given by
\be
\frac{3 E_c}{\pi^2 T}\simeq \frac{3 m_\text{ph} c_s^2}{T \pi^2} \left( \frac{\sqrt{k}}{2} + \sqrt{1+\frac{1}{k}} -\frac{1}{\sqrt{k}} \right)\,,
 \label{mc}
\ee
where $k$ is the quantity defined  in Eq.\,\eqref{eq;k}, thus it depends on $T$ as well as on $L_c$ and $c_s$. Note that the correction is nonperturbative, in the sense that it is of order $\lambda^{1/2}$, see Eqs.\,\eqref{mb} and\,\eqref{eq:E_c}. This result is similar to the one found in the calculation of the free energy of a scalar field in thermal field theory, see for instance\,\cite{bellac2000thermal}.  Such  behavior  reflects the breakdown of perturbation theory due to the presence of infrared divergences.

We can estimate the energy density reduction due to the Rabi coupling assuming that 
$L_c = \xi$, where $\xi$ is the healing length. In the setup of\,\cite{Steinhauer, MunozdeNova:2018fxv}, where the experiment is performed with $^{87}$Rb atoms, the healing length is $\xi \simeq 2\, \mu$m,    the speed of sound is $c_s\simeq 0.3 $ mm/s, and the estimated Hawking temperature is $T\simeq 0.1\, m c_s^2 \simeq 0.1$ nK, where $m$ is the mass of a $^{87}$Rb atom. 
Upon substituting these values in Eq. \eqref{eq;k}, we obtain that   $k \simeq 0.9$. 
Then, considering that $m_\text{ph} c_s^2 < T$   we find that 
\be\label{eq:estimated_red}
\frac{3 E_c}{ \pi^2 T}  \lesssim  0.2\, . 
\ee
Therefore, the effect is a reduction of the order of $20 \%$ with respect to the same value for vanishing  Rabi coupling. Note that the effect is actually proportional to $\tilde\lambda^{1/2}$ and it is independent of the mass of the used atoms.  In a similar way, we can expand  the phonon pressure
\be\label{eq:pressure2}
P_\text{ph}(m_\text{ph}) = \frac{1}{2 \pi L_c^2}  \int_{p_c}^{\infty}  \frac{E}{e^{E/T}-1} v_g^{-1} dp \,,
\ee
obtaining 
\be\label{eq:newP}
P_\text{ph}(m_\text{ph}) \simeq P_\text{ph}(0) \left(1 - \frac{6 E_c}{\pi^2 T} \right)\,,
\ee
where 
\be\label{eq:conf1} P_\text{ph}(0) = \epsilon_\text{ph}(0)\,,
\ee
indeed for vanishing phonon masses the system is scale invariant in $(1+1)$ dimensions: as already noticed phonons are mostly emitted in the direction orthogonal to the horizon.  The effect of the phonon mass is to reduce both the energy density and the pressure with respect to the values obtained with  massless phonons, in such a way that \be \label{eq:p<e} 
P_\text{ph}(m_\text{ph}) <  \epsilon_\text{ph}(m_\text{ph})\,,
\ee
where the difference between these two quantities at the leading order in $E_c/T$ is  
\be
\epsilon_\text{ph}(m_\text{ph}) -P_\text{ph}(m_\text{ph}) \simeq  \epsilon_\text{ph}(0)  \frac{3 E_c}{\pi^2 T}\,.
\ee
Thus, the same difference  is  proportional to the square root of the Rabi coupling.

\section{Viscosity and the conformal limit close to the horizon}
\label{sec:viscosity}

As a consequence of the spontaneous phonon  emission close to the acoustic horizon,   the transonic superfluid loses energy   because a nonequilibrium (relaxation) dynamics sets in\,\cite{Mannarelli:2020ebs, Mannarelli:2021olc, LuisaChiofalo:2022ykx}: the superfluid kinetic energy is transformed in heat. At the effective level, this behavior can be characterized 
by effective bulk, $\zeta_\text{eff}$, and shear, $\eta_\text{eff}$, viscous coefficients, associated to the spontaneous phonon emission. As shown in \,\cite{LuisaChiofalo:2022ykx}, the shear-to-entropy density ratio, $\eta_\text{eff}/s_\text{ph}$, of the phonon gas emitted at the acoustic horizon saturates the so-called KSS bound. First, we briefly recap the results of \,\cite{LuisaChiofalo:2022ykx}, obtained for a single superfluid and then move to the evaluation of  the ratios $\zeta_\text{eff}/s_\text{ph}$ and $\eta_\text{eff}/s_\text{ph}$ in binary superfluids.

Let us first discuss the bulk viscosity for  a fluid flowing  with no shear. Close to the acoustic horizon, the viscous stress tensor is
\be
\label{eq:sigmap}
\sigma_{ik}' = \zeta_\text{eff} \, \delta_{ix}\delta_{kx}{\bm \nabla} \cdot {\bm v} \simeq 2 \pi \, T  \, \zeta_\text{eff} \, \delta_{ix}\delta_{kx}\,,
\ee 
where we have taken into account that phonons are mostly emitted along the direction orthogonal to the horizon and on the rhs we have used Eqs.~\eqref{eq:v} and \eqref{eq:T_hawking}. Equating $\sigma_{xx}' = P_\text{ph}$, where $P_\text{ph}$ is given in Eq.\,\eqref{eq:pressure2},  we have that
\be\label{eq:zetaeff}
 2 \pi \, T  \, \zeta_\text{eff} = P_\text{ph}\,,
\ee
which relates the bulk viscosity to the pressure of the phonon gas. This equation states that  close to the acoustic  horizon the  only considered dissipative mechanism is the spontaneous phonon emission; any other viscous process is assumed  negligible. However, it is well known that  scattering processes are dissipative and contribute to the fluid viscosity\,\cite{Khalatnikov}. 
Here we consider the phonon gas in a region very close to the acoustic horizon. The extension of this region is assumed to be smaller than the typical phonon mean free path, so that phonon scattering processes can be neglected.
Then, we use the thermodynamic relation \be \epsilon_\text{ph}= - P_\text{ph}  +  T s_\text{ph}\,, \ee combined with the relation  in Eq.\,\eqref{eq:conf1}, valid for a gas of massless particles in $1+1$ dimensions,  to rewrite Eq.\,\eqref{eq:zetaeff} as
\be\label{eq:zeta}
 2 \pi T  \zeta_\text{eff} = \frac{T s_\text{ph}}{2}\,,
\ee
and so it readily follows that 
\be
\frac{ \zeta_\text{eff}}{ s_\text{ph}} = \frac{1}{4 \pi} \,.
 \label{trgen}
\ee

In a similar way, we can obtain the ratio between the shear viscosity coefficient and the entropy density. In this case, we have to assume that the flow has shear, thus that the velocity, which is  along the $x$ direction, depends on a transverse  coordinate, for instance $v_x$ depends on the $y$ coordinate.
 In this condition, there is a ''transverse" pressure\,\cite{LuisaChiofalo:2022ykx}, such that 
\be\label{eq:eta}
\sigma_{yx}' =  K\, \eta = - \frac{K}{C} P_\text{ph} \,,
\ee
where $K = \partial_y v_x$ and $C$ is given in Eq. \eqref{eq:T_hawking}. Using the same reasoning used above, we readily obtain that
\be
\frac{ \eta_\text{eff}}{ s_\text{ph}} = \frac{1}{4 \pi}\,,
\ee
corresponding to the KSS bound. 

These results can be easily extended to  binary superfluids with vanishing Rabi coupling.  In the particular conditions considered in Sec.\,\ref{sec:fluctuations_no_Rabi},  the low-energy spectrum consists of two decoupled massless phonons that we called $a$ and $b$. 
We have assumed that both fluids move with the same velocity $v$, however each phonon mode has its specific speed of sound, see Eqs.\,\eqref{eqs:soundspeed}, meaning that in general there will be two distinct sonic horizons, corresponding to the conditions $v=c_{sa}$,  and $v=c_{sb}$  for the modes $a$ and $b$, respectively. Then, as in the single-fluid case, one can determine the effective bulk and shear viscosity to entropy ratios  for each fluid.

For nonvanishing Rabi coupling, one of the two phonons becomes massive and this is expected to change the viscosity to entropy density ratios.  The reason is that for massive phonons the relation in Eq.~\eqref{eq:conf1} does not hold. The basic reason is that the phonon mass  breaks   the scale invariance. More specifically, the needed ingredients to saturate the KSS bound in the presence of an acoustic horizon are the following:
\begin{enumerate}[label=\roman*)]
    \item close to the horizon the system is effectively $(1+1)$-dimensional;
    \item the relation ${\bm \nabla} \cdot {\bm v} = 2 \pi T$;
    \item the relation $P_\text{ph} = \epsilon_\text{ph}$.
\end{enumerate}
While conditions i) and ii) hold as well for a massive phonon gas,  condition  iii) is violated by the phonon mass. As we have shown in the previous section,  a mass term  implies Eq.\,\eqref{eq:p<e}; hence  
\be 
Ts_{\rm ph}= P_{\rm ph} + \epsilon _{\rm ph } \simeq 2 P_{\rm ph}\left(1 +  \frac{3 E_c}{2 \pi^2 T} \right)\,,
\ee
and then
\be \label{eq:vilationKSS}
\frac{\eta_{\rm eff}}{s_{\rm ph}}=\frac{\zeta_{\rm eff}}{s_{\rm ph}}\simeq \frac{1}{4 \pi }\left(1-\frac{3 E_c}{2 \pi^2 T}\right) \,,
\ee
and using the estimate in Eq.\,\eqref{eq:estimated_red}, we have a reduction of about $10\%$ with respect to the KSS bound. This is the main result of the present work. 
 It has been obtained assuming that the two horizons are sufficiently distant, so that one can separately account for their contributions to  entropy and shear viscosity.   In the case where the two horizons are very close or coincide, one should consider the combined effect of the
 emission processes of massless and massive phonons.  Since the two emission processes are independent,  the total entropy and shear viscosity  are given by the sum of the corresponding  contributions. This leads to   a factor $2$ reduction of the KSS bound violation reported in Eq.\,\eqref{eq:vilationKSS}.

Let us remark on an  important aspect of the  obtained results: they hold because  the phonon mass  is not an interaction term; therefore it does not produce phonon scatterings. Its only effect is to break the scale invariance and thus the  phonon pressure must be less than the phonon energy density. This means that  the  bulk and shear viscosities can still be written as in Eqs.\,\eqref{eq:zeta} and  \,\eqref{eq:eta}, respectively, but  now the phonon pressure is given by  Eq.\,\eqref{eq:newP}. Since the  phonon mass reduces the phonon pressure, it follows that with increasing phonon mass  the effective  viscosities are reduced, as well. \\
Concerning the effect of the interactions on the KSS bound, we observe that they immediately violate the conditions {i)} and iii) (encoding scale invariance); therefore they are expected to induce deviations from the same bound. This aspect  deserves future investigation.

\section{Conclusions}
\label{sec:conclusions}

Binary ultracold gases are versatile playgrounds for producing nontrivial phases of matter~\cite{YagoMalo24}. In the present paper,
 we have determined the properties of the superfluid ground state for a two-boson mixture coupled by intraspecies and interspecies interactions, as well as by a Rabi exchange term.  Later on, the corresponding effective low-energy excitation theory was determined. Assuming a perturbative Rabi coupling, we have shown that when the number densities of the two fluids are similar, it is possible to write the total pressure of the system as the sum of two partial pressures. Moreover, we have obtained that in the same conditions, the low-energy excitations of the binary superfluid are two decoupled modes, corresponding to the relative and  the total phase oscillations. The former is massive, with a squared mass proportional to the Rabi coupling, while the latter remains massless. 

Separating the total system as the sum of two noninteracting subsystems is useful when discussing the spontaneous phonon emission close to an acoustic horizon. Indeed, we can treat the emission of the two phonon modes as  independent processes. The spontaneous phonon emission  is an irreversible process, that at the hydrodynamic level amounts to effective viscosities. In binary superfluids,  the spontaneous emission of massless phonons  is analogous to the phonon process in a single superfluid, thus the viscosity-to-entropy ratios  saturate the KSS bound~\cite{Kovtun:2004de, LuisaChiofalo:2022ykx}: it is equal to $1/4\pi$. However, the spontaneous emission of massive phonons at the acoustic horizon violates the same bound: in this case,  ${\eta_{\rm eff}}/{s_{\rm ph}}={\zeta_{\rm eff}}/{s_{\rm ph}}<1/4\pi$. 
The violation is estimated of the order of $10 \%$ using the specific setup in \,\cite{MunozdeNova:2018fxv}. In general, the magnitude of the same deviation scales with the square root of the Rabi frequency, and it is independent on the mass of the used atoms. Therefore, it seems that to enhance the deviation from the KSS bound one needs to design systems with  the highest possible Rabi frequency  and with sufficiently tunable speeds of sound and/or fluid velocity as necessary ingredients to have an acoustic horizon. Our  setup provides a versatile tool to explore the degree of KSS-bound violation and even advance the exciting quest for a universal $\eta/s$ bound, under controllable external conditions that can be engineered in current quantum technology platforms.

 In actual experiments, the  measurements of the local values of the entropy density and  viscosity are required and seem feasible. The former may be achieved by employing  local-thermometry techniques, as in\,\cite{PhysRevLett.125.113601},  which allow to measure extremely small values of the entropy by means of radio-frequency spectroscopy. Local values of the  shear viscosity  may be obtained by appropriate inversion of the cloud-averaged viscosity\,\cite{PhysRevLett.115.020401}. \\

\section*{Acknowledgments} 
The authors are pleased to thank 
Stefano Carretta, Leonardo Fallani, Stefano Liberati, Luca Salasnich,  Paolo Santini, and Andrea Trombettoni for support and fruitful discussions. 

L.L. acknowledges financial support by a project funded under the National Recovery and Resilience Plan (NRRP), Mission 4 Component 2 Investment 1.3 - Call for tender No. 341 of 15/03/2022 of Italian Ministry of University and Research funded by the European Union – NextGenerationEU, award number PE0000023, Concession Decree No. 1564 of 11/10/2022 adopted by the Italian Ministry of University and Research, CUP D93C22000940001, Project title "National Quantum Science and Technology Institute" (NQSTI), spoke 2.

This study has been also carried out within the National Centre on HPC, Big Data and Quantum Computing - SPOKE 10 (Quantum Computing) and received funding from the European Union Next-GenerationEU - National Recovery and Resilience Plan (NRRP) – MISSION 4 COMPONENT 2, INVESTMENT N. 1.4 – CUP N. I53C22000690001.
M.L.C. acknowledges the Terrestrial Very-Long-Baseline Atom
Interferometry (TVLBAI) proto-collaboration

\section*{Data availability}
No data were created or analyzed in this study.

\appendix

\section{General derivations of the acoustic metric and of the low-energy Lagrangian}
\label{sec:appendix}

We determine the effective low-energy Lagrangian for a superfluid, using two different methods. For simplicity, we consider a single-species  boson gas, but the two approaches can be straightforwardly extended  to binary superfluids. 

\subsection{Expansion around mean field of a generic Lagrangian}
In this first method we employ the same procedure used in Sec.\,\ref{sec:onebec}, but we start from an unspecified Lagrangian  for a complex scalar field in the Madelung representation,
\be
\mathcal{L}\equiv\mathcal{L}(\rho, \partial_\mu\rho,  \partial_\mu\theta)\,,
\ee
where $\rho$ is the radial field and $\theta$ is the cyclic coordinate. We assume vanishing temperature and  that the Lagrangian is invariant under a global $U(1)$ transformation. In the ground state, such a group is assumed to be  spontaneously broken.  

We include  fluctuations on  top of the mean field by the as in Eq.\,\eqref{eq:rhoexp}, hence
 $\rho$ and $\theta$ now indicate the solutions of the classical equation of motion, while $\tr$ and $\tilteta$  are the quantum fluctuations.
  Expanding in powers of the fluctuations,  and neglecting surface terms, we have that
 \begin{align}
{\cal L} =  &{\cal L}_0(\rho, \partial_\mu \rho,  \partial_\mu \theta) + \tr \left.\left(\frac{\delta \el}{\delta \rho} - \partial_\mu \frac{\delta \el}{\delta \partial_\mu\rho} \right)\right\vert_{\rho,\theta} \nonumber \\
&+\delta \partial_\mu\tilteta\left.\frac{\delta \el}{\delta \partial_\mu\theta}\right\vert_{\rho,\theta} + \el_2+ \dots
 \end{align}
 where the dots indicate cubic  and higher terms in the fluctuations. The first term on the right-hand side,  when evaluated in the stationary point, is the  background pressure: it does not include the effect of fluctuations; in the model with quartic interactions it is given in Eq.\,\eqref{eq:pressure}. 
 Since $\rho$ and $\theta$ are solutions of the classical equations of motion, it follows that  the terms linear in the perturbations vanish. Thus the first nontrivial term is the  quadratic  Lagrangian
 \begin{align}\label{eq:L2}
\el_2 =&\frac{\tr^2}{2} \left.\frac{\delta^2 \el}{\delta \rho^2}\right\vert_{\rho,\theta} + \tr \partial_\mu \tr  \left.\frac{\delta^2 \el}{\delta \rho\delta \partial_\mu \rho}\right\vert_{\rho,\theta} 
+ \frac{\partial_\mu\tr \partial_\nu\tr}{2} \left.\frac{\delta^2 \el}{\delta \partial_\mu \rho \delta\partial_\nu \rho}\right\vert_{\rho,\theta} \nonumber \\
& +  \tr \partial_\mu \tilteta  \left.\frac{\delta^2 \el}{\delta\rho \delta \partial_\mu \theta}\right\vert_{\rho,\theta} \,  +\partial_\nu \tr \partial_\mu \tilteta  \left.\frac{\delta^2 \el}{\delta\partial_\nu \rho \delta \partial_\mu \theta}\right\vert_{\rho,\theta} \nonumber \\
&+ \frac{\partial_\nu \tilteta \partial_\mu \tilteta}2  \left.\frac{\delta^2 \el}{\delta\partial_\nu \theta \delta \partial_\mu \theta}\right\vert_{\rho,\theta}\,,
 \end{align}
where we have taken into account that $\tilteta$ has only  derivative couplings.   

Let us discuss in some details the various terms in the right-hand side of Eq.\,\eqref{eq:L2}.  The first term is proportional to minus the  curvature of the potential at the minimum: 
 \be
 \label{eq:mtilde}
\frac{\tr^2}{2}\left.\frac{\delta ^2\el}{\delta \rho ^2}\right\vert_{\rho,\theta} = -\frac{\tilde m^2}2 \tr^2 \,,
 \ee
where $\tilde m$ is the effective mass of the radial field. The second term is  a total derivative and can be neglected.  The third term is 
 \be
\frac{\partial_\mu\tr \partial_\nu\tr}{2} \left.\frac{\delta^2 \el}{\delta \partial_\mu \rho \delta\partial_\nu \rho}\right\vert_{\rho,\theta} = \frac{K^{\mu \nu}}2 \partial_\mu\tr \partial_\nu\tr\,,
 \ee
where $K^{\mu \nu}$ is  a matrix depending on the background; in flat spacetime  $K^{\mu \nu} = \eta^{\mu \nu}$.
 The fourth term can be written as
\be
\tr \partial_\mu \tilteta  \left.\frac{\delta^2 \el}{\delta\rho \delta \partial_\mu \theta}\right\vert_{\rho,\theta}= V^\mu \tr \, \partial_\mu  \tilteta\,,
\ee
which is a  mixing term. Since  $V^\mu$   is by construction a four-vector, it explicitly breaks the Lorentz symmetry. By dimensional analysis it follows that it has dimension $2$. The fifth term can be written as
\be \partial_\nu \tr \partial_\mu \tilteta  \left.\frac{\delta^2 \el}{\delta\partial_\nu \rho \delta \partial_\mu \theta}\right\vert_{\rho,\theta} = A \,  \eta^{\mu\nu} \partial_\nu \tr \partial_\mu \tilteta\,,
\ee
where $A$ is a constant of dimension $1$.
It vanishes if the  interactions  in $\cal L $ stem from a potential term, thus the radial field has no derivative couplings. Finally, the last term can be written as
 \be
 \frac{\partial_\nu \tilteta \partial_\mu \tilteta}2  \left.\frac{\delta^2 \el}{\delta\partial_\nu \theta \delta \partial_\mu \theta}\right\vert_{\rho,\theta} = \frac{L^{\mu\nu}}2 \partial_\nu \tilteta \partial_\mu \tilteta\,,
 \ee
where $L^{\mu\nu}$ depends on the background; in flat spacetime  $L^{\mu\nu} = B^2 \eta^{\mu\nu}$, where $B$ has dimension $1$.

In the end, the quadratic Lagrangian reads
\begin{align}\label{eq:L2n}
\el_2 =& \frac{1}2 \partial_\nu\tr \partial^\nu\tr-\frac{\tilde m^2}2 \tr^2 + V^\mu \tr \partial_\mu \tilteta + \frac{B^2}2 \partial_\nu \tilteta \partial^\nu \tilteta+A   \partial^\mu \tr \partial_\mu \tilteta\,,
\end{align}
where $\tilteta$ is clearly the NGB associated to the spontaneous symmetry breaking of the $U(1)$ global symmetry. 

Since $\tr$ field is massive, we can integrate it out. Its equation of motion reads,
\be
(\square + \tilde m^2) \tr = V^\mu \partial_\mu \tilteta - A \square \tilteta  \,,
\ee
which we can formally invert to obtain
\be\label{eq:substitute}
\tr = \frac{1}{\square + \tilde m^2} \left( V^\mu \partial_\mu \tilteta-A \square \tilteta\right)\,,
\ee
showing a generic feature of effective-field theories: the non locality. However, higher derivative terms are suppressed as powers of $p^2/\tilde m^2$, where $p$ is the momentum of the NGB. Note that the relevant scale for the expansion is $\tilde m$ and not $m$. Since 
$\tilde m$ vanishes close to the second order phase transition point, it follows that the energy and momentum range of validity of the low-energy effective Lagrangian shrinks close to the transition to the normal phase.

Upon substituting Eq.\,\eqref{eq:substitute} in  Eq. \eqref{eq:L2n},  we obtain
\begin{align}
\el_2 =& \frac{1}2  \left(V^\mu  \partial_\mu \tilteta - A\square \tilteta\right) \frac{1}{\square + \tilde m^2} \left(V^\mu  \partial_\mu \tilteta - A\square \tilteta\right) \nonumber \\
&+ \frac{B^2 \eta^{\mu\nu}}2 \partial_\nu \tilteta \partial_\mu \tilteta \,,
\end{align}
which is the most general form of the low-energy effective Lagrangian. At the leading order in  momenta, 
we replace
\be
 \frac{1}{\square + \tilde m^2} \to  \frac{1}{\tilde m^2}\,,\quad V^\mu  \partial_\mu \tilteta - A\square \tilteta\to V^\mu  \partial_\mu \tilteta \,,
\ee
so that we can rewrite the effective low-energy Lagrangian as
\begin{align}\label{eq:lag_effective_gen}
\el_2 \simeq&  \frac{1}2 \left( B^2 \eta^{\mu\nu} + \frac{V^\mu V^\nu}{\tilde m^2}\right)  \partial_\nu \tilteta \partial_\mu \tilteta \equiv \frac{\sqrt{- g} g^{\mu\nu}}2\partial_\nu \tilteta \partial_\mu \tilteta\,,
\end{align}
where $g_{\mu\nu}$ is the acoustic metric.
The obtained  low-energy Lagrangian  is valid for momenta $p \ll \tilde m$; alternatively, it  can be  obtained neglecting the kinetic term   $\frac{1}2 \partial_\nu\tr \partial^\nu\tr$ and taking as well  $A=0$ in Eq.\,\eqref{eq:L2n}.

\subsection{Derivative expansion of the pressure}
We now present a different method to derive   the low-energy theory and  the acoustic metric, based on the observation that  the low-energy Lagrangian at the leading order in momenta of any system with a $U(1)$ broken symmetry can be determined from the knowledge of the pressure\,\cite{son2002,manuel2010}.
This method has the advantage of bypassing the integration of the radial field: we start directly with the  effective Lagrangian of the $\tilteta$ field. However, it has the disadvantage to include only the leading order  terms in the phonon momentum.  Taking the global symmetry as a local one, the low-energy Lagrangian including leading order derivatives is given  by
\be\label{eq:functional}
{\cal L} = P \left[\sqrt{D_\mu \tilteta D^\mu \tilteta} \right]\,,
\ee
where $D_\mu \tilteta = \partial_\mu \tilteta - A_\mu $ and  $P$ is a functional having the same algebraic expression of the  pressure. As in Eq.\,\eqref{eq:covder},  we take the  external field given by
\be
A_\nu = \frac{\mu}{\gamma} v_\nu\,, 
\ee
thus we assume a background  flow with four-velocity $ v_\nu$. Note that the present definition of the covariant derivative slightly differs from the expression in Eq.\,\eqref{eq:covder}, because here it acts on the phase field. For convenience we replace $\mu/\gamma \to \mu$. Upon expanding the right hand side of  Eq.\,\eqref{eq:functional} around $\mu$, we obtain that 
\begin{align}
{\cal L} =& P(\mu) - \frac{\partial P}{\partial \mu} v_\mu \partial^\mu \tilteta \nonumber \\
&+ \frac{1}2 \left[ \eta^{\mu \nu} \frac{1}{\mu}  \frac{\partial P}{\partial \mu}  + v^\mu v^\nu \left(\frac{\partial^2 P}{\partial \mu^2} - \frac{1}{\mu}  \frac{\partial P}{\partial \mu} \right)  \right] \partial_\nu \tilteta \partial_\mu \tilteta\,,
\end{align}
thus the quadratic term is
\be
{\cal L}_2 = \frac{n}{2 \mu} \left(\eta^{\mu \nu} + \left(\frac{1}{c_s^2} -1\right) v^\mu v^\nu \right)\partial_\nu \tilteta \partial_\mu  \tilteta\,,
\label{quad}
\ee
where we  used the thermodynamic relations
\be
\label{eq:thermo}
n = \frac{\partial P}{\partial \mu} \quad \text{and}  \quad c_s^2 = \frac{n}{\mu\frac{\partial^2 P}{\partial \mu^2}} \,, 
\ee
the second expression stemming from $c_s^2 = \frac{\partial P}{\partial \rho}$.
The low-energy Lagrangian can be further expanded including higher orders of  $\partial_\mu  \tilteta$; this generates interaction terms at the leading order in the momenta, see the discussion in\,\cite{manuel2010}.


\begin{thebibliography}{84}%
\makeatletter
\providecommand \@ifxundefined [1]{%
 \@ifx{#1\undefined}
}%
\providecommand \@ifnum [1]{%
 \ifnum #1\expandafter \@firstoftwo
 \else \expandafter \@secondoftwo
 \fi
}%
\providecommand \@ifx [1]{%
 \ifx #1\expandafter \@firstoftwo
 \else \expandafter \@secondoftwo
 \fi
}%
\providecommand \natexlab [1]{#1}%
\providecommand \enquote  [1]{``#1''}%
\providecommand \bibnamefont  [1]{#1}%
\providecommand \bibfnamefont [1]{#1}%
\providecommand \citenamefont [1]{#1}%
\providecommand \href@noop [0]{\@secondoftwo}%
\providecommand \href [0]{\begingroup \@sanitize@url \@href}%
\providecommand \@href[1]{\@@startlink{#1}\@@href}%
\providecommand \@@href[1]{\endgroup#1\@@endlink}%
\providecommand \@sanitize@url [0]{\catcode `\\12\catcode `\$12\catcode
  `\&12\catcode `\#12\catcode `\^12\catcode `\_12\catcode `\%12\relax}%
\providecommand \@@startlink[1]{}%
\providecommand \@@endlink[0]{}%
\providecommand \url  [0]{\begingroup\@sanitize@url \@url }%
\providecommand \@url [1]{\endgroup\@href {#1}{\urlprefix }}%
\providecommand \urlprefix  [0]{URL }%
\providecommand \Eprint [0]{\href }%
\providecommand \doibase [0]{http://dx.doi.org/}%
\providecommand \selectlanguage [0]{\@gobble}%
\providecommand \bibinfo  [0]{\@secondoftwo}%
\providecommand \bibfield  [0]{\@secondoftwo}%
\providecommand \translation [1]{[#1]}%
\providecommand \BibitemOpen [0]{}%
\providecommand \bibitemStop [0]{}%
\providecommand \bibitemNoStop [0]{.\EOS\space}%
\providecommand \EOS [0]{\spacefactor3000\relax}%
\providecommand \BibitemShut  [1]{\csname bibitem#1\endcsname}%
\let\auto@bib@innerbib\@empty
\bibitem [{\citenamefont {Barcelo}\ \emph {et~al.}(2005)\citenamefont
  {Barcelo}, \citenamefont {Liberati},\ and\ \citenamefont
  {Visser}}]{Barcelo:2005fc}%
  \BibitemOpen
  \bibfield  {author} {\bibinfo {author} {\bibfnamefont {C.}~\bibnamefont
  {Barcelo}}, \bibinfo {author} {\bibfnamefont {S.}~\bibnamefont {Liberati}}, \
  and\ \bibinfo {author} {\bibfnamefont {M.}~\bibnamefont {Visser}},\ }\href
  {\doibase 10.12942/lrr-2005-12} {\bibfield  {journal} {\bibinfo  {journal}
  {Liv. Rev. Rel.}\ }\textbf {\bibinfo {volume} {8}},\ \bibinfo {pages} {12}
  (\bibinfo {year} {2005})},\ \Eprint {http://arxiv.org/abs/gr-qc/0505065}
  {arXiv:gr-qc/0505065} \BibitemShut {NoStop}%
\bibitem [{\citenamefont {Altman}\ \emph {et~al.}(2021)\citenamefont {Altman}
  \emph {et~al.}}]{Altman:2019vbv}%
  \BibitemOpen
  \bibfield  {author} {\bibinfo {author} {\bibfnamefont {E.}~\bibnamefont
  {Altman}} \emph {et~al.},\ }\href {\doibase 10.1103/PRXQuantum.2.017003}
  {\bibfield  {journal} {\bibinfo  {journal} {PRX Quantum}\ }\textbf {\bibinfo
  {volume} {2}},\ \bibinfo {pages} {017003} (\bibinfo {year} {2021})},\ \Eprint
  {http://arxiv.org/abs/1912.06938} {arXiv:1912.06938 [quant-ph]} \BibitemShut
  {NoStop}%
\bibitem [{\citenamefont {Warszawski}\ and\ \citenamefont
  {Melatos}(2011)}]{Warszawski:2011vy}%
  \BibitemOpen
  \bibfield  {author} {\bibinfo {author} {\bibfnamefont {L.}~\bibnamefont
  {Warszawski}}\ and\ \bibinfo {author} {\bibfnamefont {A.}~\bibnamefont
  {Melatos}},\ }\href {\doibase 10.1111/j.1365-2966.2011.18803.x} {\bibfield
  {journal} {\bibinfo  {journal} {Mon. Not. Roy. Astron. Soc.}\ }\textbf
  {\bibinfo {volume} {415}},\ \bibinfo {pages} {1611} (\bibinfo {year}
  {2011})},\ \Eprint {http://arxiv.org/abs/1103.6090} {arXiv:1103.6090
  [astro-ph.SR]} \BibitemShut {NoStop}%
\bibitem [{\citenamefont {Warszawski}\ \emph {et~al.}(2012)\citenamefont
  {Warszawski}, \citenamefont {Melatos},\ and\ \citenamefont
  {Berloff}}]{Warszawski:2012ns}%
  \BibitemOpen
  \bibfield  {author} {\bibinfo {author} {\bibfnamefont {L.}~\bibnamefont
  {Warszawski}}, \bibinfo {author} {\bibfnamefont {A.}~\bibnamefont {Melatos}},
  \ and\ \bibinfo {author} {\bibfnamefont {N.}~\bibnamefont {Berloff}},\ }\href
  {\doibase 10.1103/PhysRevB.85.104503} {\bibfield  {journal} {\bibinfo
  {journal} {Phys. Rev. B}\ }\textbf {\bibinfo {volume} {85}},\ \bibinfo
  {pages} {104503} (\bibinfo {year} {2012})},\ \Eprint
  {http://arxiv.org/abs/1203.5133} {arXiv:1203.5133 [cond-mat.other]}
  \BibitemShut {NoStop}%
\bibitem [{\citenamefont {Faccio}\ \emph {et~al.}(2013)\citenamefont {Faccio},
  \citenamefont {Belgiorno}, \citenamefont {Cacciatori}, \citenamefont
  {Gorini}, \citenamefont {Liberati},\ and\ \citenamefont
  {Moschella}}]{Faccio:2013kpa}%
  \BibitemOpen
  \bibinfo {editor} {\bibfnamefont {D.}~\bibnamefont {Faccio}}, \bibinfo
  {editor} {\bibfnamefont {F.}~\bibnamefont {Belgiorno}}, \bibinfo {editor}
  {\bibfnamefont {S.}~\bibnamefont {Cacciatori}}, \bibinfo {editor}
  {\bibfnamefont {V.}~\bibnamefont {Gorini}}, \bibinfo {editor} {\bibfnamefont
  {S.}~\bibnamefont {Liberati}}, \ and\ \bibinfo {editor} {\bibfnamefont
  {U.}~\bibnamefont {Moschella}},\ eds.,\ \href {\doibase
  10.1007/978-3-319-00266-8} {\emph {\bibinfo {title} {{Analog. Grav.
  Phenomen.}}}},\ Vol.\ \bibinfo {volume} {870}\ (\bibinfo {year}
  {2013})\BibitemShut {NoStop}%
\bibitem [{\citenamefont {Poli}\ \emph {et~al.}(2023)\citenamefont {Poli},
  \citenamefont {Bland}, \citenamefont {White}, \citenamefont {Mark},
  \citenamefont {Ferlaino}, \citenamefont {Trabucco},\ and\ \citenamefont
  {Mannarelli}}]{Poli:2023vyp}%
  \BibitemOpen
  \bibfield  {author} {\bibinfo {author} {\bibfnamefont {E.}~\bibnamefont
  {Poli}}, \bibinfo {author} {\bibfnamefont {T.}~\bibnamefont {Bland}},
  \bibinfo {author} {\bibfnamefont {S.~J.~M.}\ \bibnamefont {White}}, \bibinfo
  {author} {\bibfnamefont {M.~J.}\ \bibnamefont {Mark}}, \bibinfo {author}
  {\bibfnamefont {F.}~\bibnamefont {Ferlaino}}, \bibinfo {author}
  {\bibfnamefont {S.}~\bibnamefont {Trabucco}}, \ and\ \bibinfo {author}
  {\bibfnamefont {M.}~\bibnamefont {Mannarelli}},\ }\href {\doibase
  10.1103/PhysRevLett.131.223401} {\bibfield  {journal} {\bibinfo  {journal}
  {Phys. Rev. Lett.}\ }\textbf {\bibinfo {volume} {131}},\ \bibinfo {pages}
  {223401} (\bibinfo {year} {2023})},\ \Eprint
  {http://arxiv.org/abs/2306.09698} {arXiv:2306.09698 [cond-mat.quant-gas]}
  \BibitemShut {NoStop}%
\bibitem [{\citenamefont {Bland}\ \emph {et~al.}(2024)\citenamefont {Bland},
  \citenamefont {Ferlaino}, \citenamefont {Mannarelli}, \citenamefont {Poli},\
  and\ \citenamefont {Trabucco}}]{Bland:2024klj}%
  \BibitemOpen
  \bibfield  {author} {\bibinfo {author} {\bibfnamefont {T.}~\bibnamefont
  {Bland}}, \bibinfo {author} {\bibfnamefont {F.}~\bibnamefont {Ferlaino}},
  \bibinfo {author} {\bibfnamefont {M.}~\bibnamefont {Mannarelli}}, \bibinfo
  {author} {\bibfnamefont {E.}~\bibnamefont {Poli}}, \ and\ \bibinfo {author}
  {\bibfnamefont {S.}~\bibnamefont {Trabucco}},\ }\href {\doibase
  10.1007/s00601-024-01949-7} {\bibfield  {journal} {\bibinfo  {journal} {Few
  Body Syst.}\ }\textbf {\bibinfo {volume} {65}},\ \bibinfo {pages} {81}
  (\bibinfo {year} {2024})},\ \Eprint {http://arxiv.org/abs/2407.03212}
  {arXiv:2407.03212 [cond-mat.quant-gas]} \BibitemShut {NoStop}%
\bibitem [{\citenamefont {Yago~Malo}\ \emph {et~al.}(2024)\citenamefont
  {Yago~Malo}, \citenamefont {Lepori}, \citenamefont {Gentini},\ and\
  \citenamefont {Chiofalo}}]{YagoMalo24}%
  \BibitemOpen
  \bibfield  {author} {\bibinfo {author} {\bibfnamefont {J.}~\bibnamefont
  {Yago~Malo}}, \bibinfo {author} {\bibfnamefont {L.}~\bibnamefont {Lepori}},
  \bibinfo {author} {\bibfnamefont {L.}~\bibnamefont {Gentini}}, \ and\
  \bibinfo {author} {\bibfnamefont {M.~L.}\ \bibnamefont {Chiofalo}},\ }\href
  {\doibase 10.3390/technologies12050064} {\bibfield  {journal} {\bibinfo
  {journal} {Technologies}\ }\textbf {\bibinfo {volume} {12}} (\bibinfo {year}
  {2024}),\ 10.3390/technologies12050064}\BibitemShut {NoStop}%
\bibitem [{\citenamefont {Cipriani}\ \emph {et~al.}(2024)\citenamefont
  {Cipriani}, \citenamefont {Mannarelli}, \citenamefont {Nesti},\ and\
  \citenamefont {Trabucco}}]{Cipriani:2024bcc}%
  \BibitemOpen
  \bibfield  {author} {\bibinfo {author} {\bibfnamefont {L.}~\bibnamefont
  {Cipriani}}, \bibinfo {author} {\bibfnamefont {M.}~\bibnamefont
  {Mannarelli}}, \bibinfo {author} {\bibfnamefont {F.}~\bibnamefont {Nesti}}, \
  and\ \bibinfo {author} {\bibfnamefont {S.}~\bibnamefont {Trabucco}},\ }\href
  {\doibase 10.1103/PhysRevD.110.L021301} {\bibfield  {journal} {\bibinfo
  {journal} {Phys. Rev. D}\ }\textbf {\bibinfo {volume} {110}},\ \bibinfo
  {pages} {L021301} (\bibinfo {year} {2024})},\ \Eprint
  {http://arxiv.org/abs/2403.03833} {arXiv:2403.03833 [astro-ph.CO]}
  \BibitemShut {NoStop}%
\bibitem [{\citenamefont {Coviello}\ \emph {et~al.}(2024)\citenamefont
  {Coviello}, \citenamefont {Chiofalo}, \citenamefont {Grasso}, \citenamefont
  {Liberati}, \citenamefont {Mannarelli},\ and\ \citenamefont
  {Trabucco}}]{Coviello:2024vht}%
  \BibitemOpen
  \bibfield  {author} {\bibinfo {author} {\bibfnamefont {C.}~\bibnamefont
  {Coviello}}, \bibinfo {author} {\bibfnamefont {M.~L.}\ \bibnamefont
  {Chiofalo}}, \bibinfo {author} {\bibfnamefont {D.}~\bibnamefont {Grasso}},
  \bibinfo {author} {\bibfnamefont {S.}~\bibnamefont {Liberati}}, \bibinfo
  {author} {\bibfnamefont {M.}~\bibnamefont {Mannarelli}}, \ and\ \bibinfo
  {author} {\bibfnamefont {S.}~\bibnamefont {Trabucco}},\ }\href@noop {} {\
  (\bibinfo {year} {2024})},\ \Eprint {http://arxiv.org/abs/2410.00264}
  {arXiv:2410.00264 [gr-qc]} \BibitemShut {NoStop}%
\bibitem [{\citenamefont {Magierski}\ \emph {et~al.}(2024)\citenamefont
  {Magierski}, \citenamefont {Barresi}, \citenamefont {Makowski}, \citenamefont
  {Pcak},\ and\ \citenamefont {Wlazlowski}}]{Magierski:2024gvu}%
  \BibitemOpen
  \bibfield  {author} {\bibinfo {author} {\bibfnamefont {P.}~\bibnamefont
  {Magierski}}, \bibinfo {author} {\bibfnamefont {A.}~\bibnamefont {Barresi}},
  \bibinfo {author} {\bibfnamefont {A.}~\bibnamefont {Makowski}}, \bibinfo
  {author} {\bibfnamefont {D.}~\bibnamefont {Pcak}}, \ and\ \bibinfo {author}
  {\bibfnamefont {G.}~\bibnamefont {Wlazlowski}},\ }\href {\doibase
  10.1140/epja/s10050-024-01378-4} {\bibfield  {journal} {\bibinfo  {journal}
  {Eur. Phys. J. A}\ }\textbf {\bibinfo {volume} {60}},\ \bibinfo {pages} {186}
  (\bibinfo {year} {2024})},\ \Eprint {http://arxiv.org/abs/2406.14158}
  {arXiv:2406.14158 [cond-mat.quant-gas]} \BibitemShut {NoStop}%
\bibitem [{\citenamefont {Tu}\ and\ \citenamefont {Li}(2024)}]{Tu:2024sam}%
  \BibitemOpen
  \bibfield  {author} {\bibinfo {author} {\bibfnamefont {Z.-H.}\ \bibnamefont
  {Tu}}\ and\ \bibinfo {author} {\bibfnamefont {A.}~\bibnamefont {Li}},\
  }\href@noop {} {\  (\bibinfo {year} {2024})},\ \Eprint
  {http://arxiv.org/abs/2412.09219} {arXiv:2412.09219 [nucl-th]} \BibitemShut
  {NoStop}%
\bibitem [{\citenamefont {Sch\"utzhold}(2025)}]{Schutzhold:2025qna}%
  \BibitemOpen
  \bibfield  {author} {\bibinfo {author} {\bibfnamefont {R.}~\bibnamefont
  {Sch\"utzhold}},\ }\href@noop {} {\  (\bibinfo {year} {2025})},\ \Eprint
  {http://arxiv.org/abs/2501.03785} {arXiv:2501.03785 [quant-ph]} \BibitemShut
  {NoStop}%
\bibitem [{\citenamefont {Liberati}\ \emph
  {et~al.}(2006{\natexlab{a}})\citenamefont {Liberati}, \citenamefont
  {Visser},\ and\ \citenamefont {Weinfurtner}}]{Liberati:2005pr}%
  \BibitemOpen
  \bibfield  {author} {\bibinfo {author} {\bibfnamefont {S.}~\bibnamefont
  {Liberati}}, \bibinfo {author} {\bibfnamefont {M.}~\bibnamefont {Visser}}, \
  and\ \bibinfo {author} {\bibfnamefont {S.}~\bibnamefont {Weinfurtner}},\
  }\href {\doibase 10.1103/PhysRevLett.96.151301} {\bibfield  {journal}
  {\bibinfo  {journal} {Phys. Rev. Lett.}\ }\textbf {\bibinfo {volume} {96}},\
  \bibinfo {pages} {151301} (\bibinfo {year} {2006}{\natexlab{a}})},\ \Eprint
  {http://arxiv.org/abs/gr-qc/0512139} {arXiv:gr-qc/0512139} \BibitemShut
  {NoStop}%
\bibitem [{\citenamefont {Adams}\ \emph {et~al.}(2012)\citenamefont {Adams},
  \citenamefont {Carr}, \citenamefont {Schäfer}, \citenamefont {Steinberg},\
  and\ \citenamefont {Thomas}}]{Adams_2012}%
  \BibitemOpen
  \bibfield  {author} {\bibinfo {author} {\bibfnamefont {A.}~\bibnamefont
  {Adams}}, \bibinfo {author} {\bibfnamefont {L.~D.}\ \bibnamefont {Carr}},
  \bibinfo {author} {\bibfnamefont {T.}~\bibnamefont {Schäfer}}, \bibinfo
  {author} {\bibfnamefont {P.}~\bibnamefont {Steinberg}}, \ and\ \bibinfo
  {author} {\bibfnamefont {J.~E.}\ \bibnamefont {Thomas}},\ }\href {\doibase
  10.1088/1367-2630/14/11/115009} {\bibfield  {journal} {\bibinfo  {journal}
  {New J. Phys.}\ }\textbf {\bibinfo {volume} {14}},\ \bibinfo {pages} {115009}
  (\bibinfo {year} {2012})}\BibitemShut {NoStop}%
\bibitem [{\citenamefont {Kovtun}\ \emph {et~al.}(2005)\citenamefont {Kovtun},
  \citenamefont {Son},\ and\ \citenamefont {Starinets}}]{Kovtun:2004de}%
  \BibitemOpen
  \bibfield  {author} {\bibinfo {author} {\bibfnamefont {P.}~\bibnamefont
  {Kovtun}}, \bibinfo {author} {\bibfnamefont {D.~T.}\ \bibnamefont {Son}}, \
  and\ \bibinfo {author} {\bibfnamefont {A.~O.}\ \bibnamefont {Starinets}},\
  }\href {\doibase 10.1103/PhysRevLett.94.111601} {\bibfield  {journal}
  {\bibinfo  {journal} {Phys. Rev. Lett.}\ }\textbf {\bibinfo {volume} {94}},\
  \bibinfo {pages} {111601} (\bibinfo {year} {2005})},\ \Eprint
  {http://arxiv.org/abs/hep-th/0405231} {arXiv:hep-th/0405231} \BibitemShut
  {NoStop}%
\bibitem [{\citenamefont {Cremonini}(2011)}]{Cremonini}%
  \BibitemOpen
  \bibfield  {author} {\bibinfo {author} {\bibfnamefont {S.}~\bibnamefont
  {Cremonini}},\ }\href {\doibase 10.1142/S0217984911027315} {\bibfield
  {journal} {\bibinfo  {journal} {Mod. Phys. Lett B}\ }\textbf {\bibinfo
  {volume} {25}},\ \bibinfo {pages} {1867} (\bibinfo {year} {2011})},\ \Eprint
  {http://arxiv.org/abs/https://doi.org/10.1142/S0217984911027315}
  {https://doi.org/10.1142/S0217984911027315} \BibitemShut {NoStop}%
\bibitem [{\citenamefont {Mannarelli}\ and\ \citenamefont
  {Manuel}(2010)}]{manuel2010}%
  \BibitemOpen
  \bibfield  {author} {\bibinfo {author} {\bibfnamefont {M.}~\bibnamefont
  {Mannarelli}}\ and\ \bibinfo {author} {\bibfnamefont {C.}~\bibnamefont
  {Manuel}},\ }\href {\doibase 10.1103/PhysRevD.81.043002} {\bibfield
  {journal} {\bibinfo  {journal} {Phys. Rev. D}\ }\textbf {\bibinfo {volume}
  {81}},\ \bibinfo {pages} {043002} (\bibinfo {year} {2010})}\BibitemShut
  {NoStop}%
\bibitem [{\citenamefont {L.~Chiofalo}\ \emph {et~al.}(2024)\citenamefont
  {L.~Chiofalo}, \citenamefont {Grasso}, \citenamefont {Mannarelli},\ and\
  \citenamefont {Trabucco}}]{LuisaChiofalo:2022ykx}%
  \BibitemOpen
  \bibfield  {author} {\bibinfo {author} {\bibfnamefont {M.}~\bibnamefont
  {L.~Chiofalo}}, \bibinfo {author} {\bibfnamefont {D.}~\bibnamefont {Grasso}},
  \bibinfo {author} {\bibfnamefont {M.}~\bibnamefont {Mannarelli}}, \ and\
  \bibinfo {author} {\bibfnamefont {S.}~\bibnamefont {Trabucco}},\ }\href
  {\doibase 10.1088/1367-2630/ad4628} {\bibfield  {journal} {\bibinfo
  {journal} {New J. Phys.}\ }\textbf {\bibinfo {volume} {26}},\ \bibinfo
  {pages} {053021} (\bibinfo {year} {2024})},\ \Eprint
  {http://arxiv.org/abs/2202.13790} {arXiv:2202.13790 [gr-qc]} \BibitemShut
  {NoStop}%
\bibitem [{\citenamefont {Cherman}\ \emph {et~al.}(2008)\citenamefont
  {Cherman}, \citenamefont {Cohen},\ and\ \citenamefont
  {Hohler}}]{Cherman:2007fj}%
  \BibitemOpen
  \bibfield  {author} {\bibinfo {author} {\bibfnamefont {A.}~\bibnamefont
  {Cherman}}, \bibinfo {author} {\bibfnamefont {T.~D.}\ \bibnamefont {Cohen}},
  \ and\ \bibinfo {author} {\bibfnamefont {P.~M.}\ \bibnamefont {Hohler}},\
  }\href {\doibase 10.1088/1126-6708/2008/02/026} {\bibfield  {journal}
  {\bibinfo  {journal} {JHEP}\ }\textbf {\bibinfo {volume} {02}},\ \bibinfo
  {pages} {026} (\bibinfo {year} {2008})},\ \Eprint
  {http://arxiv.org/abs/0708.4201} {arXiv:0708.4201 [hep-th]} \BibitemShut
  {NoStop}%
\bibitem [{\citenamefont {Ge}\ \emph {et~al.}(2020)\citenamefont {Ge},
  \citenamefont {Jian}, \citenamefont {Wang}, \citenamefont {Xian},\ and\
  \citenamefont {Yao}}]{Ge}%
  \BibitemOpen
  \bibfield  {author} {\bibinfo {author} {\bibfnamefont {X.-H.}\ \bibnamefont
  {Ge}}, \bibinfo {author} {\bibfnamefont {S.-K.}\ \bibnamefont {Jian}},
  \bibinfo {author} {\bibfnamefont {Y.-L.}\ \bibnamefont {Wang}}, \bibinfo
  {author} {\bibfnamefont {Z.-Y.}\ \bibnamefont {Xian}}, \ and\ \bibinfo
  {author} {\bibfnamefont {H.}~\bibnamefont {Yao}},\ }\href {\doibase
  10.1103/PhysRevResearch.2.023366} {\bibfield  {journal} {\bibinfo  {journal}
  {Phys. Rev. Res.}\ }\textbf {\bibinfo {volume} {2}},\ \bibinfo {pages}
  {023366} (\bibinfo {year} {2020})}\BibitemShut {NoStop}%
\bibitem [{\citenamefont {Sadeghi}\ and\ \citenamefont
  {Parvizi}(2016)}]{Sadeghi}%
  \BibitemOpen
  \bibfield  {author} {\bibinfo {author} {\bibfnamefont {M.}~\bibnamefont
  {Sadeghi}}\ and\ \bibinfo {author} {\bibfnamefont {S.}~\bibnamefont
  {Parvizi}},\ }\href {\doibase 10.1088/0264-9381/33/3/035005} {\bibfield
  {journal} {\bibinfo  {journal} {Class. Quant. Gravity}\ }\textbf {\bibinfo
  {volume} {33}},\ \bibinfo {pages} {035005} (\bibinfo {year}
  {2016})}\BibitemShut {NoStop}%
\bibitem [{\citenamefont {Cornell}\ and\ \citenamefont {Wieman}(2002)}]{CW}%
  \BibitemOpen
  \bibfield  {author} {\bibinfo {author} {\bibfnamefont {E.~A.}\ \bibnamefont
  {Cornell}}\ and\ \bibinfo {author} {\bibfnamefont {C.~E.}\ \bibnamefont
  {Wieman}},\ }\href {\doibase 10.1103/RevModPhys.74.875} {\bibfield  {journal}
  {\bibinfo  {journal} {Rev. Mod. Phys.}\ }\textbf {\bibinfo {volume} {74}},\
  \bibinfo {pages} {875} (\bibinfo {year} {2002})}\BibitemShut {NoStop}%
\bibitem [{\citenamefont {Recati}\ and\ \citenamefont
  {Stringari}(2022)}]{recati2022}%
  \BibitemOpen
  \bibfield  {author} {\bibinfo {author} {\bibfnamefont {A.}~\bibnamefont
  {Recati}}\ and\ \bibinfo {author} {\bibfnamefont {S.}~\bibnamefont
  {Stringari}},\ }\href {\doibase
  https://doi.org/10.1146/annurev-conmatphys-031820-121316} {\bibfield
  {journal} {\bibinfo  {journal} {Ann. Rev. Cond. Matt. Phys.}\ }\textbf
  {\bibinfo {volume} {13}},\ \bibinfo {pages} {407} (\bibinfo {year}
  {2022})}\BibitemShut {NoStop}%
\bibitem [{\citenamefont {Baroni}\ \emph {et~al.}(2024)\citenamefont {Baroni},
  \citenamefont {Lamporesi},\ and\ \citenamefont {Zaccanti}}]{Baroni_2024}%
  \BibitemOpen
  \bibfield  {author} {\bibinfo {author} {\bibfnamefont {C.}~\bibnamefont
  {Baroni}}, \bibinfo {author} {\bibfnamefont {G.}~\bibnamefont {Lamporesi}}, \
  and\ \bibinfo {author} {\bibfnamefont {M.}~\bibnamefont {Zaccanti}},\ }\href
  {\doibase 10.1038/s42254-024-00773-6} {\bibfield  {journal} {\bibinfo
  {journal} {Nat. Rev. Phys.}\ } (\bibinfo {year} {2024}),\
  10.1038/s42254-024-00773-6}\BibitemShut {NoStop}%
\bibitem [{\citenamefont {Son}\ and\ \citenamefont
  {Stephanov}(2002{\natexlab{a}})}]{Son:2001td}%
  \BibitemOpen
  \bibfield  {author} {\bibinfo {author} {\bibfnamefont {D.~T.}\ \bibnamefont
  {Son}}\ and\ \bibinfo {author} {\bibfnamefont {M.~A.}\ \bibnamefont
  {Stephanov}},\ }\href {\doibase 10.1103/PhysRevA.65.063621} {\bibfield
  {journal} {\bibinfo  {journal} {Phys. Rev. A}\ }\textbf {\bibinfo {volume}
  {65}},\ \bibinfo {pages} {063621} (\bibinfo {year} {2002}{\natexlab{a}})},\
  \Eprint {http://arxiv.org/abs/cond-mat/0103451} {arXiv:cond-mat/0103451}
  \BibitemShut {NoStop}%
\bibitem [{\citenamefont {Kevrekidis}\ \emph {et~al.}(2004)\citenamefont
  {Kevrekidis}, \citenamefont {Nistazakis}, \citenamefont {Frantzeskakis},
  \citenamefont {Malomed},\ and\ \citenamefont
  {Carretero-Gonzalez}}]{Kevrekidis_2004}%
  \BibitemOpen
  \bibfield  {author} {\bibinfo {author} {\bibfnamefont {P.~G.}\ \bibnamefont
  {Kevrekidis}}, \bibinfo {author} {\bibfnamefont {H.~E.}\ \bibnamefont
  {Nistazakis}}, \bibinfo {author} {\bibfnamefont {D.~J.}\ \bibnamefont
  {Frantzeskakis}}, \bibinfo {author} {\bibfnamefont {B.~A.}\ \bibnamefont
  {Malomed}}, \ and\ \bibinfo {author} {\bibfnamefont {R.}~\bibnamefont
  {Carretero-Gonzalez}},\ }\href {\doibase 10.1140/epjd/e2003-00311-6}
  {\bibfield  {journal} {\bibinfo  {journal} {Eur. Phys. J. D}\ }\textbf
  {\bibinfo {volume} {28}},\ \bibinfo {pages} {181–185} (\bibinfo {year}
  {2004})}\BibitemShut {NoStop}%
\bibitem [{\citenamefont {Bakkali-Hassani}\ \emph {et~al.}(2021)\citenamefont
  {Bakkali-Hassani}, \citenamefont {Maury}, \citenamefont {Zou}, \citenamefont
  {Le~Cerf}, \citenamefont {Saint-Jalm}, \citenamefont {Castilho},
  \citenamefont {Nascimbene}, \citenamefont {Dalibard},\ and\ \citenamefont
  {Beugnon}}]{Bakkali_Hassani_2021}%
  \BibitemOpen
  \bibfield  {author} {\bibinfo {author} {\bibfnamefont {B.}~\bibnamefont
  {Bakkali-Hassani}}, \bibinfo {author} {\bibfnamefont {C.}~\bibnamefont
  {Maury}}, \bibinfo {author} {\bibfnamefont {Y.-Q.}\ \bibnamefont {Zou}},
  \bibinfo {author} {\bibfnamefont {E.}~\bibnamefont {Le~Cerf}}, \bibinfo
  {author} {\bibfnamefont {R.}~\bibnamefont {Saint-Jalm}}, \bibinfo {author}
  {\bibfnamefont {P.}~\bibnamefont {Castilho}}, \bibinfo {author}
  {\bibfnamefont {S.}~\bibnamefont {Nascimbene}}, \bibinfo {author}
  {\bibfnamefont {J.}~\bibnamefont {Dalibard}}, \ and\ \bibinfo {author}
  {\bibfnamefont {J.}~\bibnamefont {Beugnon}},\ }\href {\doibase
  10.1103/physrevlett.127.023603} {\bibfield  {journal} {\bibinfo  {journal}
  {Phys. Rev. Lett.}\ }\textbf {\bibinfo {volume} {127}} (\bibinfo {year}
  {2021}),\ 10.1103/physrevlett.127.023603}\BibitemShut {NoStop}%
\bibitem [{\citenamefont {Romero-Ros}\ \emph {et~al.}(2024)\citenamefont
  {Romero-Ros}, \citenamefont {Katsimiga}, \citenamefont {Mistakidis},
  \citenamefont {Mossman}, \citenamefont {Biondini}, \citenamefont
  {Schmelcher}, \citenamefont {Engels},\ and\ \citenamefont
  {Kevrekidis}}]{Romero-Ros:2023mgp}%
  \BibitemOpen
  \bibfield  {author} {\bibinfo {author} {\bibfnamefont {A.}~\bibnamefont
  {Romero-Ros}}, \bibinfo {author} {\bibfnamefont {G.~C.}\ \bibnamefont
  {Katsimiga}}, \bibinfo {author} {\bibfnamefont {S.~I.}\ \bibnamefont
  {Mistakidis}}, \bibinfo {author} {\bibfnamefont {S.}~\bibnamefont {Mossman}},
  \bibinfo {author} {\bibfnamefont {G.}~\bibnamefont {Biondini}}, \bibinfo
  {author} {\bibfnamefont {P.}~\bibnamefont {Schmelcher}}, \bibinfo {author}
  {\bibfnamefont {P.}~\bibnamefont {Engels}}, \ and\ \bibinfo {author}
  {\bibfnamefont {P.~G.}\ \bibnamefont {Kevrekidis}},\ }\href {\doibase
  10.1103/PhysRevLett.132.033402} {\bibfield  {journal} {\bibinfo  {journal}
  {Phys. Rev. Lett.}\ }\textbf {\bibinfo {volume} {132}},\ \bibinfo {pages}
  {033402} (\bibinfo {year} {2024})},\ \Eprint
  {http://arxiv.org/abs/2304.05951} {arXiv:2304.05951 [nlin.PS]} \BibitemShut
  {NoStop}%
\bibitem [{\citenamefont {Hamner}\ \emph {et~al.}(2011)\citenamefont {Hamner},
  \citenamefont {Chang}, \citenamefont {Engels},\ and\ \citenamefont
  {Hoefer}}]{Hamner_2011}%
  \BibitemOpen
  \bibfield  {author} {\bibinfo {author} {\bibfnamefont {C.}~\bibnamefont
  {Hamner}}, \bibinfo {author} {\bibfnamefont {J.~J.}\ \bibnamefont {Chang}},
  \bibinfo {author} {\bibfnamefont {P.}~\bibnamefont {Engels}}, \ and\ \bibinfo
  {author} {\bibfnamefont {M.~A.}\ \bibnamefont {Hoefer}},\ }\href {\doibase
  10.1103/physrevlett.106.065302} {\bibfield  {journal} {\bibinfo  {journal}
  {Phys. Rev. Lett.}\ }\textbf {\bibinfo {volume} {106}} (\bibinfo {year}
  {2011}),\ 10.1103/physrevlett.106.065302}\BibitemShut {NoStop}%
\bibitem [{\citenamefont {Farolfi}\ \emph {et~al.}(2020)\citenamefont
  {Farolfi}, \citenamefont {Trypogeorgos}, \citenamefont {Mordini},
  \citenamefont {Lamporesi},\ and\ \citenamefont {Ferrari}}]{Farolfi_2020}%
  \BibitemOpen
  \bibfield  {author} {\bibinfo {author} {\bibfnamefont {A.}~\bibnamefont
  {Farolfi}}, \bibinfo {author} {\bibfnamefont {D.}~\bibnamefont
  {Trypogeorgos}}, \bibinfo {author} {\bibfnamefont {C.}~\bibnamefont
  {Mordini}}, \bibinfo {author} {\bibfnamefont {G.}~\bibnamefont {Lamporesi}},
  \ and\ \bibinfo {author} {\bibfnamefont {G.}~\bibnamefont {Ferrari}},\ }\href
  {\doibase 10.1103/physrevlett.125.030401} {\bibfield  {journal} {\bibinfo
  {journal} {Phys. Rev. Lett.}\ }\textbf {\bibinfo {volume} {125}} (\bibinfo
  {year} {2020}),\ 10.1103/physrevlett.125.030401}\BibitemShut {NoStop}%
\bibitem [{\citenamefont {Richaud}\ \emph {et~al.}(2023)\citenamefont
  {Richaud}, \citenamefont {Lamporesi}, \citenamefont {Capone},\ and\
  \citenamefont {Recati}}]{Richaud_2023}%
  \BibitemOpen
  \bibfield  {author} {\bibinfo {author} {\bibfnamefont {A.}~\bibnamefont
  {Richaud}}, \bibinfo {author} {\bibfnamefont {G.}~\bibnamefont {Lamporesi}},
  \bibinfo {author} {\bibfnamefont {M.}~\bibnamefont {Capone}}, \ and\ \bibinfo
  {author} {\bibfnamefont {A.}~\bibnamefont {Recati}},\ }\href {\doibase
  10.1103/physreva.107.053317} {\bibfield  {journal} {\bibinfo  {journal}
  {Phys. Rev. A}\ }\textbf {\bibinfo {volume} {107}} (\bibinfo {year} {2023}),\
  10.1103/physreva.107.053317}\BibitemShut {NoStop}%
\bibitem [{\citenamefont {Petrov}(2015)}]{petrov2015}%
  \BibitemOpen
  \bibfield  {author} {\bibinfo {author} {\bibfnamefont {D.~S.}\ \bibnamefont
  {Petrov}},\ }\href {\doibase 10.1103/PhysRevLett.115.155302} {\bibfield
  {journal} {\bibinfo  {journal} {Phys. Rev. Lett.}\ }\textbf {\bibinfo
  {volume} {115}},\ \bibinfo {pages} {155302} (\bibinfo {year}
  {2015})}\BibitemShut {NoStop}%
\bibitem [{\citenamefont {Semeghini}\ \emph {et~al.}(2018)\citenamefont
  {Semeghini}, \citenamefont {Ferioli}, \citenamefont {Masi}, \citenamefont
  {Mazzinghi}, \citenamefont {Wolswijk}, \citenamefont {Minardi}, \citenamefont
  {Modugno}, \citenamefont {Modugno}, \citenamefont {Inguscio},\ and\
  \citenamefont {Fattori}}]{Semeghini_2018}%
  \BibitemOpen
  \bibfield  {author} {\bibinfo {author} {\bibfnamefont {G.}~\bibnamefont
  {Semeghini}}, \bibinfo {author} {\bibfnamefont {G.}~\bibnamefont {Ferioli}},
  \bibinfo {author} {\bibfnamefont {L.}~\bibnamefont {Masi}}, \bibinfo {author}
  {\bibfnamefont {C.}~\bibnamefont {Mazzinghi}}, \bibinfo {author}
  {\bibfnamefont {L.}~\bibnamefont {Wolswijk}}, \bibinfo {author}
  {\bibfnamefont {F.}~\bibnamefont {Minardi}}, \bibinfo {author} {\bibfnamefont
  {M.}~\bibnamefont {Modugno}}, \bibinfo {author} {\bibfnamefont
  {G.}~\bibnamefont {Modugno}}, \bibinfo {author} {\bibfnamefont
  {M.}~\bibnamefont {Inguscio}}, \ and\ \bibinfo {author} {\bibfnamefont
  {M.}~\bibnamefont {Fattori}},\ }\href {\doibase
  10.1103/physrevlett.120.235301} {\bibfield  {journal} {\bibinfo  {journal}
  {Phys. Rev. Lett.}\ }\textbf {\bibinfo {volume} {120}} (\bibinfo {year}
  {2018}),\ 10.1103/physrevlett.120.235301}\BibitemShut {NoStop}%
\bibitem [{\citenamefont {Myatt}\ \emph {et~al.}(1997)\citenamefont {Myatt},
  \citenamefont {Burt}, \citenamefont {Ghrist}, \citenamefont {Cornell},\ and\
  \citenamefont {Wieman}}]{Myatt:1997zz}%
  \BibitemOpen
  \bibfield  {author} {\bibinfo {author} {\bibfnamefont {C.~J.}\ \bibnamefont
  {Myatt}}, \bibinfo {author} {\bibfnamefont {E.~A.}\ \bibnamefont {Burt}},
  \bibinfo {author} {\bibfnamefont {R.~W.}\ \bibnamefont {Ghrist}}, \bibinfo
  {author} {\bibfnamefont {E.~A.}\ \bibnamefont {Cornell}}, \ and\ \bibinfo
  {author} {\bibfnamefont {C.~E.}\ \bibnamefont {Wieman}},\ }\href {\doibase
  10.1103/PhysRevLett.78.586} {\bibfield  {journal} {\bibinfo  {journal} {Phys.
  Rev. Lett.}\ }\textbf {\bibinfo {volume} {78}},\ \bibinfo {pages} {586}
  (\bibinfo {year} {1997})}\BibitemShut {NoStop}%
\bibitem [{\citenamefont {Tuoriniemi}\ \emph {et~al.}(2002)\citenamefont
  {Tuoriniemi}, \citenamefont {Martikainen}, \citenamefont {Pentti},
  \citenamefont {Sebedash}, \citenamefont {Boldarev},\ and\ \citenamefont
  {Pickett}}]{Tuoriniemi2002}%
  \BibitemOpen
  \bibfield  {author} {\bibinfo {author} {\bibfnamefont {J.}~\bibnamefont
  {Tuoriniemi}}, \bibinfo {author} {\bibfnamefont {J.}~\bibnamefont
  {Martikainen}}, \bibinfo {author} {\bibfnamefont {E.}~\bibnamefont {Pentti}},
  \bibinfo {author} {\bibfnamefont {A.}~\bibnamefont {Sebedash}}, \bibinfo
  {author} {\bibfnamefont {S.}~\bibnamefont {Boldarev}}, \ and\ \bibinfo
  {author} {\bibfnamefont {G.}~\bibnamefont {Pickett}},\ }\href {\doibase
  10.1023/A:1021468614550} {\bibfield  {journal} {\bibinfo  {journal} {J. Low
  Temp. Phys.}\ }\textbf {\bibinfo {volume} {129}},\ \bibinfo {pages} {531}
  (\bibinfo {year} {2002})}\BibitemShut {NoStop}%
\bibitem [{\citenamefont {Rysti}\ \emph {et~al.}(2012)\citenamefont {Rysti},
  \citenamefont {Tuoriniemi},\ and\ \citenamefont
  {Salmela}}]{PhysRevB.85.134529}%
  \BibitemOpen
  \bibfield  {author} {\bibinfo {author} {\bibfnamefont {J.}~\bibnamefont
  {Rysti}}, \bibinfo {author} {\bibfnamefont {J.}~\bibnamefont {Tuoriniemi}}, \
  and\ \bibinfo {author} {\bibfnamefont {A.}~\bibnamefont {Salmela}},\ }\href
  {\doibase 10.1103/PhysRevB.85.134529} {\bibfield  {journal} {\bibinfo
  {journal} {Phys. Rev. B}\ }\textbf {\bibinfo {volume} {85}},\ \bibinfo
  {pages} {134529} (\bibinfo {year} {2012})}\BibitemShut {NoStop}%
\bibitem [{\citenamefont {Lewenstein}\ \emph {et~al.}(2012)\citenamefont
  {Lewenstein}, \citenamefont {Sanpera},\ and\ \citenamefont
  {Ahufinger}}]{annabook}%
  \BibitemOpen
  \bibfield  {author} {\bibinfo {author} {\bibfnamefont {M.}~\bibnamefont
  {Lewenstein}}, \bibinfo {author} {\bibfnamefont {A.}~\bibnamefont {Sanpera}},
  \ and\ \bibinfo {author} {\bibfnamefont {V.}~\bibnamefont {Ahufinger}},\
  }\href {https://books.google.it/books?id=Wpl91RDxV5IC} {\emph {\bibinfo
  {title} {Ultracold Atoms in Optical Lattices: Simulating quantum many-body
  systems}}}\ (\bibinfo  {publisher} {OUP Oxford},\ \bibinfo {year}
  {2012})\BibitemShut {NoStop}%
\bibitem [{\citenamefont {Inguscio}\ and\ \citenamefont
  {Fallani}(2013)}]{fallanibook}%
  \BibitemOpen
  \bibfield  {author} {\bibinfo {author} {\bibfnamefont {M.}~\bibnamefont
  {Inguscio}}\ and\ \bibinfo {author} {\bibfnamefont {L.}~\bibnamefont
  {Fallani}},\ }\href {https://books.google.it/books?id=8zZoAgAAQBAJ} {\emph
  {\bibinfo {title} {Atomic Physics: Precise Measurements and Ultracold
  Matter}}}\ (\bibinfo  {publisher} {OUP Oxford},\ \bibinfo {year}
  {2013})\BibitemShut {NoStop}%
\bibitem [{\citenamefont {{Ferrier-Barbut}}\ \emph {et~al.}(2014)\citenamefont
  {{Ferrier-Barbut}}, \citenamefont {{Delehaye}}, \citenamefont {{Laurent}},
  \citenamefont {{Grier}}, \citenamefont {{Pierce}}, \citenamefont {{Rem}},
  \citenamefont {{Chevy}},\ and\ \citenamefont
  {{Salomon}}}]{2014Sci...345.1035F}%
  \BibitemOpen
  \bibfield  {author} {\bibinfo {author} {\bibfnamefont {I.}~\bibnamefont
  {{Ferrier-Barbut}}}, \bibinfo {author} {\bibfnamefont {M.}~\bibnamefont
  {{Delehaye}}}, \bibinfo {author} {\bibfnamefont {S.}~\bibnamefont
  {{Laurent}}}, \bibinfo {author} {\bibfnamefont {A.~T.}\ \bibnamefont
  {{Grier}}}, \bibinfo {author} {\bibfnamefont {M.}~\bibnamefont {{Pierce}}},
  \bibinfo {author} {\bibfnamefont {B.~S.}\ \bibnamefont {{Rem}}}, \bibinfo
  {author} {\bibfnamefont {F.}~\bibnamefont {{Chevy}}}, \ and\ \bibinfo
  {author} {\bibfnamefont {C.}~\bibnamefont {{Salomon}}},\ }\href {\doibase
  10.1126/science.1255380} {\bibfield  {journal} {\bibinfo  {journal}
  {Science}\ }\textbf {\bibinfo {volume} {345}},\ \bibinfo {pages} {1035}
  (\bibinfo {year} {2014})},\ \Eprint {http://arxiv.org/abs/1404.2548}
  {arXiv:1404.2548 [cond-mat.quant-gas]} \BibitemShut {NoStop}%
\bibitem [{\citenamefont {Lepori}\ \emph {et~al.}(2015)\citenamefont {Lepori},
  \citenamefont {Trombettoni},\ and\ \citenamefont {Vinci}}]{lvm2015}%
  \BibitemOpen
  \bibfield  {author} {\bibinfo {author} {\bibfnamefont {L.}~\bibnamefont
  {Lepori}}, \bibinfo {author} {\bibfnamefont {A.}~\bibnamefont {Trombettoni}},
  \ and\ \bibinfo {author} {\bibfnamefont {W.}~\bibnamefont {Vinci}},\ }\href
  {http://stacks.iop.org/0295-5075/109/i=5/a=50002} {\bibfield  {journal}
  {\bibinfo  {journal} {Europhys. Lett.}\ }\textbf {\bibinfo {volume} {109}},\
  \bibinfo {pages} {50002} (\bibinfo {year} {2015})}\BibitemShut {NoStop}%
\bibitem [{\citenamefont {Lepori}\ and\ \citenamefont
  {Mannarelli}(2019)}]{Lepori:2019vec}%
  \BibitemOpen
  \bibfield  {author} {\bibinfo {author} {\bibfnamefont {L.}~\bibnamefont
  {Lepori}}\ and\ \bibinfo {author} {\bibfnamefont {M.}~\bibnamefont
  {Mannarelli}},\ }\href {\doibase 10.1103/PhysRevD.99.096011} {\bibfield
  {journal} {\bibinfo  {journal} {Phys. Rev. D}\ }\textbf {\bibinfo {volume}
  {99}},\ \bibinfo {pages} {096011} (\bibinfo {year} {2019})},\ \Eprint
  {http://arxiv.org/abs/1901.07488} {arXiv:1901.07488 [hep-ph]} \BibitemShut
  {NoStop}%
\bibitem [{\citenamefont {Nespolo}\ \emph {et~al.}(2017)\citenamefont
  {Nespolo}, \citenamefont {Astrakharchik},\ and\ \citenamefont
  {Recati}}]{Nespolo_2017}%
  \BibitemOpen
  \bibfield  {author} {\bibinfo {author} {\bibfnamefont {J.}~\bibnamefont
  {Nespolo}}, \bibinfo {author} {\bibfnamefont {G.~E.}\ \bibnamefont
  {Astrakharchik}}, \ and\ \bibinfo {author} {\bibfnamefont {A.}~\bibnamefont
  {Recati}},\ }\href {\doibase 10.1088/1367-2630/aa93a0} {\bibfield  {journal}
  {\bibinfo  {journal} {New J. Phys.}\ }\textbf {\bibinfo {volume} {19}},\
  \bibinfo {pages} {125005} (\bibinfo {year} {2017})}\BibitemShut {NoStop}%
\bibitem [{\citenamefont {Fischer}\ and\ \citenamefont
  {Schutzhold}(2004)}]{Fischer:2004bf}%
  \BibitemOpen
  \bibfield  {author} {\bibinfo {author} {\bibfnamefont {U.~R.}\ \bibnamefont
  {Fischer}}\ and\ \bibinfo {author} {\bibfnamefont {R.}~\bibnamefont
  {Schutzhold}},\ }\href {\doibase 10.1103/PhysRevA.70.063615} {\bibfield
  {journal} {\bibinfo  {journal} {Phys. Rev. A}\ }\textbf {\bibinfo {volume}
  {70}},\ \bibinfo {pages} {063615} (\bibinfo {year} {2004})},\ \Eprint
  {http://arxiv.org/abs/cond-mat/0406470} {arXiv:cond-mat/0406470} \BibitemShut
  {NoStop}%
\bibitem [{\citenamefont {Visser}\ and\ \citenamefont
  {Weinfurtner}(2005)}]{Visser:2005ss}%
  \BibitemOpen
  \bibfield  {author} {\bibinfo {author} {\bibfnamefont {M.}~\bibnamefont
  {Visser}}\ and\ \bibinfo {author} {\bibfnamefont {S.}~\bibnamefont
  {Weinfurtner}},\ }\href {\doibase 10.1103/PhysRevD.72.044020} {\bibfield
  {journal} {\bibinfo  {journal} {Phys. Rev. D}\ }\textbf {\bibinfo {volume}
  {72}},\ \bibinfo {pages} {044020} (\bibinfo {year} {2005})},\ \Eprint
  {http://arxiv.org/abs/gr-qc/0506029} {arXiv:gr-qc/0506029} \BibitemShut
  {NoStop}%
\bibitem [{\citenamefont {Liberati}\ \emph
  {et~al.}(2006{\natexlab{b}})\citenamefont {Liberati}, \citenamefont
  {Visser},\ and\ \citenamefont {Weinfurtner}}]{Liberati_2006}%
  \BibitemOpen
  \bibfield  {author} {\bibinfo {author} {\bibfnamefont {S.}~\bibnamefont
  {Liberati}}, \bibinfo {author} {\bibfnamefont {M.}~\bibnamefont {Visser}}, \
  and\ \bibinfo {author} {\bibfnamefont {S.}~\bibnamefont {Weinfurtner}},\
  }\href {\doibase 10.1088/0264-9381/23/9/023} {\bibfield  {journal} {\bibinfo
  {journal} {Class. Quant. Grav.}\ }\textbf {\bibinfo {volume} {23}},\ \bibinfo
  {pages} {3129–3154} (\bibinfo {year} {2006}{\natexlab{b}})}\BibitemShut
  {NoStop}%
\bibitem [{\citenamefont {Berti}\ \emph {et~al.}(2024)\citenamefont {Berti},
  \citenamefont {Fernandes}, \citenamefont {Butera}, \citenamefont {Recati},
  \citenamefont {Wouters},\ and\ \citenamefont {Carusotto}}]{berti2024}%
  \BibitemOpen
  \bibfield  {author} {\bibinfo {author} {\bibfnamefont {A.}~\bibnamefont
  {Berti}}, \bibinfo {author} {\bibfnamefont {L.}~\bibnamefont {Fernandes}},
  \bibinfo {author} {\bibfnamefont {S.~G.}\ \bibnamefont {Butera}}, \bibinfo
  {author} {\bibfnamefont {A.}~\bibnamefont {Recati}}, \bibinfo {author}
  {\bibfnamefont {M.}~\bibnamefont {Wouters}}, \ and\ \bibinfo {author}
  {\bibfnamefont {I.}~\bibnamefont {Carusotto}},\ }\href
  {https://arxiv.org/abs/2408.17292} {} (\bibinfo {year} {2024}),\ \Eprint
  {http://arxiv.org/abs/2408.17292} {arXiv:2408.17292 [cond-mat.quant-gas]}
  \BibitemShut {NoStop}%
\bibitem [{\citenamefont {Zenesini}\ \emph {et~al.}(2024)\citenamefont
  {Zenesini}, \citenamefont {Berti}, \citenamefont {Cominotti}, \citenamefont
  {Rogora}, \citenamefont {Moss}, \citenamefont {Billam}, \citenamefont
  {Carusotto}, \citenamefont {Lamporesi}, \citenamefont {Recati},\ and\
  \citenamefont {Ferrari}}]{Zenesini:2023afv}%
  \BibitemOpen
  \bibfield  {author} {\bibinfo {author} {\bibfnamefont {A.}~\bibnamefont
  {Zenesini}}, \bibinfo {author} {\bibfnamefont {A.}~\bibnamefont {Berti}},
  \bibinfo {author} {\bibfnamefont {R.}~\bibnamefont {Cominotti}}, \bibinfo
  {author} {\bibfnamefont {C.}~\bibnamefont {Rogora}}, \bibinfo {author}
  {\bibfnamefont {I.~G.}\ \bibnamefont {Moss}}, \bibinfo {author}
  {\bibfnamefont {T.~P.}\ \bibnamefont {Billam}}, \bibinfo {author}
  {\bibfnamefont {I.}~\bibnamefont {Carusotto}}, \bibinfo {author}
  {\bibfnamefont {G.}~\bibnamefont {Lamporesi}}, \bibinfo {author}
  {\bibfnamefont {A.}~\bibnamefont {Recati}}, \ and\ \bibinfo {author}
  {\bibfnamefont {G.}~\bibnamefont {Ferrari}},\ }\href {\doibase
  10.1038/s41567-023-02345-4} {\bibfield  {journal} {\bibinfo  {journal} {Nat.
  Phys.}\ }\textbf {\bibinfo {volume} {20}},\ \bibinfo {pages} {558} (\bibinfo
  {year} {2024})},\ \Eprint {http://arxiv.org/abs/2305.05225} {arXiv:2305.05225
  [hep-ph]} \BibitemShut {NoStop}%
\bibitem [{\citenamefont {Shapiro}\ and\ \citenamefont
  {Teukolsky}(1983)}]{Shapiro:1983du}%
  \BibitemOpen
  \bibfield  {author} {\bibinfo {author} {\bibfnamefont {S.~L.}\ \bibnamefont
  {Shapiro}}\ and\ \bibinfo {author} {\bibfnamefont {S.~A.}\ \bibnamefont
  {Teukolsky}},\ }\href@noop {} {\emph {\bibinfo {title} {{Black holes, white
  dwarfs, and neutron stars: The physics of compact objects}}}}\ (\bibinfo
  {year} {1983})\BibitemShut {NoStop}%
\bibitem [{\citenamefont {Alford}\ \emph {et~al.}(1999)\citenamefont {Alford},
  \citenamefont {Rajagopal},\ and\ \citenamefont {Wilczek}}]{Alford:1998mk}%
  \BibitemOpen
  \bibfield  {author} {\bibinfo {author} {\bibfnamefont {M.~G.}\ \bibnamefont
  {Alford}}, \bibinfo {author} {\bibfnamefont {K.}~\bibnamefont {Rajagopal}}, \
  and\ \bibinfo {author} {\bibfnamefont {F.}~\bibnamefont {Wilczek}},\ }\href
  {\doibase 10.1016/S0550-3213(98)00668-3} {\bibfield  {journal} {\bibinfo
  {journal} {Nucl. Phys.}\ }\textbf {\bibinfo {volume} {B 537}},\ \bibinfo
  {pages} {443} (\bibinfo {year} {1999})},\ \Eprint
  {http://arxiv.org/abs/hep-ph/9804403} {arXiv:hep-ph/9804403 [hep-ph]}
  \BibitemShut {NoStop}%
\bibitem [{\citenamefont {Bedaque}\ and\ \citenamefont
  {Schafer}(2002)}]{Bedaque:2001je}%
  \BibitemOpen
  \bibfield  {author} {\bibinfo {author} {\bibfnamefont {P.~F.}\ \bibnamefont
  {Bedaque}}\ and\ \bibinfo {author} {\bibfnamefont {T.}~\bibnamefont
  {Schafer}},\ }\href {\doibase 10.1016/S0375-9474(01)01272-6} {\bibfield
  {journal} {\bibinfo  {journal} {Nucl. Phys.}\ }\textbf {\bibinfo {volume} {A
  697}},\ \bibinfo {pages} {802} (\bibinfo {year} {2002})},\ \Eprint
  {http://arxiv.org/abs/hep-ph/0105150} {arXiv:hep-ph/0105150 [hep-ph]}
  \BibitemShut {NoStop}%
\bibitem [{\citenamefont {Kaplan}\ and\ \citenamefont
  {Reddy}(2002)}]{Kaplan:2001qk}%
  \BibitemOpen
  \bibfield  {author} {\bibinfo {author} {\bibfnamefont {D.}~\bibnamefont
  {Kaplan}}\ and\ \bibinfo {author} {\bibfnamefont {S.}~\bibnamefont {Reddy}},\
  }\href {\doibase 10.1103/PhysRevD.65.054042} {\bibfield  {journal} {\bibinfo
  {journal} {Phys. Rev.}\ }\textbf {\bibinfo {volume} {D 65}},\ \bibinfo
  {pages} {054042} (\bibinfo {year} {2002})},\ \Eprint
  {http://arxiv.org/abs/hep-ph/0107265} {arXiv:hep-ph/0107265 [hep-ph]}
  \BibitemShut {NoStop}%
\bibitem [{\citenamefont {Alford}\ \emph {et~al.}(2008)\citenamefont {Alford},
  \citenamefont {Schmitt}, \citenamefont {Rajagopal},\ and\ \citenamefont
  {Schafer}}]{Alford:2007xm}%
  \BibitemOpen
  \bibfield  {author} {\bibinfo {author} {\bibfnamefont {M.~G.}\ \bibnamefont
  {Alford}}, \bibinfo {author} {\bibfnamefont {A.}~\bibnamefont {Schmitt}},
  \bibinfo {author} {\bibfnamefont {K.}~\bibnamefont {Rajagopal}}, \ and\
  \bibinfo {author} {\bibfnamefont {T.}~\bibnamefont {Schafer}},\ }\href
  {\doibase 10.1103/RevModPhys.80.1455} {\bibfield  {journal} {\bibinfo
  {journal} {Rev. Mod. Phys.}\ }\textbf {\bibinfo {volume} {80}},\ \bibinfo
  {pages} {1455} (\bibinfo {year} {2008})},\ \Eprint
  {http://arxiv.org/abs/0709.4635} {arXiv:0709.4635 [hep-ph]} \BibitemShut
  {NoStop}%
\bibitem [{\citenamefont {Anglani}\ \emph {et~al.}(2014)\citenamefont
  {Anglani}, \citenamefont {Casalbuoni}, \citenamefont {Ciminale},
  \citenamefont {Ippolito}, \citenamefont {Gatto} \emph
  {et~al.}}]{Anglani:2013gfu}%
  \BibitemOpen
  \bibfield  {author} {\bibinfo {author} {\bibfnamefont {R.}~\bibnamefont
  {Anglani}}, \bibinfo {author} {\bibfnamefont {R.}~\bibnamefont {Casalbuoni}},
  \bibinfo {author} {\bibfnamefont {M.}~\bibnamefont {Ciminale}}, \bibinfo
  {author} {\bibfnamefont {N.}~\bibnamefont {Ippolito}}, \bibinfo {author}
  {\bibfnamefont {R.}~\bibnamefont {Gatto}},  \emph {et~al.},\ }\href {\doibase
  10.1103/RevModPhys.86.509} {\bibfield  {journal} {\bibinfo  {journal} {Rev.
  Mod. Phys.}\ }\textbf {\bibinfo {volume} {86}},\ \bibinfo {pages} {509}
  (\bibinfo {year} {2014})},\ \Eprint {http://arxiv.org/abs/1302.4264}
  {arXiv:1302.4264 [hep-ph]} \BibitemShut {NoStop}%
\bibitem [{\citenamefont {Annett}(2004)}]{annett}%
  \BibitemOpen
  \bibfield  {author} {\bibinfo {author} {\bibfnamefont {J.}~\bibnamefont
  {Annett}},\ }\href {https://books.google.it/books?id=WZcXmBrZIc8C} {\emph
  {\bibinfo {title} {Superconductivity, Superfluids and Condensates}}},\ Oxford
  Master Series in Physics\ (\bibinfo  {publisher} {OUP Oxford},\ \bibinfo
  {year} {2004})\BibitemShut {NoStop}%
\bibitem [{\citenamefont {Pethick}\ and\ \citenamefont
  {Smith}(2008)}]{pethick}%
  \BibitemOpen
  \bibfield  {author} {\bibinfo {author} {\bibfnamefont {C.}~\bibnamefont
  {Pethick}}\ and\ \bibinfo {author} {\bibfnamefont {H.}~\bibnamefont
  {Smith}},\ }\href {https://books.google.it/books?id=HobTUdxBoFcC} {\emph
  {\bibinfo {title} {Bose-Einstein Condensation in Dilute Gases}}}\ (\bibinfo
  {publisher} {Cambridge University Press},\ \bibinfo {year}
  {2008})\BibitemShut {NoStop}%
\bibitem [{\citenamefont {Lepori}\ \emph {et~al.}(2018)\citenamefont {Lepori},
  \citenamefont {Maraga}, \citenamefont {Celi}, \citenamefont {Dell’Anna},\
  and\ \citenamefont {Trombettoni}}]{lepori2018}%
  \BibitemOpen
  \bibfield  {author} {\bibinfo {author} {\bibfnamefont {L.}~\bibnamefont
  {Lepori}}, \bibinfo {author} {\bibfnamefont {A.}~\bibnamefont {Maraga}},
  \bibinfo {author} {\bibfnamefont {A.}~\bibnamefont {Celi}}, \bibinfo {author}
  {\bibfnamefont {L.}~\bibnamefont {Dell’Anna}}, \ and\ \bibinfo {author}
  {\bibfnamefont {A.}~\bibnamefont {Trombettoni}},\ }\href {\doibase
  10.3390/condmat3020014} {\bibfield  {journal} {\bibinfo  {journal} {Cond.
  Matt.}\ }\textbf {\bibinfo {volume} {3}} (\bibinfo {year} {2018}),\
  10.3390/condmat3020014}\BibitemShut {NoStop}%
\bibitem [{\citenamefont {Pu}\ and\ \citenamefont
  {Bigelow}(1998{\natexlab{a}})}]{PhysRevLett.80.1130}%
  \BibitemOpen
  \bibfield  {author} {\bibinfo {author} {\bibfnamefont {H.}~\bibnamefont
  {Pu}}\ and\ \bibinfo {author} {\bibfnamefont {N.~P.}\ \bibnamefont
  {Bigelow}},\ }\href {\doibase 10.1103/PhysRevLett.80.1130} {\bibfield
  {journal} {\bibinfo  {journal} {Phys. Rev. Lett.}\ }\textbf {\bibinfo
  {volume} {80}},\ \bibinfo {pages} {1130} (\bibinfo {year}
  {1998}{\natexlab{a}})}\BibitemShut {NoStop}%
\bibitem [{\citenamefont {Pu}\ and\ \citenamefont
  {Bigelow}(1998{\natexlab{b}})}]{PhysRevLett.80.1134}%
  \BibitemOpen
  \bibfield  {author} {\bibinfo {author} {\bibfnamefont {H.}~\bibnamefont
  {Pu}}\ and\ \bibinfo {author} {\bibfnamefont {N.~P.}\ \bibnamefont
  {Bigelow}},\ }\href {\doibase 10.1103/PhysRevLett.80.1134} {\bibfield
  {journal} {\bibinfo  {journal} {Phys. Rev. Lett.}\ }\textbf {\bibinfo
  {volume} {80}},\ \bibinfo {pages} {1134} (\bibinfo {year}
  {1998}{\natexlab{b}})}\BibitemShut {NoStop}%
\bibitem [{\citenamefont {Hall}\ \emph {et~al.}(1998)\citenamefont {Hall},
  \citenamefont {Matthews}, \citenamefont {Wieman},\ and\ \citenamefont
  {Cornell}}]{Hall:1998zz}%
  \BibitemOpen
  \bibfield  {author} {\bibinfo {author} {\bibfnamefont {D.~S.}\ \bibnamefont
  {Hall}}, \bibinfo {author} {\bibfnamefont {M.~R.}\ \bibnamefont {Matthews}},
  \bibinfo {author} {\bibfnamefont {C.~E.}\ \bibnamefont {Wieman}}, \ and\
  \bibinfo {author} {\bibfnamefont {E.~A.}\ \bibnamefont {Cornell}},\ }\href
  {\doibase 10.1103/PhysRevLett.81.1543} {\bibfield  {journal} {\bibinfo
  {journal} {Phys. Rev. Lett.}\ }\textbf {\bibinfo {volume} {81}},\ \bibinfo
  {pages} {1543} (\bibinfo {year} {1998})},\ \Eprint
  {http://arxiv.org/abs/cond-mat/9805327} {arXiv:cond-mat/9805327} \BibitemShut
  {NoStop}%
\bibitem [{\citenamefont {Unruh}(1981)}]{Unruh:1980cg}%
  \BibitemOpen
  \bibfield  {author} {\bibinfo {author} {\bibfnamefont {W.}~\bibnamefont
  {Unruh}},\ }\href {\doibase 10.1103/PhysRevLett.46.1351} {\bibfield
  {journal} {\bibinfo  {journal} {Phys.Rev.Lett.}\ }\textbf {\bibinfo {volume}
  {46}},\ \bibinfo {pages} {1351} (\bibinfo {year} {1981})}\BibitemShut
  {NoStop}%
\bibitem [{\citenamefont {Bilic}(1999)}]{Bilic:1999sq}%
  \BibitemOpen
  \bibfield  {author} {\bibinfo {author} {\bibfnamefont {N.}~\bibnamefont
  {Bilic}},\ }\href {\doibase 10.1088/0264-9381/16/12/312} {\bibfield
  {journal} {\bibinfo  {journal} {Class. Quant. Grav.}\ }\textbf {\bibinfo
  {volume} {16}},\ \bibinfo {pages} {3953} (\bibinfo {year} {1999})},\ \Eprint
  {http://arxiv.org/abs/gr-qc/9908002} {arXiv:gr-qc/9908002} \BibitemShut
  {NoStop}%
\bibitem [{\citenamefont {Visser}\ and\ \citenamefont
  {Molina-Paris}(2010)}]{Visser:2010xv}%
  \BibitemOpen
  \bibfield  {author} {\bibinfo {author} {\bibfnamefont {M.}~\bibnamefont
  {Visser}}\ and\ \bibinfo {author} {\bibfnamefont {C.}~\bibnamefont
  {Molina-Paris}},\ }\href {\doibase 10.1088/1367-2630/12/9/095014} {\bibfield
  {journal} {\bibinfo  {journal} {New J. Phys.}\ }\textbf {\bibinfo {volume}
  {12}},\ \bibinfo {pages} {095014} (\bibinfo {year} {2010})},\ \Eprint
  {http://arxiv.org/abs/1001.1310} {arXiv:1001.1310 [gr-qc]} \BibitemShut
  {NoStop}%
\bibitem [{\citenamefont {Brout}\ \emph {et~al.}(1995)\citenamefont {Brout},
  \citenamefont {Massar}, \citenamefont {Parentani},\ and\ \citenamefont
  {Spindel}}]{Brout:1995rd}%
  \BibitemOpen
  \bibfield  {author} {\bibinfo {author} {\bibfnamefont {R.}~\bibnamefont
  {Brout}}, \bibinfo {author} {\bibfnamefont {S.}~\bibnamefont {Massar}},
  \bibinfo {author} {\bibfnamefont {R.}~\bibnamefont {Parentani}}, \ and\
  \bibinfo {author} {\bibfnamefont {P.}~\bibnamefont {Spindel}},\ }\href
  {\doibase 10.1016/0370-1573(95)00008-5} {\bibfield  {journal} {\bibinfo
  {journal} {Phys. Rept.}\ }\textbf {\bibinfo {volume} {260}},\ \bibinfo
  {pages} {329} (\bibinfo {year} {1995})},\ \Eprint
  {http://arxiv.org/abs/0710.4345} {arXiv:0710.4345 [gr-qc]} \BibitemShut
  {NoStop}%
\bibitem [{\citenamefont {Volovik}(2003)}]{Volovik:2003ga}%
  \BibitemOpen
  \bibfield  {author} {\bibinfo {author} {\bibfnamefont {G.~E.}\ \bibnamefont
  {Volovik}},\ }\href {\doibase 10.1023/A:1023762013553} {\bibfield  {journal}
  {\bibinfo  {journal} {Found. Phys.}\ }\textbf {\bibinfo {volume} {33}},\
  \bibinfo {pages} {349} (\bibinfo {year} {2003})},\ \Eprint
  {http://arxiv.org/abs/gr-qc/0301043} {arXiv:gr-qc/0301043} \BibitemShut
  {NoStop}%
\bibitem [{\citenamefont {Carusotto}\ \emph {et~al.}(2008)\citenamefont
  {Carusotto}, \citenamefont {Fagnocchi}, \citenamefont {Recati}, \citenamefont
  {Balbinot},\ and\ \citenamefont {Fabbri}}]{Carusotto}%
  \BibitemOpen
  \bibfield  {author} {\bibinfo {author} {\bibfnamefont {I.}~\bibnamefont
  {Carusotto}}, \bibinfo {author} {\bibfnamefont {S.}~\bibnamefont
  {Fagnocchi}}, \bibinfo {author} {\bibfnamefont {A.}~\bibnamefont {Recati}},
  \bibinfo {author} {\bibfnamefont {R.}~\bibnamefont {Balbinot}}, \ and\
  \bibinfo {author} {\bibfnamefont {A.}~\bibnamefont {Fabbri}},\ }\href
  {\doibase 10.1088/1367-2630/10/10/103001} {\bibfield  {journal} {\bibinfo
  {journal} {New J. Phys.}\ }\textbf {\bibinfo {volume} {10}},\ \bibinfo
  {pages} {103001} (\bibinfo {year} {2008})},\ \Eprint
  {http://arxiv.org/abs/0803.0507} {0803.0507} \BibitemShut {NoStop}%
\bibitem [{\citenamefont {Steinhauer}(2016)}]{Steinhauer}%
  \BibitemOpen
  \bibfield  {author} {\bibinfo {author} {\bibfnamefont {J.}~\bibnamefont
  {Steinhauer}},\ }\href {\doibase 10.1038/nphys3863} {\bibfield  {journal}
  {\bibinfo  {journal} {Nat. Phys.}\ }\textbf {\bibinfo {volume} {12}},\
  \bibinfo {pages} {959–965} (\bibinfo {year} {2016})}\BibitemShut {NoStop}%
\bibitem [{\citenamefont {Son}(2002)}]{son2002}%
  \BibitemOpen
  \bibfield  {author} {\bibinfo {author} {\bibfnamefont {D.~T.}\ \bibnamefont
  {Son}},\ }\href {https://arxiv.org/abs/hep-ph/0204199} {} (\bibinfo {year}
  {2002}),\ \Eprint {http://arxiv.org/abs/hep-ph/0204199} {arXiv:hep-ph/0204199
  [hep-ph]} \BibitemShut {NoStop}%
\bibitem [{\citenamefont {Dalfovo}\ \emph {et~al.}(1999)\citenamefont
  {Dalfovo}, \citenamefont {Giorgini}, \citenamefont {Pitaevskii},\ and\
  \citenamefont {Stringari}}]{Dalfovo:1999zz}%
  \BibitemOpen
  \bibfield  {author} {\bibinfo {author} {\bibfnamefont {F.}~\bibnamefont
  {Dalfovo}}, \bibinfo {author} {\bibfnamefont {S.}~\bibnamefont {Giorgini}},
  \bibinfo {author} {\bibfnamefont {L.~P.}\ \bibnamefont {Pitaevskii}}, \ and\
  \bibinfo {author} {\bibfnamefont {S.}~\bibnamefont {Stringari}},\ }\href
  {\doibase 10.1103/RevModPhys.71.463} {\bibfield  {journal} {\bibinfo
  {journal} {Rev. Mod. Phys.}\ }\textbf {\bibinfo {volume} {71}},\ \bibinfo
  {pages} {463} (\bibinfo {year} {1999})},\ \Eprint
  {http://arxiv.org/abs/cond-mat/9806038} {arXiv:cond-mat/9806038} \BibitemShut
  {NoStop}%
\bibitem [{\citenamefont {Pitaevskii}\ and\ \citenamefont
  {Stringari}(2003)}]{pitstring}%
  \BibitemOpen
  \bibfield  {author} {\bibinfo {author} {\bibfnamefont {L.}~\bibnamefont
  {Pitaevskii}}\ and\ \bibinfo {author} {\bibfnamefont {S.}~\bibnamefont
  {Stringari}},\ }\href {https://books.google.it/books?id=rIobbOxC4j4C} {\emph
  {\bibinfo {title} {Bose-Einstein Condensation}}},\ International Series of
  Monographs on Physics\ (\bibinfo  {publisher} {Clarendon Press},\ \bibinfo
  {year} {2003})\BibitemShut {NoStop}%
\bibitem [{\citenamefont {Timmermans}(1998)}]{PhysRevLett.81.5718}%
  \BibitemOpen
  \bibfield  {author} {\bibinfo {author} {\bibfnamefont {E.}~\bibnamefont
  {Timmermans}},\ }\href {\doibase 10.1103/PhysRevLett.81.5718} {\bibfield
  {journal} {\bibinfo  {journal} {Phys. Rev. Lett.}\ }\textbf {\bibinfo
  {volume} {81}},\ \bibinfo {pages} {5718} (\bibinfo {year}
  {1998})}\BibitemShut {NoStop}%
\bibitem [{\citenamefont {Li}\ and\ \citenamefont {Pagels}(1971)}]{Li:1971vr}%
  \BibitemOpen
  \bibfield  {author} {\bibinfo {author} {\bibfnamefont {L.-F.}\ \bibnamefont
  {Li}}\ and\ \bibinfo {author} {\bibfnamefont {H.}~\bibnamefont {Pagels}},\
  }\href {\doibase 10.1103/PhysRevLett.26.1204} {\bibfield  {journal} {\bibinfo
   {journal} {Phys. Rev. Lett.}\ }\textbf {\bibinfo {volume} {26}},\ \bibinfo
  {pages} {1204} (\bibinfo {year} {1971})}\BibitemShut {NoStop}%
\bibitem [{\citenamefont {Son}\ and\ \citenamefont
  {Stephanov}(2002{\natexlab{b}})}]{PhysRevA.65.063621}%
  \BibitemOpen
  \bibfield  {author} {\bibinfo {author} {\bibfnamefont {D.~T.}\ \bibnamefont
  {Son}}\ and\ \bibinfo {author} {\bibfnamefont {M.~A.}\ \bibnamefont
  {Stephanov}},\ }\href {\doibase 10.1103/PhysRevA.65.063621} {\bibfield
  {journal} {\bibinfo  {journal} {Phys. Rev. A}\ }\textbf {\bibinfo {volume}
  {65}},\ \bibinfo {pages} {063621} (\bibinfo {year}
  {2002}{\natexlab{b}})}\BibitemShut {NoStop}%
\bibitem [{\citenamefont {Qu}\ \emph {et~al.}(2017)\citenamefont {Qu},
  \citenamefont {Tylutki}, \citenamefont {Stringari},\ and\ \citenamefont
  {Pitaevskii}}]{PhysRevA.95.033614}%
  \BibitemOpen
  \bibfield  {author} {\bibinfo {author} {\bibfnamefont {C.}~\bibnamefont
  {Qu}}, \bibinfo {author} {\bibfnamefont {M.}~\bibnamefont {Tylutki}},
  \bibinfo {author} {\bibfnamefont {S.}~\bibnamefont {Stringari}}, \ and\
  \bibinfo {author} {\bibfnamefont {L.~P.}\ \bibnamefont {Pitaevskii}},\ }\href
  {\doibase 10.1103/PhysRevA.95.033614} {\bibfield  {journal} {\bibinfo
  {journal} {Phys. Rev. A}\ }\textbf {\bibinfo {volume} {95}},\ \bibinfo
  {pages} {033614} (\bibinfo {year} {2017})}\BibitemShut {NoStop}%
\bibitem [{\citenamefont {Takayama}\ and\ \citenamefont
  {Oka}(1993)}]{Takayama:1992eu}%
  \BibitemOpen
  \bibfield  {author} {\bibinfo {author} {\bibfnamefont {K.}~\bibnamefont
  {Takayama}}\ and\ \bibinfo {author} {\bibfnamefont {M.}~\bibnamefont {Oka}},\
  }\href {\doibase 10.1016/0375-9474(93)90270-8} {\bibfield  {journal}
  {\bibinfo  {journal} {Nucl. Phys. A}\ }\textbf {\bibinfo {volume} {551}},\
  \bibinfo {pages} {637} (\bibinfo {year} {1993})}\BibitemShut {NoStop}%
\bibitem [{\citenamefont {Weinfurtner}\ \emph {et~al.}(2007)\citenamefont
  {Weinfurtner}, \citenamefont {Liberati},\ and\ \citenamefont
  {Visser}}]{Weinfurtner:2006wt}%
  \BibitemOpen
  \bibfield  {author} {\bibinfo {author} {\bibfnamefont {S.}~\bibnamefont
  {Weinfurtner}}, \bibinfo {author} {\bibfnamefont {S.}~\bibnamefont
  {Liberati}}, \ and\ \bibinfo {author} {\bibfnamefont {M.}~\bibnamefont
  {Visser}},\ }\href {\doibase 10.1007/3-540-70859-6_6} {\bibfield  {journal}
  {\bibinfo  {journal} {Lect. Notes Phys.}\ }\textbf {\bibinfo {volume}
  {718}},\ \bibinfo {pages} {115} (\bibinfo {year} {2007})},\ \Eprint
  {http://arxiv.org/abs/gr-qc/0605121} {arXiv:gr-qc/0605121} \BibitemShut
  {NoStop}%
\bibitem [{\citenamefont {Parikh}\ and\ \citenamefont
  {Wilczek}(2000)}]{Parikh:1999mf}%
  \BibitemOpen
  \bibfield  {author} {\bibinfo {author} {\bibfnamefont {M.~K.}\ \bibnamefont
  {Parikh}}\ and\ \bibinfo {author} {\bibfnamefont {F.}~\bibnamefont
  {Wilczek}},\ }\href {\doibase 10.1103/PhysRevLett.85.5042} {\bibfield
  {journal} {\bibinfo  {journal} {Phys. Rev. Lett.}\ }\textbf {\bibinfo
  {volume} {85}},\ \bibinfo {pages} {5042} (\bibinfo {year} {2000})},\ \Eprint
  {http://arxiv.org/abs/hep-th/9907001} {arXiv:hep-th/9907001} \BibitemShut
  {NoStop}%
\bibitem [{\citenamefont {Mu\~noz~de Nova}\ \emph {et~al.}(2019)\citenamefont
  {Mu\~noz~de Nova}, \citenamefont {Golubkov}, \citenamefont {Kolobov},\ and\
  \citenamefont {Steinhauer}}]{MunozdeNova:2018fxv}%
  \BibitemOpen
  \bibfield  {author} {\bibinfo {author} {\bibfnamefont {J.~R.}\ \bibnamefont
  {Mu\~noz~de Nova}}, \bibinfo {author} {\bibfnamefont {K.}~\bibnamefont
  {Golubkov}}, \bibinfo {author} {\bibfnamefont {V.~I.}\ \bibnamefont
  {Kolobov}}, \ and\ \bibinfo {author} {\bibfnamefont {J.}~\bibnamefont
  {Steinhauer}},\ }\href {\doibase 10.1038/s41586-019-1241-0} {\bibfield
  {journal} {\bibinfo  {journal} {Nature}\ }\textbf {\bibinfo {volume} {569}},\
  \bibinfo {pages} {688} (\bibinfo {year} {2019})},\ \Eprint
  {http://arxiv.org/abs/1809.00913} {arXiv:1809.00913 [gr-qc]} \BibitemShut
  {NoStop}%
\bibitem [{\citenamefont {Mannarelli}\ \emph
  {et~al.}(2021{\natexlab{a}})\citenamefont {Mannarelli}, \citenamefont
  {Grasso}, \citenamefont {Trabucco},\ and\ \citenamefont
  {Chiofalo}}]{Mannarelli:2020ebs}%
  \BibitemOpen
  \bibfield  {author} {\bibinfo {author} {\bibfnamefont {M.}~\bibnamefont
  {Mannarelli}}, \bibinfo {author} {\bibfnamefont {D.}~\bibnamefont {Grasso}},
  \bibinfo {author} {\bibfnamefont {S.}~\bibnamefont {Trabucco}}, \ and\
  \bibinfo {author} {\bibfnamefont {M.~L.}\ \bibnamefont {Chiofalo}},\ }\href
  {\doibase 10.1103/PhysRevD.103.076001} {\bibfield  {journal} {\bibinfo
  {journal} {Phys. Rev. D}\ }\textbf {\bibinfo {volume} {103}},\ \bibinfo
  {pages} {076001} (\bibinfo {year} {2021}{\natexlab{a}})},\ \Eprint
  {http://arxiv.org/abs/2011.00019} {arXiv:2011.00019 [gr-qc]} \BibitemShut
  {NoStop}%
\bibitem [{\citenamefont {Mannarelli}\ \emph
  {et~al.}(2021{\natexlab{b}})\citenamefont {Mannarelli}, \citenamefont
  {Grasso}, \citenamefont {Trabucco},\ and\ \citenamefont
  {Chiofalo}}]{Mannarelli:2021olc}%
  \BibitemOpen
  \bibfield  {author} {\bibinfo {author} {\bibfnamefont {M.}~\bibnamefont
  {Mannarelli}}, \bibinfo {author} {\bibfnamefont {D.}~\bibnamefont {Grasso}},
  \bibinfo {author} {\bibfnamefont {S.}~\bibnamefont {Trabucco}}, \ and\
  \bibinfo {author} {\bibfnamefont {M.~L.}\ \bibnamefont {Chiofalo}},\
  }\href@noop {} {\  (\bibinfo {year} {2021}{\natexlab{b}})},\ \Eprint
  {http://arxiv.org/abs/2109.11831} {arXiv:2109.11831 [gr-qc]} \BibitemShut
  {NoStop}%
\bibitem [{\citenamefont {Le~Bellac}(2000)}]{bellac2000thermal}%
  \BibitemOpen
  \bibfield  {author} {\bibinfo {author} {\bibfnamefont {M.}~\bibnamefont
  {Le~Bellac}},\ }\href {http://books.google.it/books?id=00\_x6GR8GXoC} {\emph
  {\bibinfo {title} {Thermal Field Theory}}},\ Cambridge Monographs on
  Mathematical Physics\ (\bibinfo  {publisher} {Cambridge University Press},\
  \bibinfo {year} {2000})\BibitemShut {NoStop}%
\bibitem [{\citenamefont {{Khalatnikov}}(1965)}]{Khalatnikov}%
  \BibitemOpen
  \bibfield  {author} {\bibinfo {author} {\bibfnamefont {I.~M.}\ \bibnamefont
  {{Khalatnikov}}},\ }\href@noop {} {\emph {\bibinfo {title} {{Introduction to
  the theory of superfluidity}}}}\ (\bibinfo  {publisher} {W. A. Benjamin, New
  York},\ \bibinfo {year} {1965})\BibitemShut {NoStop}%
\bibitem [{\citenamefont {Hartke}\ \emph {et~al.}(2020)\citenamefont {Hartke},
  \citenamefont {Oreg}, \citenamefont {Jia},\ and\ \citenamefont
  {Zwierlein}}]{PhysRevLett.125.113601}%
  \BibitemOpen
  \bibfield  {author} {\bibinfo {author} {\bibfnamefont {T.}~\bibnamefont
  {Hartke}}, \bibinfo {author} {\bibfnamefont {B.}~\bibnamefont {Oreg}},
  \bibinfo {author} {\bibfnamefont {N.}~\bibnamefont {Jia}}, \ and\ \bibinfo
  {author} {\bibfnamefont {M.}~\bibnamefont {Zwierlein}},\ }\href {\doibase
  10.1103/PhysRevLett.125.113601} {\bibfield  {journal} {\bibinfo  {journal}
  {Phys. Rev. Lett.}\ }\textbf {\bibinfo {volume} {125}},\ \bibinfo {pages}
  {113601} (\bibinfo {year} {2020})}\BibitemShut {NoStop}%
\bibitem [{\citenamefont {Joseph}\ \emph {et~al.}(2015)\citenamefont {Joseph},
  \citenamefont {Elliott},\ and\ \citenamefont
  {Thomas}}]{PhysRevLett.115.020401}%
  \BibitemOpen
  \bibfield  {author} {\bibinfo {author} {\bibfnamefont {J.~A.}\ \bibnamefont
  {Joseph}}, \bibinfo {author} {\bibfnamefont {E.}~\bibnamefont {Elliott}}, \
  and\ \bibinfo {author} {\bibfnamefont {J.~E.}\ \bibnamefont {Thomas}},\
  }\href {\doibase 10.1103/PhysRevLett.115.020401} {\bibfield  {journal}
  {\bibinfo  {journal} {Phys. Rev. Lett.}\ }\textbf {\bibinfo {volume} {115}},\
  \bibinfo {pages} {020401} (\bibinfo {year} {2015})}\BibitemShut {NoStop}%
\end{thebibliography}
\end{document}